\documentclass{jfm}
\usepackage{graphicx}
\usepackage{epstopdf, epsfig}
\usepackage[usenames, dvipsnames]{color}
\usepackage{subfigure}
\usepackage{mathtools}
\usepackage{verbatim}

\title{SPOD and resolvent analysis of near-wall coherent structures in turbulent pipe flows}

\author{Leandra I. Abreu\aff{1,2}
	\corresp{\email{leandra.abreu@unesp.br}},
	Andr\'{e} V. G. Cavalieri\aff{1},
	Philipp Schlatter\aff{3},
	Ricardo Vinuesa\aff{3}
	\and Dan S. Henningson\aff{3}}

\affiliation{\aff{1}Divis\~{a}o de Engenharia Aeron\'{a}utica, Instituto Tecnol\'{o}gico de Aeron\'{a}utica (ITA), 12228-900, S\~{a}o Jos\'{e} dos Campos, SP, Brazil
	\aff{2}Universidade Estadual Paulista (UNESP), 13876-750, C\^{a}mpus de S\~{a}o Jo\~{a}o da Boa Vista, SP, Brazil
	\aff{3}Linn\'{e} FLOW Centre, KTH Mechanics, Stockholm, SE-100-44, Sweden}

\begin{document}
\maketitle

\begin{abstract}
	
	Direct numerical simulations, performed with a high-order spectral-element method, are used to study coherent structures in turbulent pipe flow at friction Reynolds numbers $Re_{\tau} = 180$ and $550$. The database was analysed using spectral proper orthogonal decomposition (SPOD) to identify energetically dominant coherent structures, most of which turn out to be streaks and quasi-streamwise vortices. To understand how such structures can be modelled, the linear flow responses to harmonic forcing were computed using the singular value decomposition of the resolvent operator, using the mean field as a base flow. The SPOD and resolvent analysis were calculated for several combinations of frequencies and wavenumbers, allowing to map out the similarities between SPOD modes and optimal responses for a wide range of relevant scales in turbulent pipe flows. In order to explore physical reasons behind the agreement between both methods, an indicator of lift-up mechanism in the resolvent analysis was introduced, activated when optimal forcing represents quasi-streamwise vortices and associated response corresponds to streaks. Good agreement between leading SPOD and resolvent modes is observed in a large region of parameter space. In this region, a significant gain separation is found in resolvent analysis, which may be attributed to the strong amplification associated with the lift-up mechanism. For both Reynolds numbers, the observed concordances were generally for structures with large energy in the buffer layer. The results highlight resolvent analysis as a pertinent reduced-order model for coherent structures in wall-bounded turbulence, particularly for streamwise elongated structures corresponding to near-wall streamwise vortices and streaks.

\end{abstract}

\begin{keywords}
	SPOD, resolvent analysis, wall-bounded turbulence, coherent structures.
\end{keywords}

\section{Introduction}\label{sec:introduction}


In turbulent wall-bounded flows, such as straight pipes, channels and boundary layers, the most typically observed coherent structures are near-wall streaks, which are elongated structures in the streamwise direction. The near-wall streaks have shown to be extremely relevant for sustaining wall-bounded turbulence \citep{kline1967structure, gupta1971spatial, hamilton1995regeneration}. Such structures have regions of alternating low and high momentum located in the viscous and buffer layers with a characteristic spanwise spacing of about $100 \nu /u_{\tau}$, where $u_{\tau}$ is the friction velocity and $\nu$ is the kinematic viscosity of the fluid \citep{kline1967structure,smith1983characteristics,marusic2017scaling}. For higher wall-normal positions, in the logarithmic layer, larger structures are observed, with similar streaky shape, i.e. also elongated structures in the streamwise direction \citep{hutchins2007evidence}. For low and moderate Reynolds numbers, most of the turbulent production of a wall-bounded turbulent flow is located in the region close to the wall. In turn, for large Reynolds numbers, the turbulent production and dissipation contribution from the logarithmic layer could be as significant as the one from the buffer layer \citep{jimenez2018coherent}. In any case, the pursuit for more effective methods to model and characterise near-wall coherent structures is a very relevant problem for efficient modeling in the industry, and for a better understanding of wall-bounded turbulence dynamics.

For that matter, the use of statistical methods in flow databases can be convenient to identify coherent structures present in turbulent flow. A useful data-driven approach is proper orthogonal decomposition (POD), first introduced in the context of turbulence by \cite{lumley1967structure,lumley1970stochastic}. POD consists in finding among a zero-mean stochastic process, given by an ensemble of realizations of the flow field, a number of orthonormal basis functions, called POD modes, that maximise the mean square energy. The extension of POD to the frequency domain is referred to as spectral proper orthogonal decomposition (SPOD), terminology introduced by \cite{picard2000pressure}. The SPOD method involves decomposition of the cross-spectral density tensor (CSD) and leads to orthonormal modes oscillating at a specific frequency, which optimally represent the second-order space-time flow statistics \citep{towne2018spectral}. Each SPOD mode thus represents a structure that develops coherently in space and time. This is a useful method to explore the flow dynamics, since the SPOD modes dissociate flow phenomena at different time scales.

A strong connection between linearized models and coherent structures has been provided by resolvent analysis, also called input/output analysis. In this context, the non-linear terms from Navier--Stokes equations are treated as external forcing, and the component-wise input-output approach is applied. Following early studies of forced transitional flows \citep{trefethen1993hydrodynamic,farrell1993stochastic,jovanovic2005componentwise}, resolvent analysis considers flows in the frequency domain, and searches for forcings that lead to the most amplified flow responses. Such linearized responses from resolvent anlysis can often be related to results of hydrodynamic stability theory, with modes corresponding to instability waves or to non-modal mechanisms such as lift-up \citep{jovanovic2005componentwise}. Resolvent analysis has gained attention in the past decade for wall-bounded turbulent flows \citep{mckeon2010critical,hwang2010linear,mckeon2013experimental}. An important result is that if the forcing can be modeled as spatial white noise, a direct correspondence between SPOD and resolvent modes is expected \citep{towne2018spectral}. Moreover, for a flow with a dominant optimal forcing, leading to a gain much larger than that of suboptimal ones, the CSD will often be dominated by the leading response obtained in resolvent analysis \citep{beneddine2016conditions,cavalieri2019wave}. This indicates that SPOD is a pertinent signal-processing method for comparison of numerical or experimental databases with predictions from resolvent analysis.

In the present study a combined analysis of the flow, with SPOD on the one hand used to decompose turbulent fluctuations and resolvent analysis on the other hand as a theoretical framework, is explored for the case of the canonical turbulent pipe flow. The direct relation between both methods indicates a path to find a reduced-order model based on the linearized equations, where the focus is on the highest amplification between forcing and response. Considering a given base flow (usually the velocity profile averaged in time and homogeneous directions), the resolvent operator may be used to discern the dominant linear mechanisms in a turbulent flow, which facilitates analysis, and opens possibilities for flow estimation and control approaches aiming at the said mechanisms \citep{towne2020resolvent}. A number of previous studies regarding turbulent pipe flows have dealt with POD without frequency decomposition \citep{hellstrom2016self}, and resolvent analysis \citep{mckeon2010critical}, but the ability of the latter to model SPOD modes has not been addressed quantitatively by a thorough comparison involving the range of relevant wavenumbers and frequencies. Thus, this is the main goal of the present paper, namely to map out the similarities of SPOD and resolvent modes for a turbulent pipe flow.

This paper is organized as follows. In~\S\ref{sec:simulation}, the simulation database obtained by \cite{el2013direct} is briefly described. This section includes some results in terms of turbulence statistics and spectral analysis, with focus on the structures present in the near-wall region. Section~\S\ref{sec:liftup} briefly describes the lift-up mechanism present in shear flows, which is relevant for the ensuing analysis. Section~\S\ref{sec:methods} proceeds with a description of the methods of SPOD and resolvent analysis, and the relationship between both approaches. A detailed comparison between SPOD and resolvent modes is presented and discussed in section~\S\ref{sec:results}. The paper is completed with conclusions in~\S\ref{sec:conclusions}.

\section{Description of the employed database}
\label{sec:simulation}

The direct numerical simulation (DNS) database employed in this work was obtained by \cite{el2013direct}. The simulations were carried out for the fully developed turbulent flow inside a smooth, circular straight pipe. The pressure-driven incompressible flow of a viscous Newtonian fluid was considered, where the governing equations are the time-dependent Navier--Stokes equations. The code used to solve these equations is \texttt{Nek5000}, developed by \cite{fischer}, which is a computational fluid dynamics solver based on the spectral-element method. This specific discretisation method is characterised by spectral accuracy, favourable diffusion and dispersion properties and efficient parallelisation.


The simulation domain consists of a circular pipe with radius $R$ and length $25R$ with the pipe axis taken along the streamwise $x$-direction. In this study we consider two different Reynolds numbers, \textit{i.e.} $Re_{\tau}=180$ and $550$, where $Re_{\tau}$ is the friction Reynolds number based on $u_{\tau}$ and $R$. Snapshots of the flow quantities within the whole computational domain are saved in a database, with non-dimensional time interval of $\delta t = 4$ (in terms of bulk velocity and pipe radius); a total of $205$ snapshots for $Re_\tau = 180$, and $\delta t = 1$ and $260$ snapshots for $Re_\tau = 550$ were considered for the present analysis. A streamwise pressure gradient drives the flow in the streamwise direction. Additional details of the numerical setup can be found in the work by \cite{el2013direct}.

\begin{figure}
	\centering
	\subfigure[$Re_\tau = 180$]{\includegraphics[width=0.45\textwidth,trim = 30 0 30 0,clip]{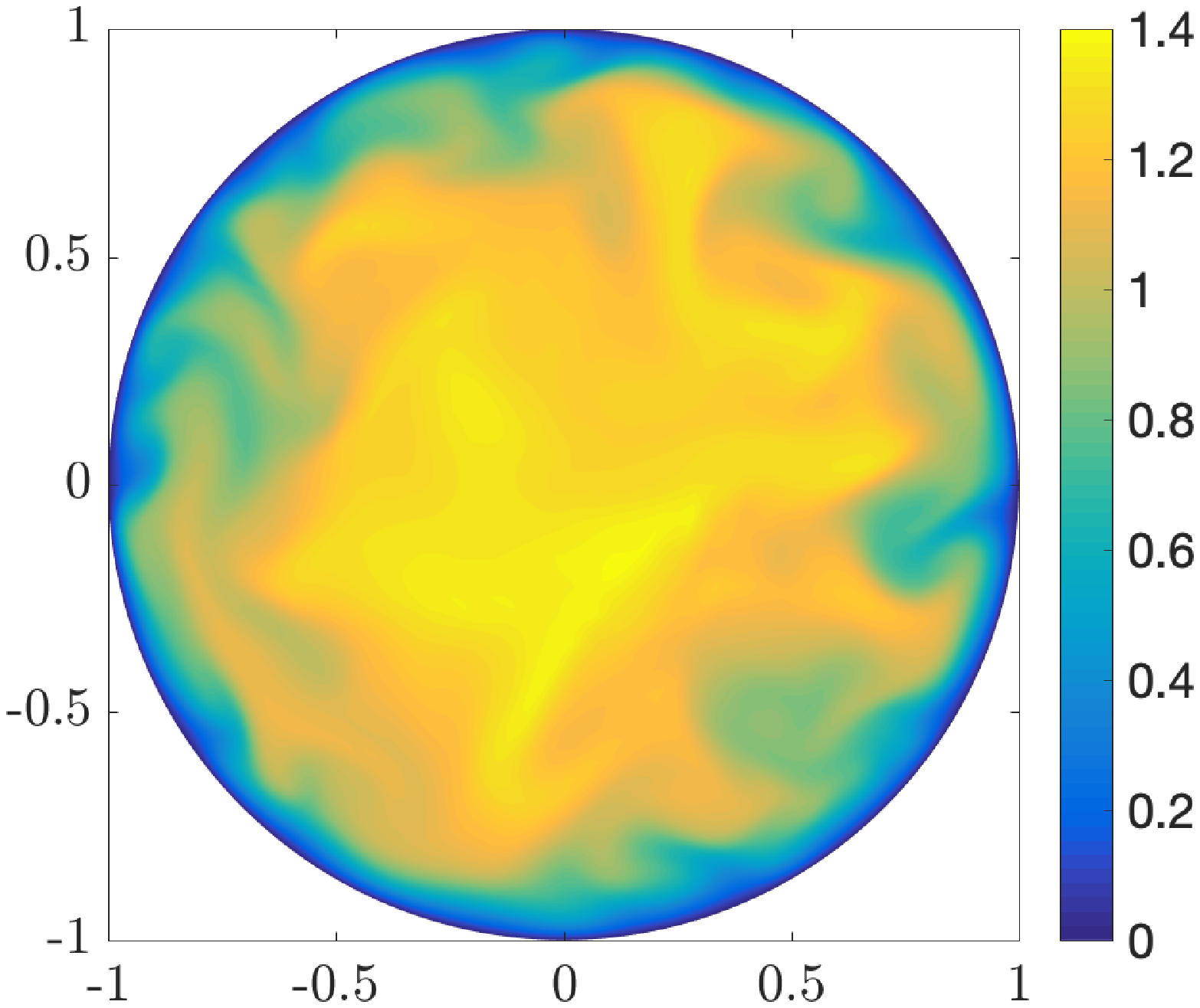}}
	\subfigure[$Re_\tau = 550$]{\includegraphics[width=0.45\textwidth,trim = 30 0 30 0,clip]{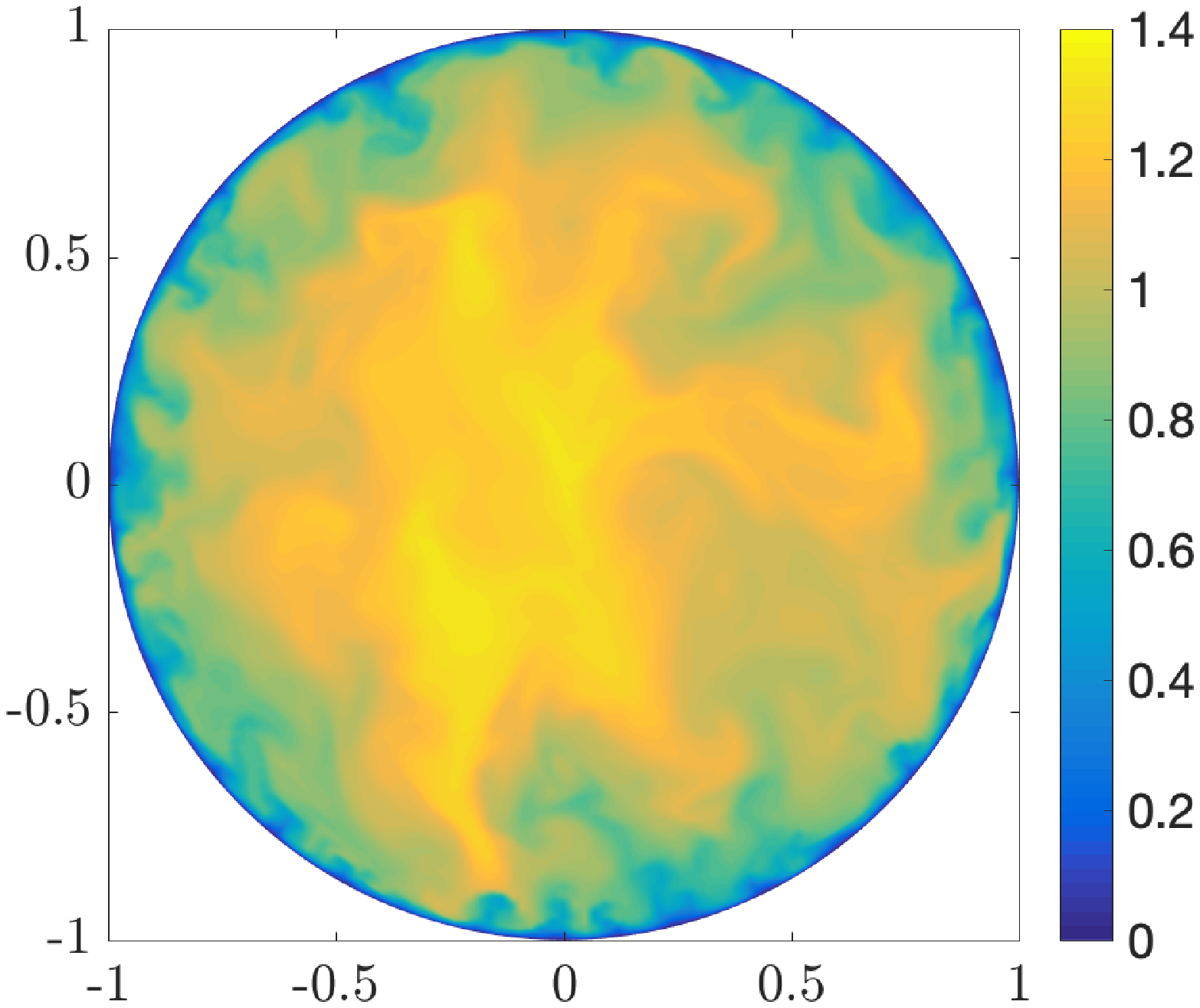}}
	\caption{Instantaneous streamwise velocity component for both friction Reynolds numbers. The axis tick labels are scaled with the pipe radius.}
	\label{fig:simulation}
\end{figure}

The resultant database from the DNS is in cylindrical coordinate system and the origin is located on the axis of the pipe. The flow is periodic in the streamwise direction $x$. The velocity vector is given by $q = [u,v,w]$ in streamwise, radial and azimuthal coordinates $[x,r,\theta]$, respectively, where $r$ is the radial coordinate. Thus the non-dimensional wall distance can be obtained by $(1-r)$. Figures \ref{fig:simulation} (a) and (b) show the instantaneous streamwise velocity component $u$ from the simulation for both $Re_\tau = 180$ and $550$, respectively, where the axis tick labels are scaled with the pipe radius.

The inner velocity scaling is defined by the friction velocity $u_{\tau}$, and the length by the viscous length scale $\nu / u_{\tau}$, and is denoted by the superscript $+$. In the following we will use $z^+ = r^+ \theta$ as a pseudo-spanwise coordinate for simple comparison with results for planar flows such as boundary layers and channels. Thus the azimuthal wavelength $\lambda_\theta$ can be associated with a pseudo-spanwise wavenumber $\lambda_{z}^{+} = r^+ \lambda_\theta$. The resolvent and SPOD modes are a function of the azimuthal wavenumber $m$, or wavelength $\lambda_\theta=2\pi/m$, and thus cannot be expressed as a function of a single pseudo-spanwise wavelength $\lambda_z^+$. For such modes, we define $\tilde{\lambda}_z^+ = (Re_\tau - 15)\lambda_\theta$, such that $\tilde{\lambda}_z^+$ represents the pseudo-spanwise wavelength at a reference position in the buffer layer at a distance of $15$ viscous units from the wall. This allows for a comparison with similar structures found in other (planar) wall-bounded flows.

\subsection{Turbulence statistics and spectral analysis}
\label{sec:spectral}

\begin{figure}
	\centering
	\subfigure[]{\includegraphics[width=0.48\textwidth,trim = 0 0 -10 0,clip]{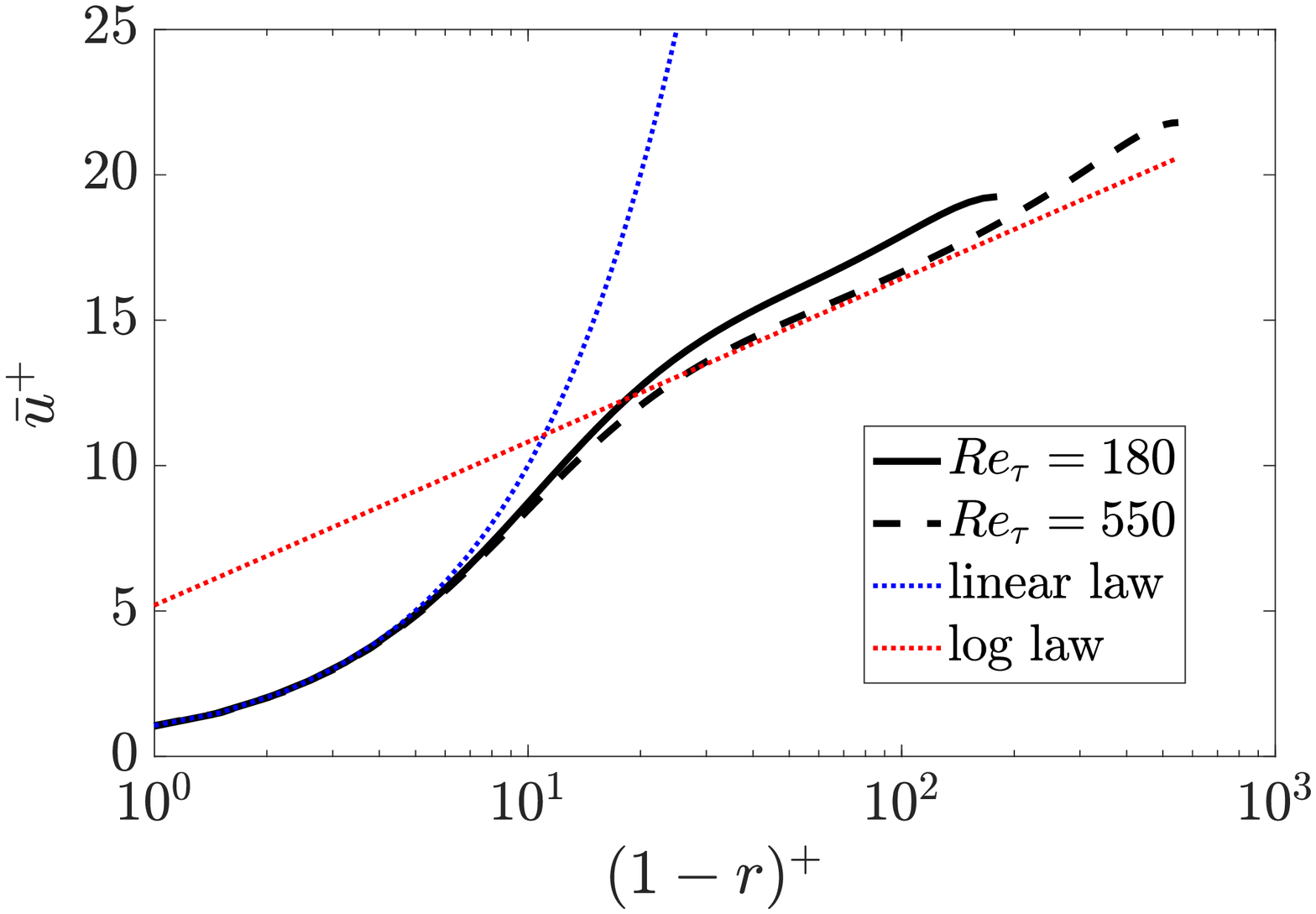}}
	\subfigure[]{\includegraphics[width=0.48\textwidth,trim = -10 0 0 0,clip]{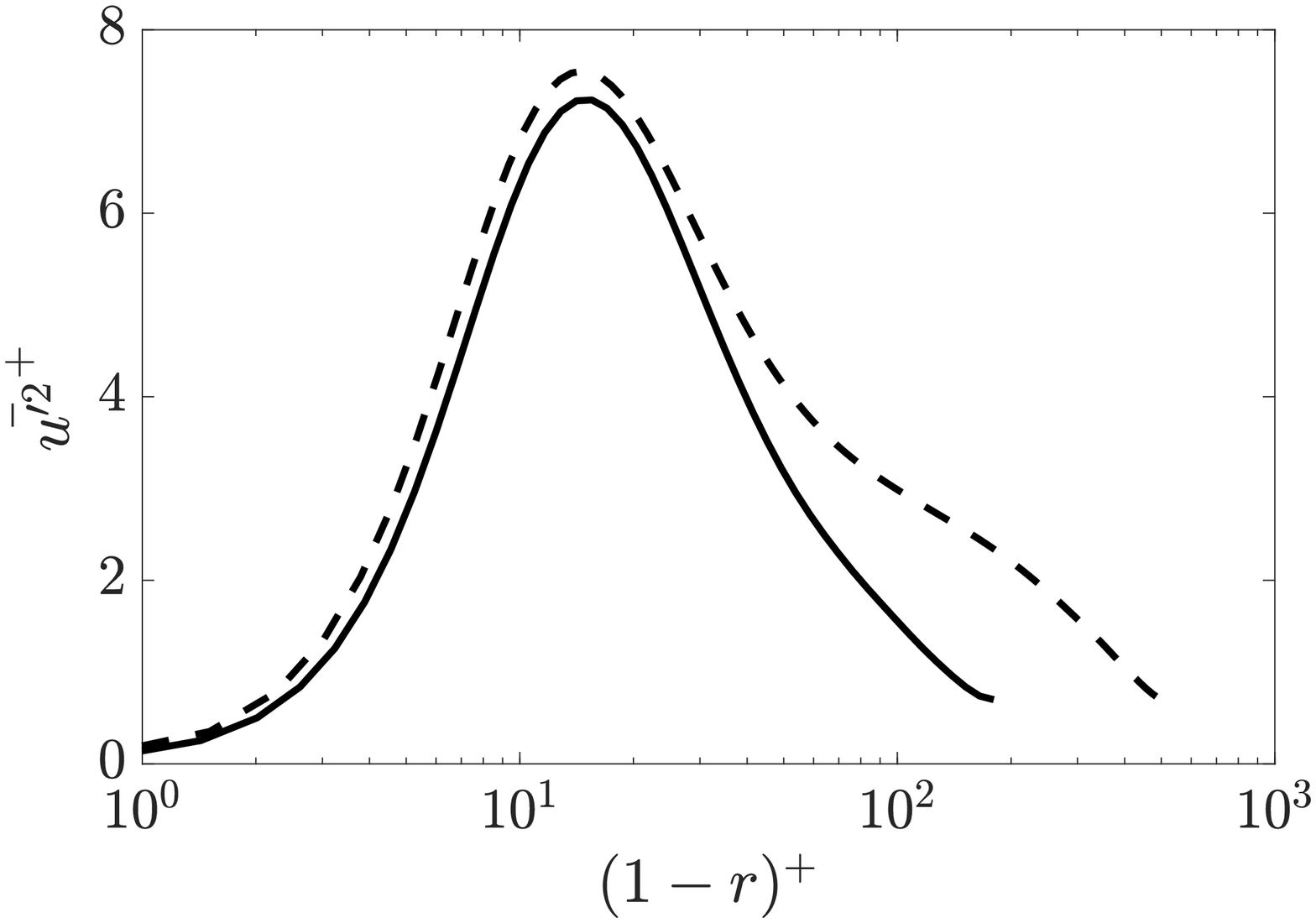}}
	\caption{(a) The mean flow and (b) streamwise velocity fluctuations at $Re_{\tau}=180$ (black solid line) and $550$ (black dashed line), both scaled in viscous units. In (a) the blue dashed line represents the linear law, $\bar{u}^+ = (1-r)^+$; and the red dashed line is the log law, $\bar{u}^+ = (1/\kappa) \ln(1-r)^+ + B$, where $\kappa = 0.41$ and $B=5.2$. }
	\label{fig:base_flow}
\end{figure}

Using standard Reynolds decomposition $u= \bar{u} + u'$, the mean streamwise velocity profiles $\bar{u}^+$ and the variance profile of the streamwise velocity fluctuations $\bar{u'^2}^{+}$ in inner scaling are shown in Figures \ref{fig:base_flow} (a) and (b) respectively, for $Re_{\tau}=180$ and $550$. The mean velocity profiles show the expected shape of wall-bounded turbulent flows when plotted as a function of wall distance where the variable $y^{+} = (1-r)^{+}$ denotes the inner-scaled wall distance. Variance profiles also have the expected pattern characteristic of wall-bounded turbulent flows, with a near-wall peak in the buffer layer at $(1-r)^{+} \approx 15$, which increases its magnitude with the Reynolds number, as expected \cite{eitel2014simulation}.

\begin{figure}
	\centering
	\subfigure[$Re_\tau=180$ at $(1-r)^{+} \approx 15$]{\includegraphics[width=0.48\textwidth,trim = 0 45 20 60,clip]{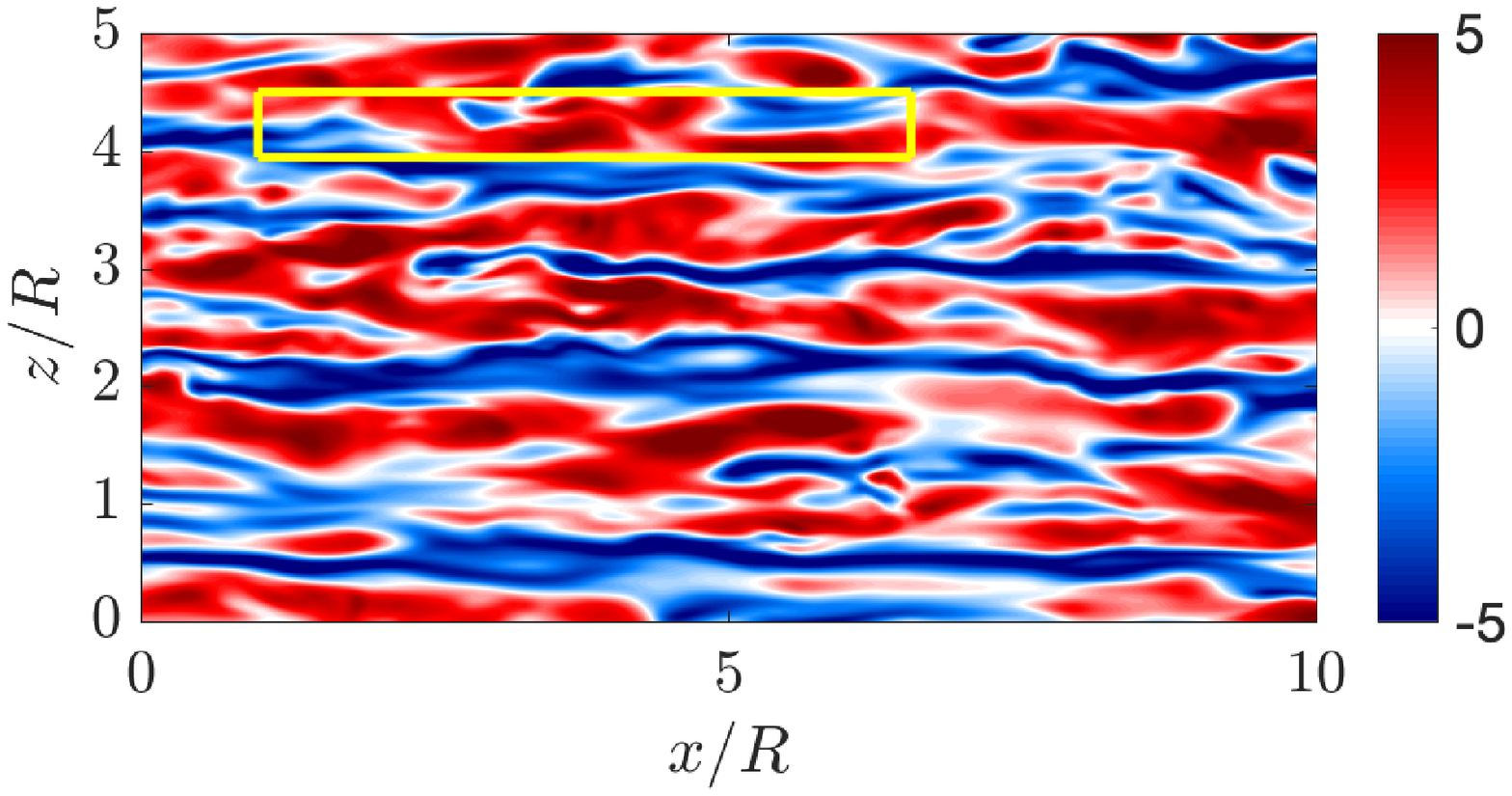}}
	\subfigure[$Re_\tau=550$ at $(1-r)^{+} \approx 15$]{\includegraphics[width=0.48\textwidth,trim = 0 45 20 60,clip]{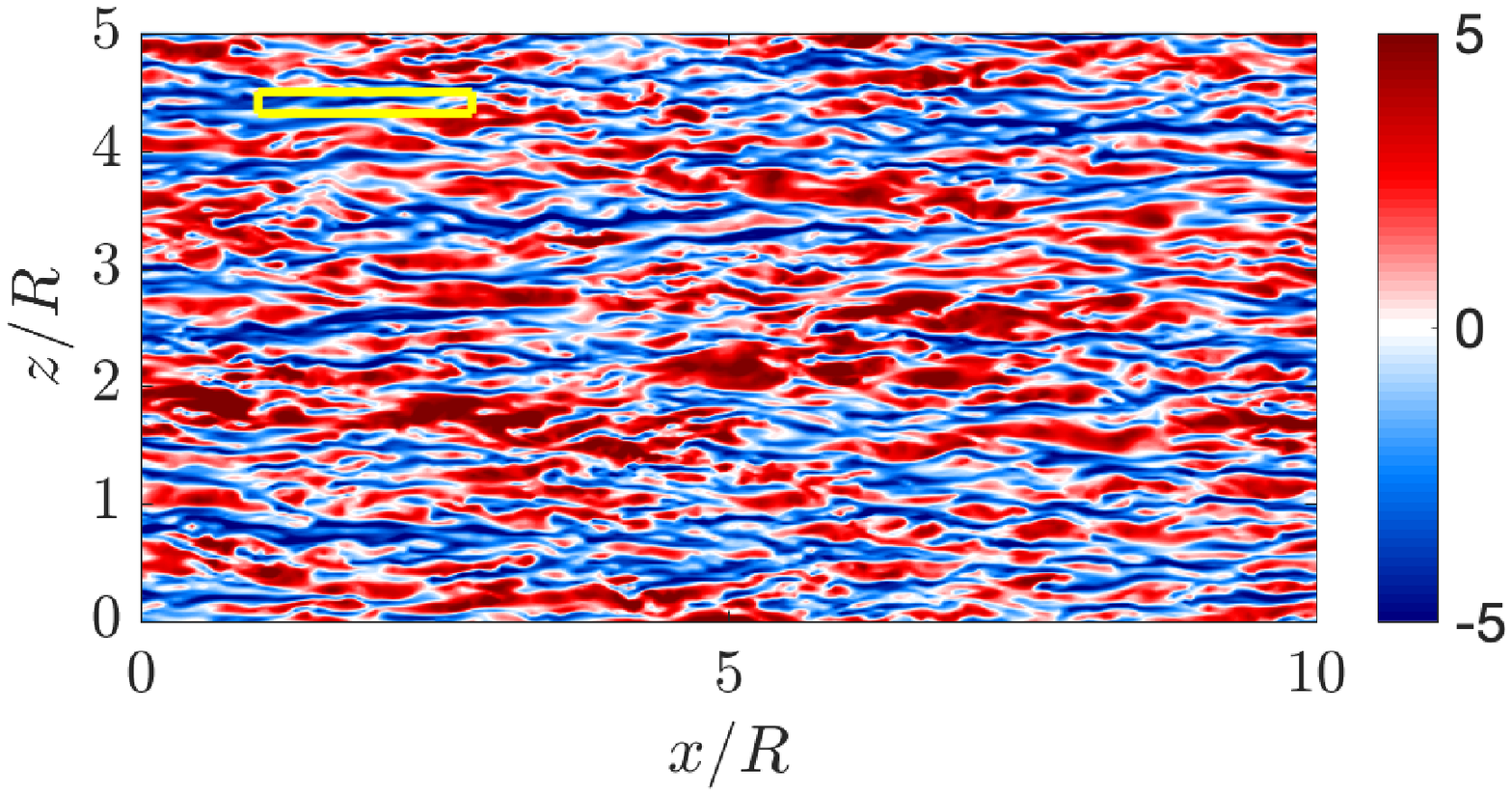}}
	\subfigure[$Re_\tau=180$ at $(1-r)^{+} \approx 100$]{\includegraphics[width=0.48\textwidth,trim = 0 90 20 120,clip]{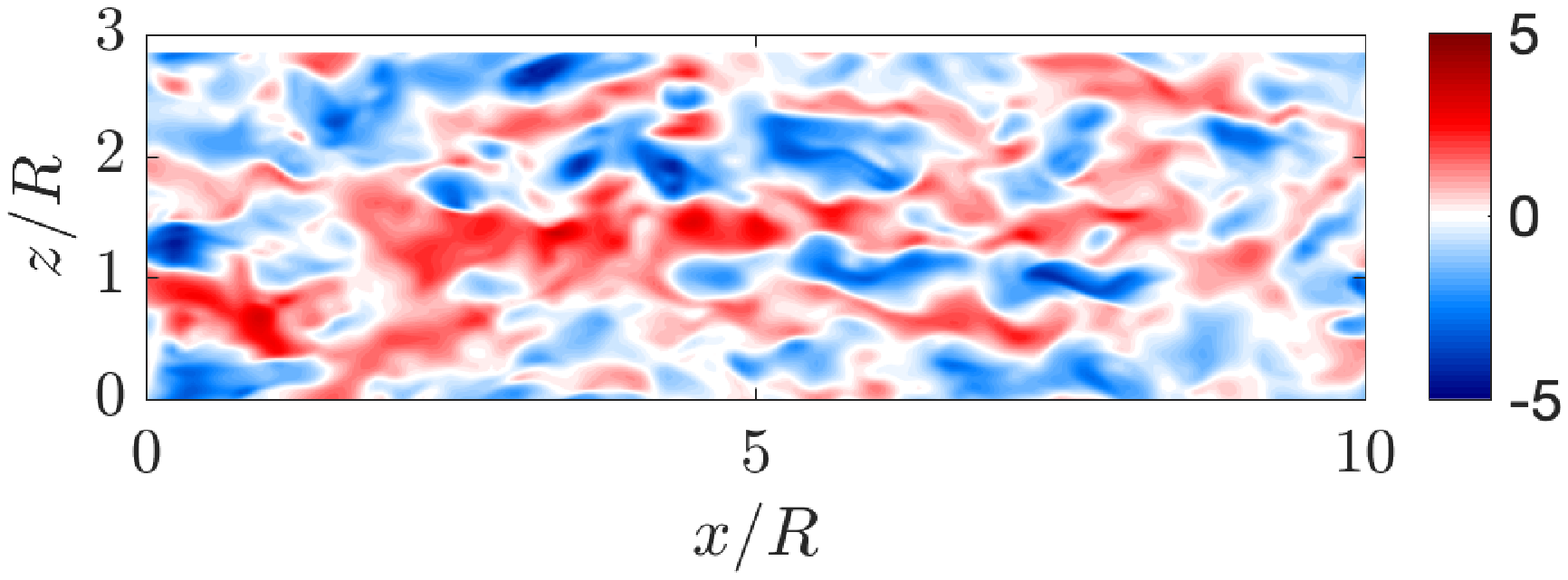}}
	\subfigure[$Re_\tau=550$ at $(1-r)^{+} \approx 100$]{\includegraphics[width=0.48\textwidth,trim = 0 90 20 120,clip]{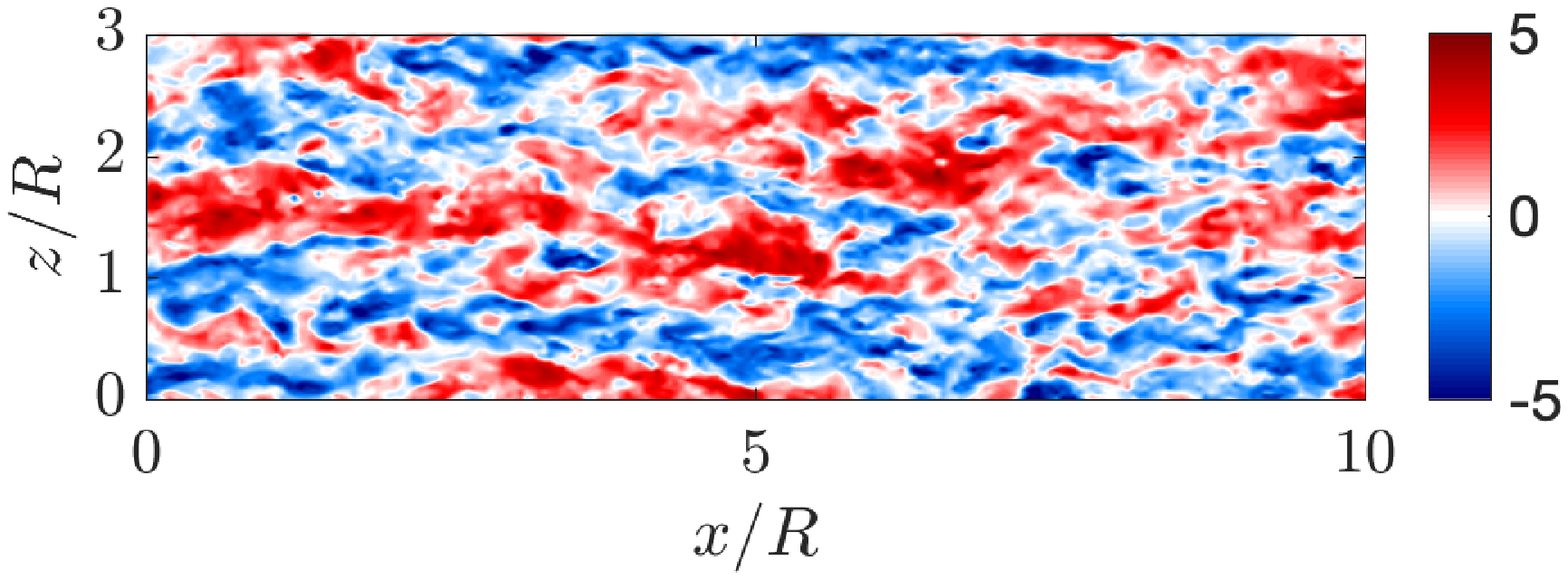}}

	\caption{Instantaneous streamwise velocity fluctuation ($u'^{+}$) field in the wall-parallel plane for both friction Reynolds numbers, where (a,b) show the results in the buffer layer $(1-r)^{+} \approx 15$, and the yellow rectangle represents a box with $(\lambda_{x}^{+},\lambda_{z}^{+}) \approx (1000,100)$; (c,d) show the results in the logarithmic layer $(1-r)^{+} \approx 100$.}
	\label{fig:streaks1}
\end{figure}

In order to visualise turbulent structures present in the buffer layer, Figures \ref{fig:streaks1} (a) and (b) show snapshots of the streamwise velocity fluctuations ($u'^{+}$) in a wall-parallel station at $(1-r)^{+} \approx 15$, for $Re_{\tau}=180$ and $550$, respectively. The same analysis is shown for the logarithmic layer in Figures \ref{fig:streaks1} (c) and (d) at $(1-r)^{+} \approx 100$, for $Re_{\tau}=180$ and $550$, respectively. These figures show that the dominant structures are elongated in the streamwise direction in all cases, with higher amplitudes near the wall in the buffer layer, in agreement with the near-wall structures found in the flow visualisations by \cite{kline1967structure}. Such long and narrow structures of the streamwise velocity component $u$ are the well-known streaks, which exhibit a range of sizes and are found between the near-wall region and the pipe core. The near-wall streaks have an amplitude peak at $(1-r)^{+} \approx 15$, in the buffer-layer, and have a characteristic peak for the axial and azimuthal wavelengths of $(\lambda_{x}^{+},\lambda_{z}^{+}) \approx (1000,100)$, represented by the yellow rectangle in Figures \ref{fig:streaks1} (a) and (b). In what follows we denote as \emph{streaky} structures, the structures elongated in the streamwise direction, with an aspect ratio such that $\lambda_{x}^{+} > 2 \lambda_{z}^{+}$ at least. Although this is a somewhat arbitrary choice, it will focus the analysis on the dominant elongated structures visible in Figure \ref{fig:streaks1}.


\begin{figure}
	\centering
	\subfigure[$Re_{\tau}=180$]{\includegraphics[width=0.48\textwidth,trim = 0 0 19 10,clip]{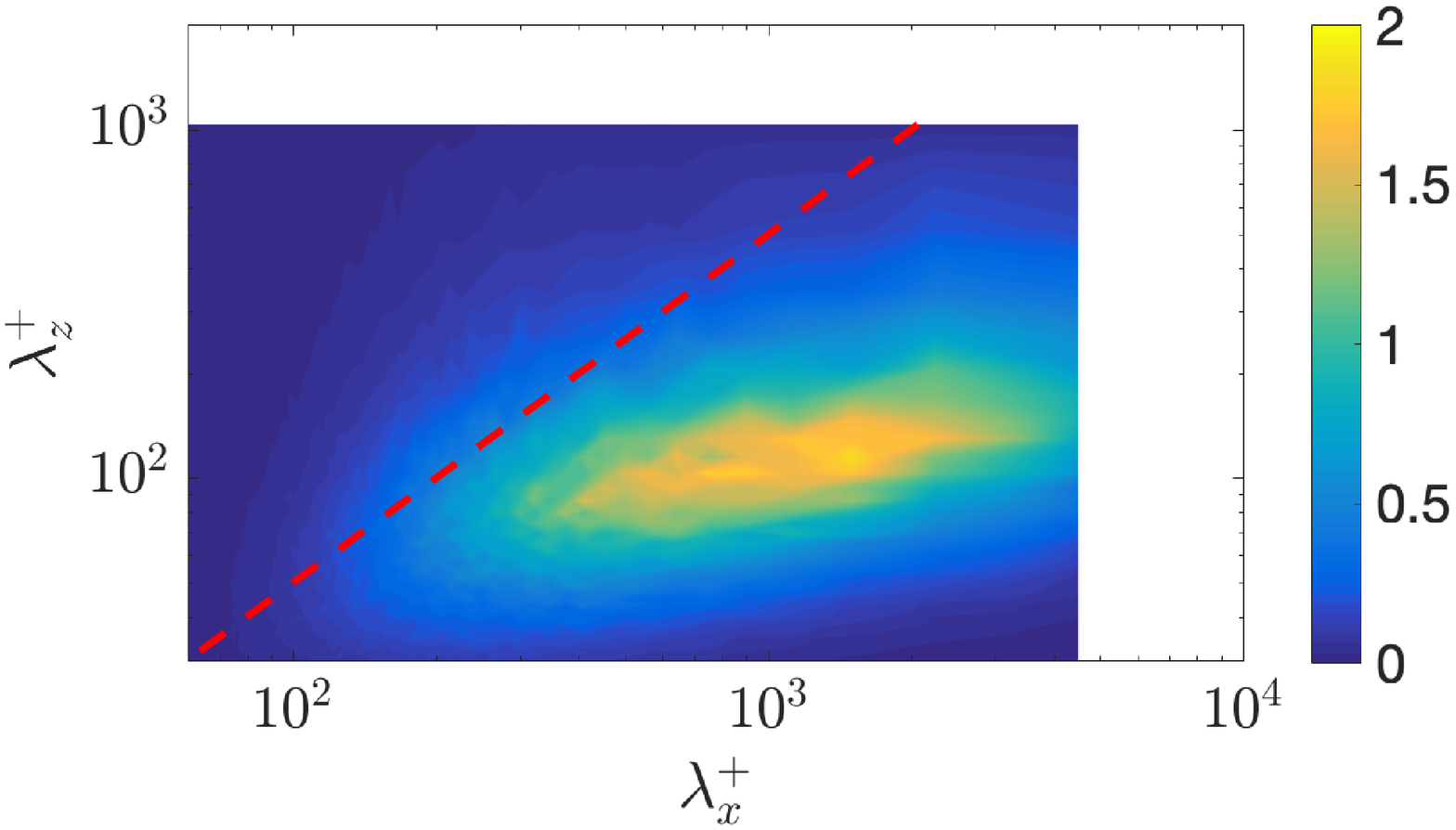}}
	\subfigure[$Re_{\tau}=550$]{\includegraphics[width=0.48\textwidth,trim = 0 0 19 10,clip]{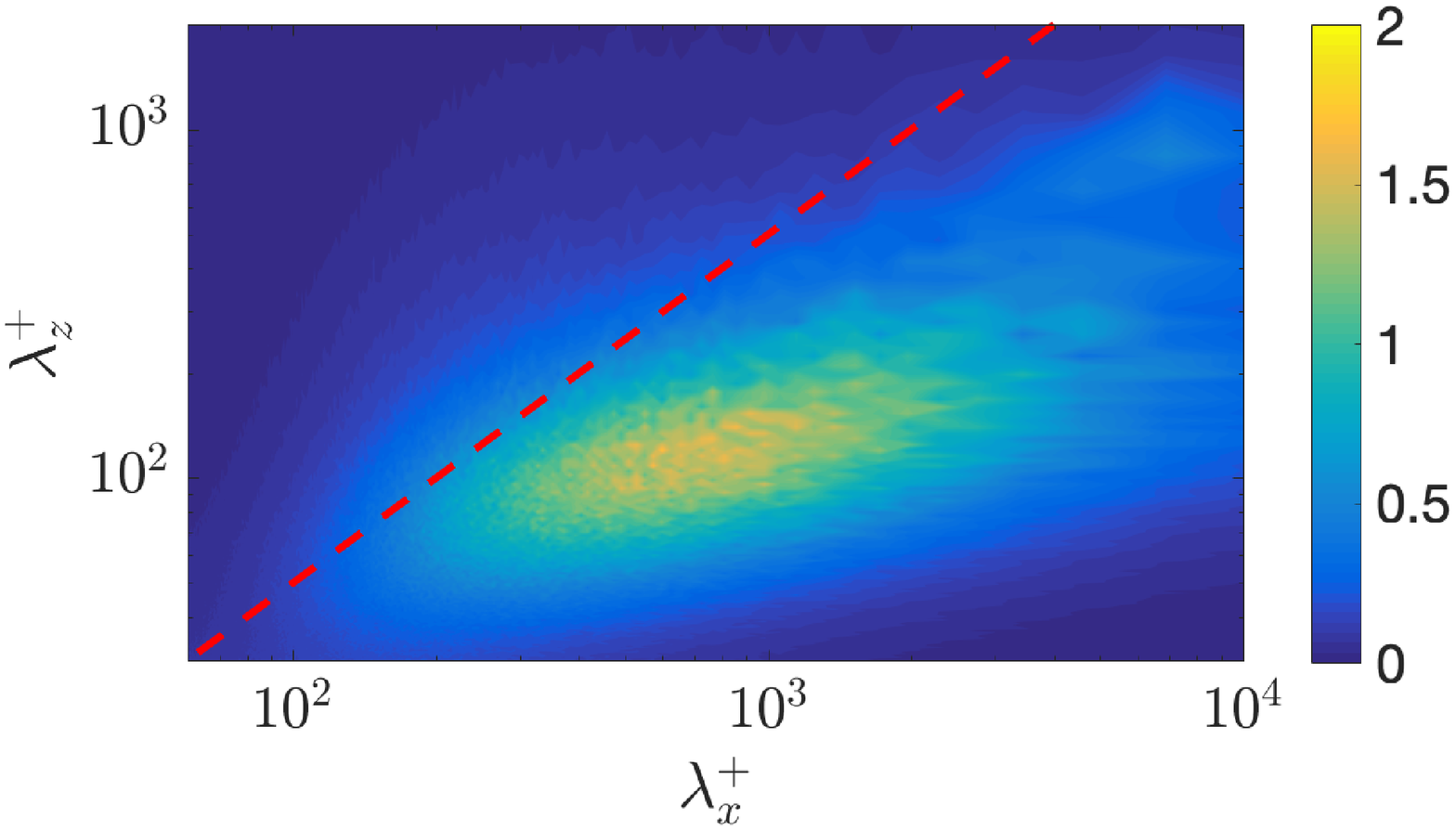}}
	\caption{Two-dimensional inner-scaled premultiplied power-spectral density of the streamwise velocity $k_{x} k_{z} E_{uu}^{+}$, at $(1-r)^{+} \approx 15$, for (a) $Re_{\tau}=180$ and (b) $Re_{\tau}=550$. The red dashed line represents $\lambda_{x}^{+} = 2 \lambda_{z}^{+}$.}
	\label{fig:energyspec}
\end{figure}

The two-dimensional inner-scaled premultiplied power-spectral density of streamwise velocity fluctuations $k_{x} k_{z} E_{uu}^{+}$ at $(1-r)^{+} \approx 15$ are shown in Figures \ref{fig:energyspec} (a) and (b) for $Re_\tau=180$ and $550$, respectively, where $k_{x}$ and $k_{z}$ refer respectively to streamwise and azimuthal wavenumbers. For both Reynolds numbers, the results in Figure \ref{fig:energyspec} show a highly energetic peak located in the near-wall region $(1-r)^{+} \approx 15$ for the wavelength combination $(\lambda_{x}^{+},\lambda_{z}^{+}) \approx (1000,100)$, which is representative of the signature of the near-wall cycle of streaks and quasi-streamwise vortices, which has been discuss at length in many studies across a range of Reynolds numbers and flow types (see for instance \cite{hoyas2006scaling,monty2009comparison,smits2011high}). Figure \ref{fig:energyspec} shows that most of the fluctuation energy is related to the aforementioned streaks, since the pre-multiplied spectrum has most of its content for $\lambda_{x}^{+} > 2 \lambda_{z}^{+}$ (below the red dashed line).

The visualizations shown in Figure \ref{fig:streaks1} and the spectra in Figure \ref{fig:energyspec} indicate that the turbulent pipe flow is dominated by near-wall streaks for the Reynolds numbers considered here. In the next section we review the lift-up mechanism, which is a key aspect in the dynamics of such streaky structures.

\section{The lift-up mechanism}\label{sec:liftup}

The lift-up mechanism was introduced in the work by \cite{ellingsen1975stability}, who identified a linear mechanism responsible for the amplification of fluctuations in shear flows, followed by the work by \cite{landahl1980note}. \cite{ellingsen1975stability} concluded that a finite disturbance independent of the streamwise coordinate leads to algebraic growth of disturbances in shear flows, even though the basic velocity does not possess any inflection point. The concept of lift-up effect has been recently explored in detail by \cite{brandt2014lift}.

We briefly outline here the derivations of the equations highlighting the lift-up effect. We consider a parallel velocity profile as $\mathbf{\bar{u}} = (\bar{u},\bar{v},\bar{w}) = (\bar{u}(y),0,0)$ where the overbar denotes averaging in the homogeneous directions and time. In its simplest derivation, the lift-up effect may be explored by considering Cartesian coordinates, but similar effects are obtained in polar coordinates \citep{ellingsen1975stability}. Thus, considering this flow as inviscid and incompressible, bounded by two parallel planes, subject to disturbances independent of the streamwise coordinate $x$, the equation for the streamwise component of the velocity fluctuations and for the streamwise vorticity, $\xi$, reduce respectively to

\begin{equation}
\frac{Du'}{Dt}=0; \\ \frac{D \xi}{Dt}=0.
\end{equation}

Introducing the streamfunction $\Psi$ for the cross-stream components:
\begin{equation}
v'=\frac{\partial \Psi}{\partial z}; \\ w'= - \frac{ \partial \Psi}{\partial y},
\end{equation}
and upon linearization, we obtain for the streamwise component:
\begin{equation}
\frac{\partial u'}{\partial t} + v' \frac{{\rm d} \bar{u}}{{\rm d} y} = 0,
\label{eq:stream}
\end{equation}
and for the cross-stream flow, i.e. $y$ and $z$ components, we obtain:
\begin{equation}
\frac{\partial \nabla^2 \Psi }{\partial t} = 0,
\label{eq:cross}
\end{equation}
where $\nabla^2$ is the two-dimensional Laplacian. We can see from equation (\ref{eq:cross}) that the cross-stream velocity components are independent of time, i.e. a streamwise-independent perturbation $v^\prime$ will not grow or decay in an inviscid flow. Equation (\ref{eq:stream}) can be integrated in time:
\begin{equation}
u^\prime = u'(0) - v' \frac{{\rm d} \bar{u}}{{\rm d} y} t,
\label{eq:ulin}
\end{equation}
to show that the perturbation $u'$ grows linearly in time, from which also the name of algebraic inviscid instability. An addition of viscous effects limits the algebraic growth in equation (\ref{eq:ulin}), which nonetheless may be of some orders of magnitude for higher Reynolds numbers \citep{brandt2014lift}.

The term $v'{\rm d}\bar{u}/{\rm d}y$ in equation (\ref{eq:stream}) is responsible for the lift-up effect, and represents the deformation of the mean velocity profile by the spanwise variations of $v'$. This is one of the terms responsible for non-orthogonal eigenvectors of the evolution operator \citep{jimenez2018coherent}. Thus, the streamwise vortices, $(v',w')$, lead to the formation of low- and high-momentum streaks, since lift-up acts most strongly on long narrow features. Lift-up works by moving low-velocity fluid from the wall upwards, creating low-velocity streaks $-u'$, and vice versa \citep{landahl1980note,brandt2014lift,jimenez2018coherent}.



The above analysis is based on the linearized equations for streamwise-independent disturbances. Non-linear effects can be considered in the resolvent framework, with non-linear terms considered as forcing, as described below. The lift-up effect is also present in this case \citep{jovanovic2005componentwise}, with optimal forcing dominated by $y$ and $z$ components shaped as a streamwise vortex \citep{hwang2010linear}, creating $v$ and $w$ components in the flow, streamwise vortices that lead to highliy amplified streaks of streamwise velocity. This effect plays an important role in the near-wall cycle described by \cite{hamilton1995regeneration,hall2010streamwise,farrell2012dynamics}, with a self-sustained cycle where streamwise vortices generate streaks, which, once high amplitudes are attained, break down due to instabilities. Subsequent non-linear interactions among streamwise oscillatory modes generated by streak instability lead to streamwise vortex regeneration, thus restarting the cycle. Due to the importance of lift-up to wall-bounded turbulent flows, we focus our analysis on the coherent structures involved in such mechanism, such as streaks and streamwise vortices.

\section{Methodologies}\label{sec:methods}
\subsection{Spectral proper orthogonal decomposition}

In the present study, SPOD follows the procedure outlined by \cite{towne2018spectral}. The method is applied to the velocity fluctuation components $u'$, $v'$ and $w'$ to characterise the turbulent kinetic energy. We first apply a fast Fourier transform (FFT) to the velocity fields in the homogeneous directions $x$ and $z$ to obtain the field for specific wavenumbers $k_{x}$ and $m$, respectively. We also perform a FFT to the velocity fields in time to obtain the field for a specific frequency $\omega$, so $\mathbf{\hat{q}}=\mathbf{\hat{q}}(k_x,r,m,\omega)$, where hats denote Fourier-transformed quantities and $\mathbf{\hat{q}} = [\mathbf{\hat{u}}; \mathbf{\hat{v}}; \mathbf{\hat{w}}]$ are the state variables; we then apply the SPOD to this transformed field, which is equivalent to solving the integral equation:
\begin{equation}
\int_{\mathbf{r'}} \mathbf{C}(\mathbf{r},\mathbf{r'},\omega) \boldsymbol{\Psi}(\mathbf{r'},\omega) \mathbf{r'} {\rm d} \mathbf{r'} = \lambda(\omega) \boldsymbol{\Psi}(\mathbf{r},\omega),
\end{equation}
where $\boldsymbol{\Psi}$ are basis functions, or SPOD modes, $\lambda$ is the corresponding eigenvalue and $\mathbf{C}$ is the two-point cross-spectral density between the three velocity components, whose dimension is $3N_r x 3N_r$, where $N_r$ is the number of points in the radial coordinates. Note that $\mathbf{C}$ is Hermitian, and thus eigenvalues are real and eigenfunctions are orthogonal. The decreasing ordering of the eigenvalues ensures that the most energetic modes in terms of kinetic energy are the first ones.
Since the number of grid points is high, we use the snapshot method, originally introduced by \cite{sirovich1987turbulence}, which is more effective to compute the SPOD numerically, as presented by \cite{towne2018spectral} and \cite{schmidt2020guide}.


The short-time fast Fourier transform (FFTs) required to solve the SPOD was taken considering blocks containing 32 snapshots with $75\%$ overlap, which leads to $22$ and $29$ blocks for $Re_{\tau}=180$ and $550$, respectively. SPOD was also evaluated using blocks containing 48 snapshots with $50\%$ and $75\%$ of overlap, and the changes in the results were not significant. Changes in leading eigenvalues did not exceed $0.1\%$ in most of the frequency/wavenumber combinations, indicating that the SPOD results are reliable and can be meaningfully analysed.

\begin{figure}
	\centering
	\subfigure[$Re_{\tau}=180$]{\includegraphics[width=0.4\textwidth,trim = 0 0 0 0,clip]{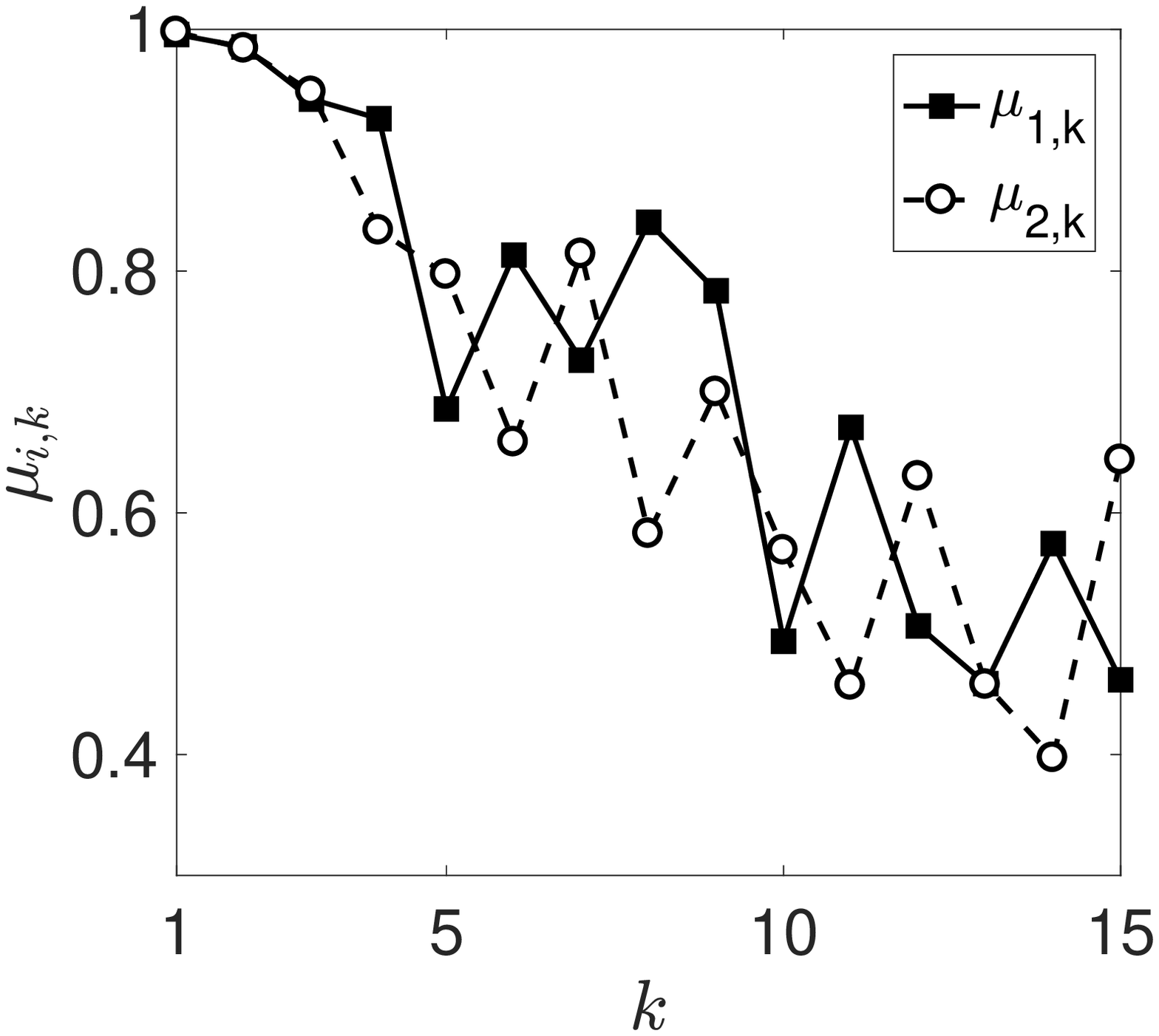}}
	\subfigure[$Re_{\tau}=550$]{\includegraphics[width=0.4\textwidth,trim = 0 0 0 0,clip]{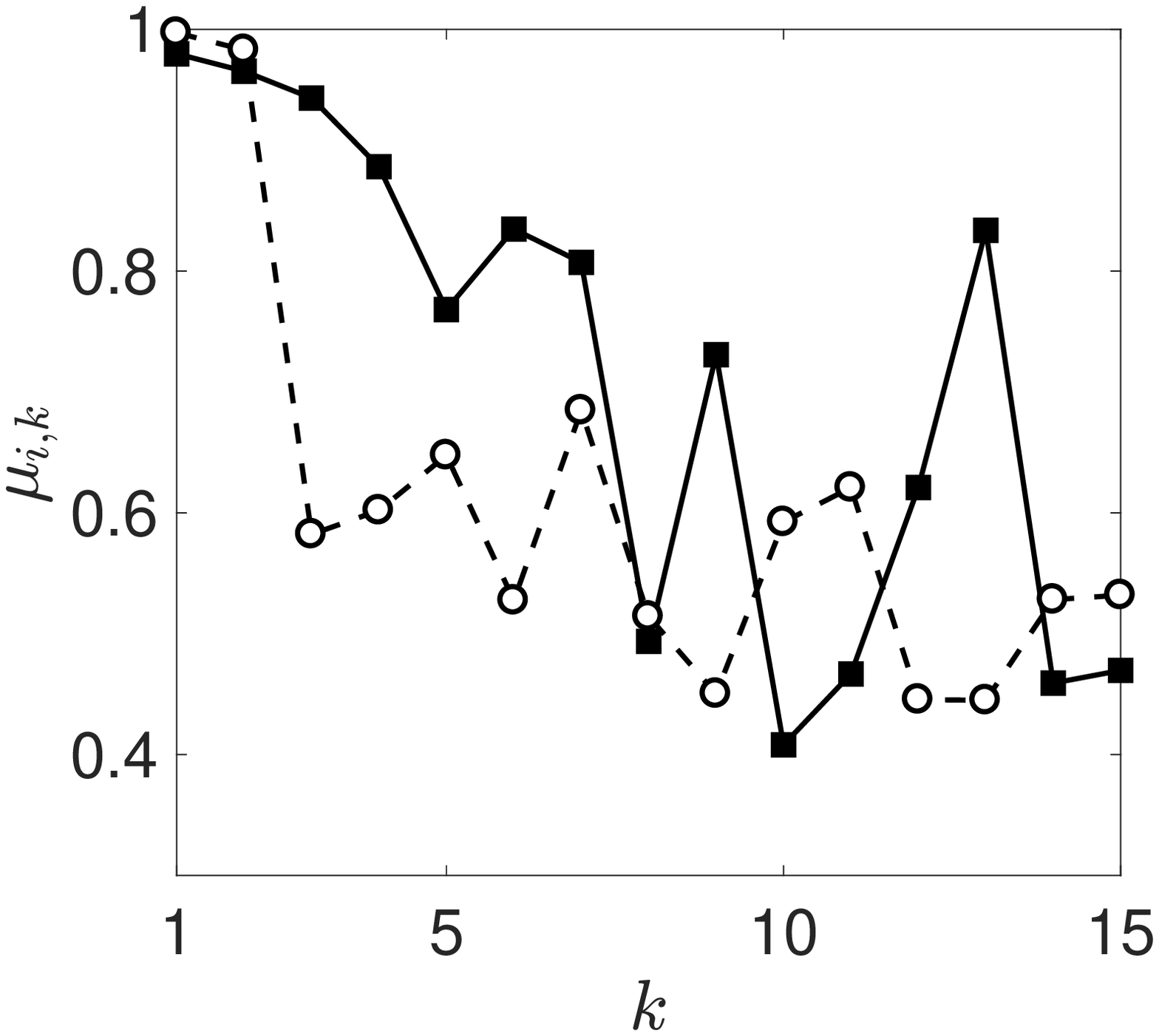}}
	\caption{Correlation coefficient $\mu_{i,k}$ to quantify the statistical convergence of the SPOD modes, considering $(\lambda_{x}^{+},\lambda_{z}^{+},\lambda_{t}^{+}) \approx (1000,100,100)$ for both Reynolds numbers.}
	\label{fig:convergence}
\end{figure}

In order to further verify the reliability of the computed SPOD modes, we carry out a convergence analysis, by dividing the total dataset into two equal parts each corresponding to $75$\% of the original dataset, and performing the SPOD on each part, so a normalised inner product is given by:

\begin{equation}
 \mu_{i,k} = \frac{\langle \Psi_{k} , \Psi_{i,k} \rangle}{\sqrt{ |\Psi_{k}|^{2} \cdot  |\Psi_{i,k}|^{2}}},
 \label{eq:mu}
\end{equation}
where $\langle, \rangle$ denotes the standard $L_2$ inner product considering the three velocity components, $i=(1;2)$ indicates each subset and $k$ each SPOD mode. This kind of analysis was also performed by \cite{lesshafft2019resolvent} and \cite{abreu2017coherent}. Figures \ref{fig:convergence} (a) and (b) show the normalised inner products of equation (\ref{eq:mu}) considering $(\lambda_{x}^{+},\lambda_{z}^{+},\lambda_{t}^{+}) \approx (1000,100,100)$ for $Re_{\tau}=180$ and $550$, respectively. Results show that the less energetic SPOD modes show discrepancies, which can be explained by the differences in the order in which the modes appear on each subset. However, for $Re_{\tau}=180$ we can observe that the first three SPOD modes exhibit a correlation coefficient close to one and can be considered as converged, and for $Re_{\tau}=550$ the first two modes are reasonably converged.

The convergence analysis of the SPOD modes is sensitive to the amount of data used \citep{lesshafft2019resolvent}. The DNS databases analysed here contain a limited number of snapshots, but for all analysed cases of $(\lambda_{x}^{+},\lambda_{z}^{+},\lambda_{t}^{+})$ studied in this paper, the first two SPOD modes were found to be converged for both Reynolds numbers, with $\mu_{i,k} \geq 0.95$.


\subsection{Resolvent analysis} \label{sec:resolvent}

To perform the resolvent analysis we follow the formulation described by \cite{mckeon2010critical}, also for a turbulent pipe flow. The linearized Navier--Stokes system for fully developed incompressible pipe flow can be written in operator notation as:
\begin{equation}
\mathbf{\hat{q}}= \mathcal{C} (i\omega \mathbf{I} - \mathcal{L})^{-1} \mathcal{B} \mathbf{\hat{f}},
\end{equation}
where $\mathcal{L}$ is the linearized Navier--Stokes operator considering the mean profile $\bar{u}$ (averaged in time, streamwise and azimuthal coordinates) as a base flow; $\omega$ is the analysed frequency; $\mathbf{\hat{q}} = [\hat{u};\hat{v};\hat{w}]$ is a vector containing all velocity components; $\mathbf{\hat{f}} = [\hat{f_x};\hat{f_y};\hat{f_z}]$ is the forcing term, or non-linear terms from Navier--Stokes equations, in streamwise, wall-normal and azimuthal directions; and the linear operators $\mathbf{\mathcal{B}}$ and $\mathbf{\mathcal{C}}$ are filters that impose restrictions both the forcing and in the output quantities of interest, respectively. Note that $\mathbf{\mathcal{B}}$ guarantees that no force will be applied in the continuity equation, and $\mathbf{\mathcal{C}}$ amounts to observation of only velocity fluctuations. The input is thus taken as the forcing term in the momentum equation, interpreted as the non-linear terms not considered in the linearisation of the Navier--Stokes equation, also labelled as generalised Reynolds stresses (for a given frequency-wavenumber combination). The output is based on a norm involving solely velocity fluctuations, and its maximisation of the output thus leads to maximal turbulent kinetic energy. The linearized Navier--Stokes operator here involves only the molecular viscosity, as in \cite{mckeon2010critical}. Even though it has been shown in some works that inclusion of an eddy viscosity improves the agreement between resolvent modes and flow statistics \citep{morra2019realizable,pickering2019eddy}, the present choice is motivated by a more straightforward interpretation of the forcing $f$ as the non-linear terms; more detailed analysis of such forcing terms are currentely being carried out by our group in related works \citep{nogueira2020forcing}. All variables here are Fourier transformed in time, streamwise and azimuthal directions. 

It should be pointed out that the choice of base flow is not unique. A number of works (e.g. \cite{waleffe1997self,schoppa2002coherent,farrell2012dynamics}) carry out analysis using a base flow resulting from averaging on streamwise direction. The resulting flow is $x$- and $t$-dependent, and may be subject to streak instability and transient growth. In this work, we have chosen to use the mean flow with averaging in all homogeneous directions. This facilitates comparison with SPOD modes taken from data, as such modes display better statistical convergence once preliminary Fourier decompositions are applied to the database. This allows a broad comparison between resolvent modes and DNS data, with the caveat that mechanisms such as streak instability and transient growth cannot be detected in the analysis in straightforward manner. 

The resolvent operator is defined as $\mathcal{R} = \mathcal{C} (i\omega \mathbf{I} - \mathcal{L})^{-1} \mathcal{B}$, and its singular value decomposition leads to optimal forcing modes, causing maximum amplification between input and output,

\begin{equation}
\mathcal{R} = U \Sigma V^{\dagger},
\label{eq:R}
\end{equation}
where $\dagger$ superscript indicates the Hermitian of the matrix. The above equation decomposes $\mathcal{R}$ into two orthonormal bases $U$ and $V$, where $U^{\dagger} U = I$ and $V^{\dagger} V = I$. Here $U$ is the output and $V$ is the input bases, so the size of $U$ is $N_{q} \mathrm{x} N_{modes}$ and the size of $V$ is $N_{f} \mathrm{x} N_{modes}$, where $N_{q}$ is the size of the output $\mathbf{\hat{q}}$, $N_f$ if the size of the input $\mathbf{\hat{f}}$ and $N_{modes}$ is the number of resolvent modes. The matrix $\Sigma$ is diagonal, with real, positive values, in decreasing order $\sigma_1 \geq \sigma_2 \geq ... \geq \sigma_n$.


This approach leads to identification of modes that optimally describe the linear amplification mechanisms in stable systems. In particular, resolvent analysis evaluates the flow response to time-periodic forcing. The method provides two orthonormal bases, one for forcing and the other one for the associated flow responses, and each pair of forcing and response modes is related by a gain. Response modes with high gains are expected to be dominant in the flow, as will be described next. Here we investigate how the response modes obtained using resolvent analysis are able to model dominant structures in turbulent pipe flow, which can be extracted from the DNS database. More details about the present resolvent formulation can be seen in \cite{mckeon2010critical}, and the relationship between SPOD and resolvent modes is documented in \cite{towne2018spectral} and \cite{lesshafft2019resolvent}.

The mean velocity profiles used to compute the resolvent analysis, $\bar{u}=\bar{u}(1-r)$, were extracted from the simulation considering azimuthal, streamwise and temporal averages. The base flow in wall units is shown in Figure \ref{fig:base_flow} (a) for both Reynolds numbers.

\subsection{SPOD vs. resolvent analysis}
\label{sec:SPODres}

Recent works have explored the connection between SPOD modes and the flow responses to stochastic forcing successfully \citep{abreu2017coherent, lesshafft2019resolvent, towne2018spectral}. To relate mathematically SPOD and resolvent analysis we can write the relation between the flow realisations $\bf{\hat{q}}$ and the resolvent operator $\mathbf{\mathcal{R}}$ for a problem with harmonic forcing $\mathbf{\hat{f}}$ as

\begin{equation}
\mathbf{\hat{q}} =  \mathbf{\mathcal{R}} \mathbf{\hat{f}}.
\label{eq:SPODres}
\end{equation}
Analysis of stochastic fields require a formulation in terms of two-point statistics. This can be obtained by mutiplying eq. \ref{eq:SPODres} by its Hermitian and taking the expected value $\mathcal{E}()$ of the forcing. This leads to

\begin{equation}
\mathcal{E}(\mathbf{\hat{q}} \mathbf{\hat{q}}^{\dagger}) =  \mathbf{\mathcal{R}} \mathcal{E}( \mathbf{\hat{f}} \mathbf{\hat{f}}^{\dagger} )  \mathbf{\mathcal{R}}^{\dagger}.
\label{eq.ex}
\end{equation}
If the forcing is white noise in space, $ \mathcal{E}( \mathbf{\hat{f}} \mathbf{\hat{f}}^{\dagger}) = \mathbf{\mathcal{I}}$, eq. \ref{eq.ex} becomes 
\begin{equation}
\mathbf{\Psi} \mathbf{\Lambda} \mathbf{\Psi}^{\dagger} =  \mathbf{U} \mathbf{\Sigma}^{2}  \mathbf{U}^{\dagger},
\end{equation}
meaning that the SPOD modes are simply the response modes from resolvent analysis, with SPOD eigenvalues equal to the square of resolvent gains.

It is clear that the forcing is not white noise in space, as, for instance, non-linear terms in the Navier--Stokes system vanish on the wall. The analysis of \cite{nogueira2020forcing} for a minimal flow unit suggests that the forcing statistics are spatially coherent. In this case, the response statistics depend on the details of the forcing in eq. \ref{eq.ex}, such that

\begin{equation}
\mathbf{\Psi} \mathbf{\Lambda} \mathbf{\Psi}^{\dagger} = \mathbf{U} \mathbf{\Sigma} \mathbf{V}^{\dagger} \mathcal{E}( \mathbf{\hat{f}} \mathbf{\hat{f}}^{\dagger}) \mathbf{V} \mathbf{\Sigma} \mathbf{U}^{\dagger}.
\end{equation}

However, if the resolvent operator has a dominant amplification mechanism, such that $\sigma_1 \ll \sigma_2$, the response CSD will be dominated by the contribution of the leading response mode \citep{beneddine2016conditions,towne2018spectral,cavalieri2019wave}. To see this, neglecting the high-order gains leads to

\begin{equation}
\mathcal{E}( \mathbf{\hat{q}} \mathbf{\hat{q}}^{\dagger}) \approx  u_1 \sigma_1 v_1^\dagger \mathcal{E}( \mathbf{\hat{f}} \mathbf{\hat{f}}^{\dagger}) v_1 \sigma_1 u_1^\dagger
\end{equation}
and hence the flow response is approximately given by the projection of the forcing statistics onto the optimal forcing mode $v_1$, amplified by the leading gain $\sigma_1$ and taking the shape of the most amplified response $u_1$. In this case the leading SPOD mode may be close to the optimal flow response, even though the forcing is not white in space.

The expressions above consider an Euclidean inner product, which is appropriate for matrices. The non-uniform grids used in this work require the use of integration weights for the discretisation of the inner product. Resolvent analysis and SPOD should be modified so as to account for integration weigths; appropriate expressions are presented by \cite{towne2018spectral,lesshafft2019resolvent,cavalieri2019wave} and are not repeated here for brevity.

\section{Results and discussions} \label{sec:results}

\begin{figure}
	\centering
	\subfigure[SPOD mode 1]{\includegraphics[width=0.47\textwidth,trim = 35 58 0 60,clip]{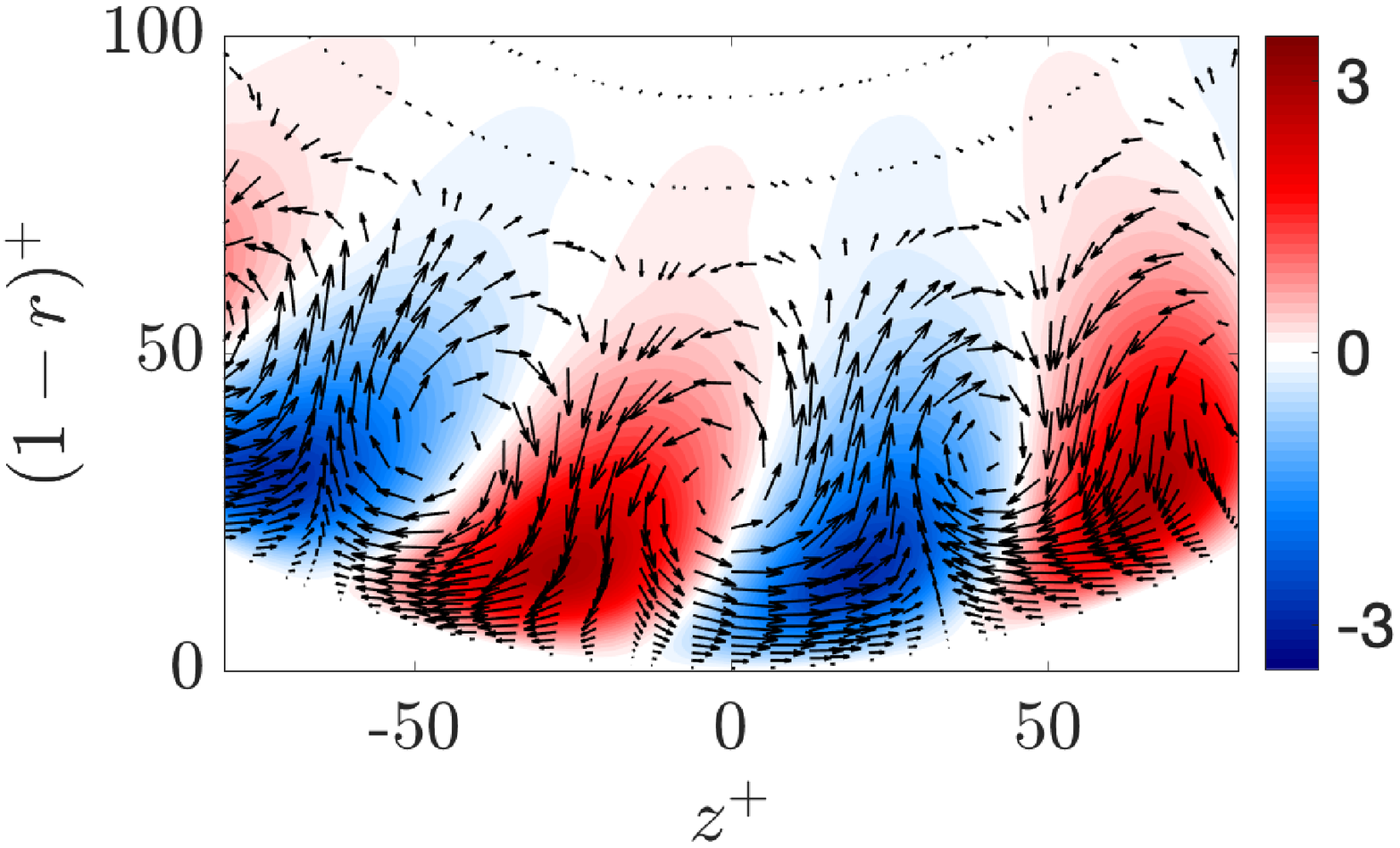}}
	\subfigure[Resolvent mode 1]{\includegraphics[width=0.47\textwidth,trim = 35 58 0 60,clip]{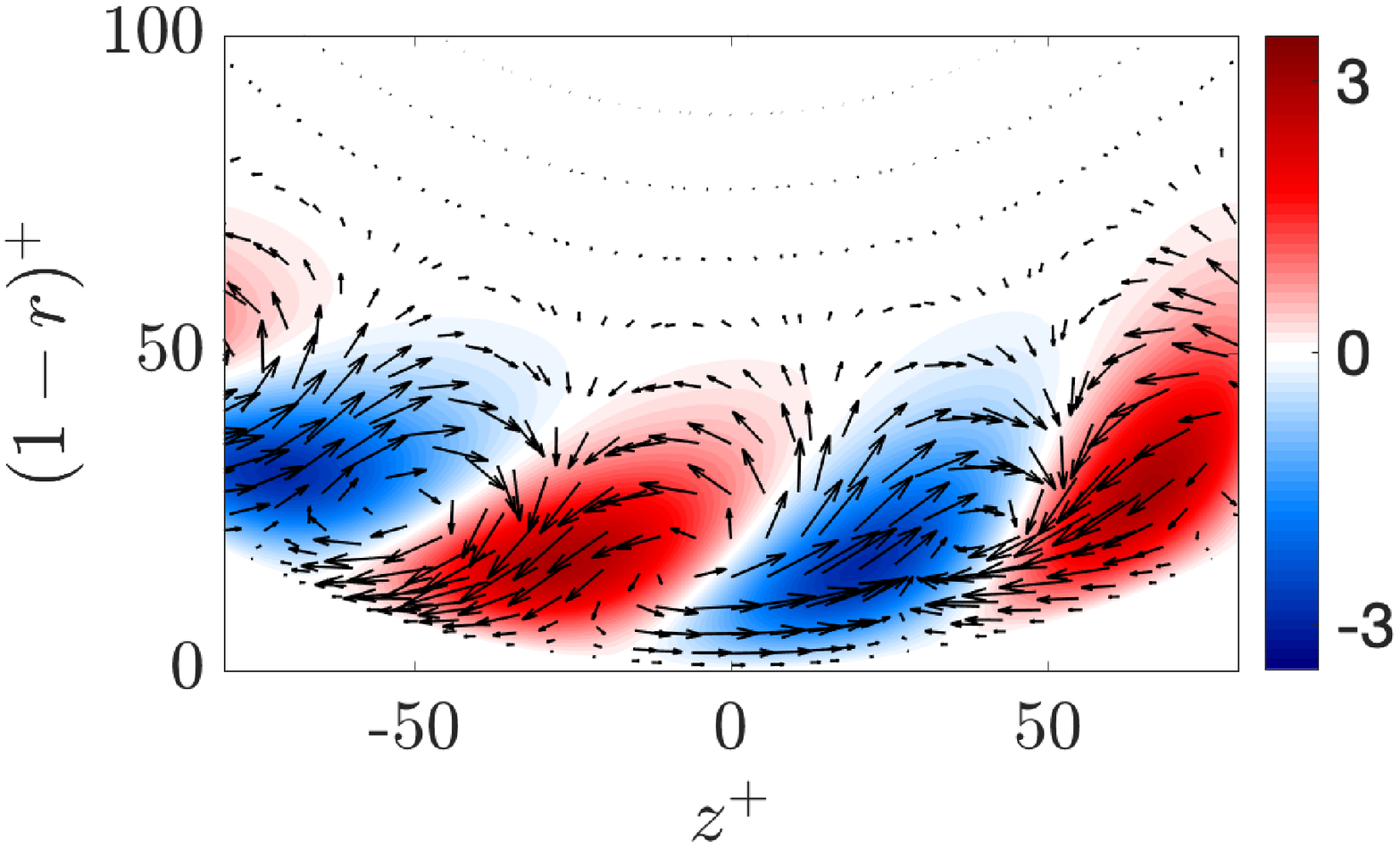}}
	
	\subfigure[SPOD mode 2]{\includegraphics[width=0.47\textwidth,trim = 35 58 0 60,clip]{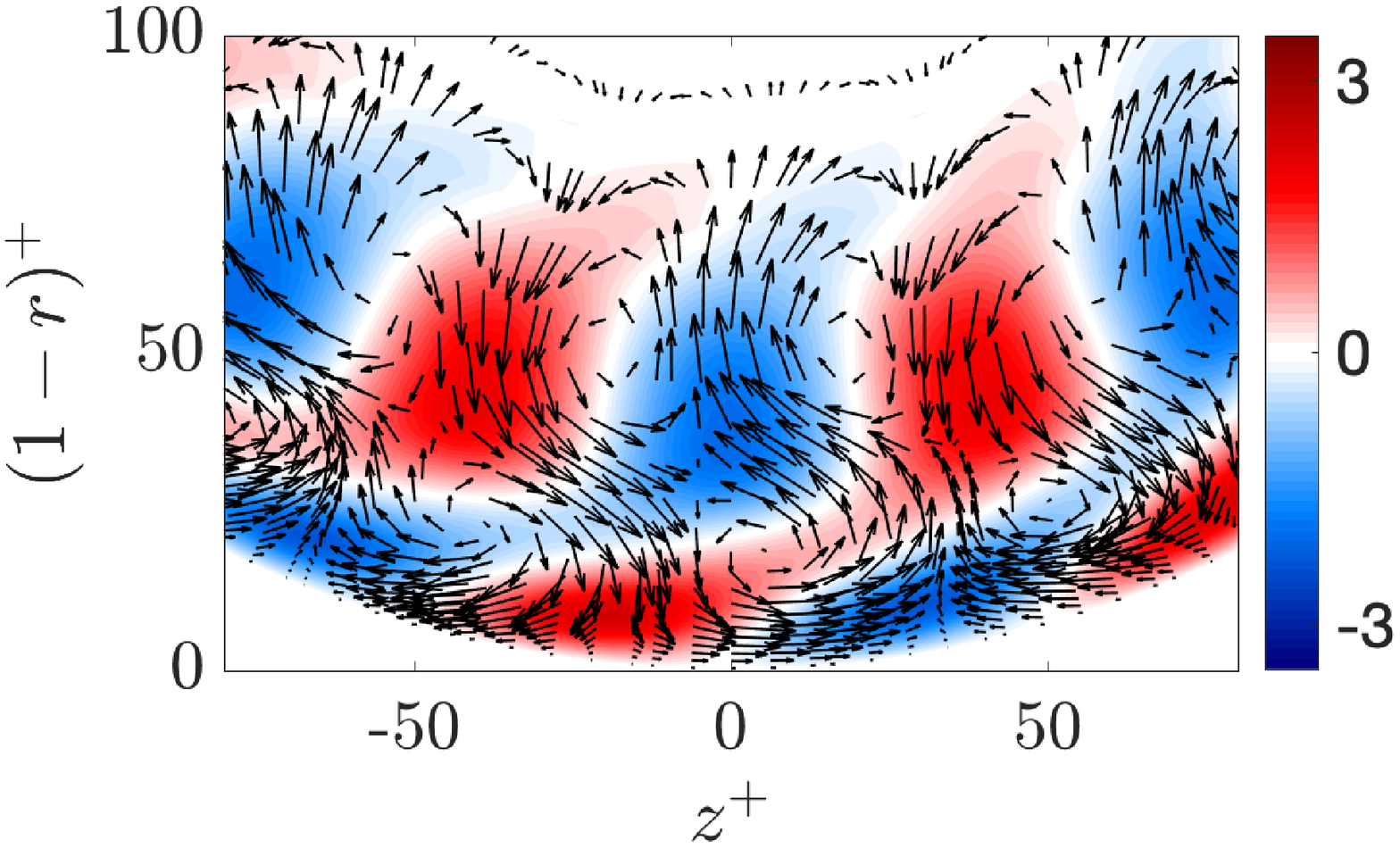}}
	\subfigure[Resolvent mode 2]{\includegraphics[width=0.47\textwidth,trim = 35 58 0 60,clip]{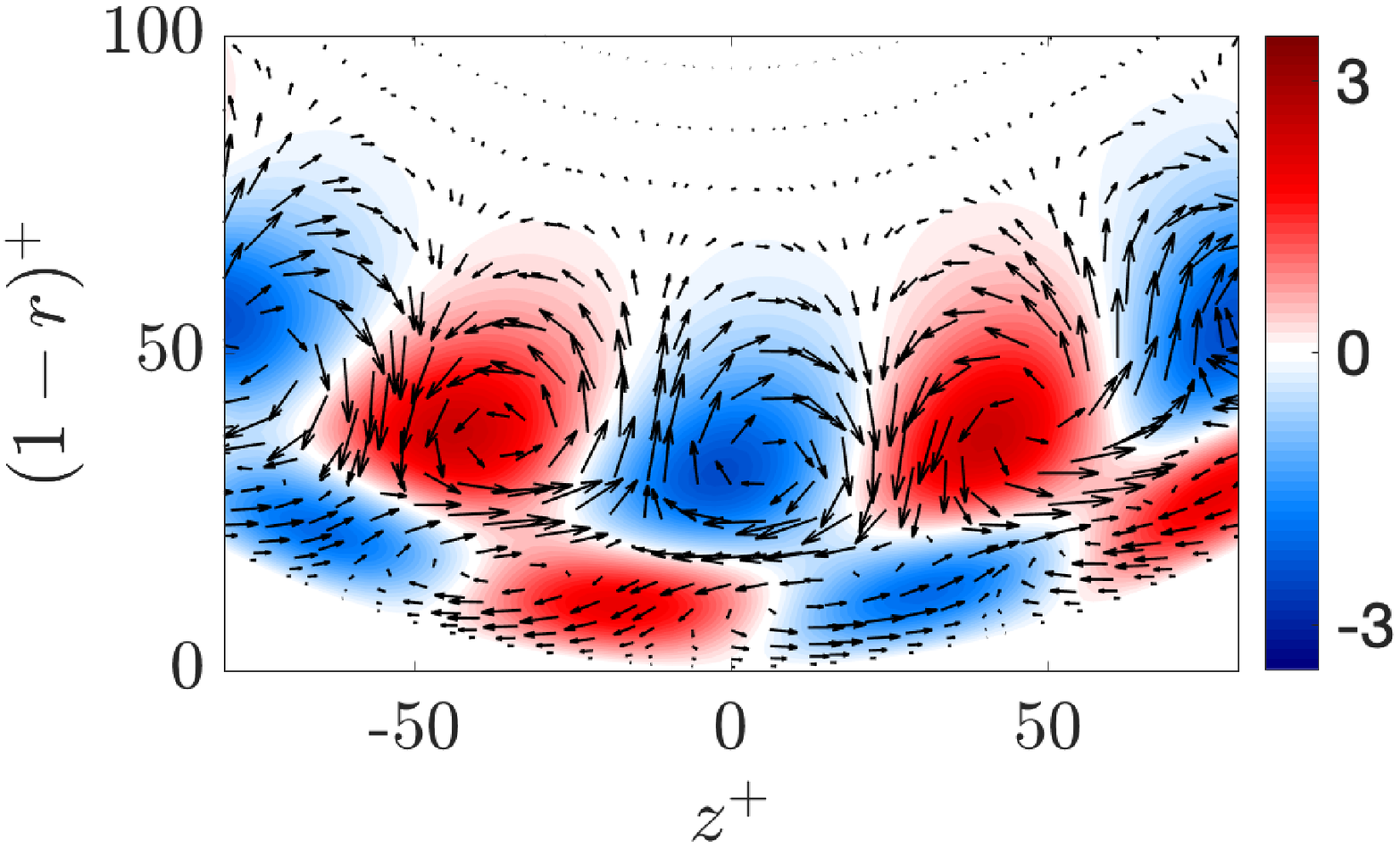}}
	
	\caption{Comparison between the first two SPOD modes and optimal and suboptimal responses from resolvent analysis using cross-stream view of the $v-w$ components of the vortices (arrows) and the $u$ component of the streak (red and blue contours) for $(\lambda_{x}^{+},\lambda_{z}^{+},\lambda_{t}^{+})\approx (1000,100,100)$ at $Re_{\tau}=180$. The axis ticks labels are scaled in inner units.}
	\label{fig:uvw_2D_180}
\end{figure}

\begin{figure}
	\centering
	\subfigure[SPOD mode 1]{\includegraphics[width=0.47\textwidth,trim = 35 58 0 60,clip]{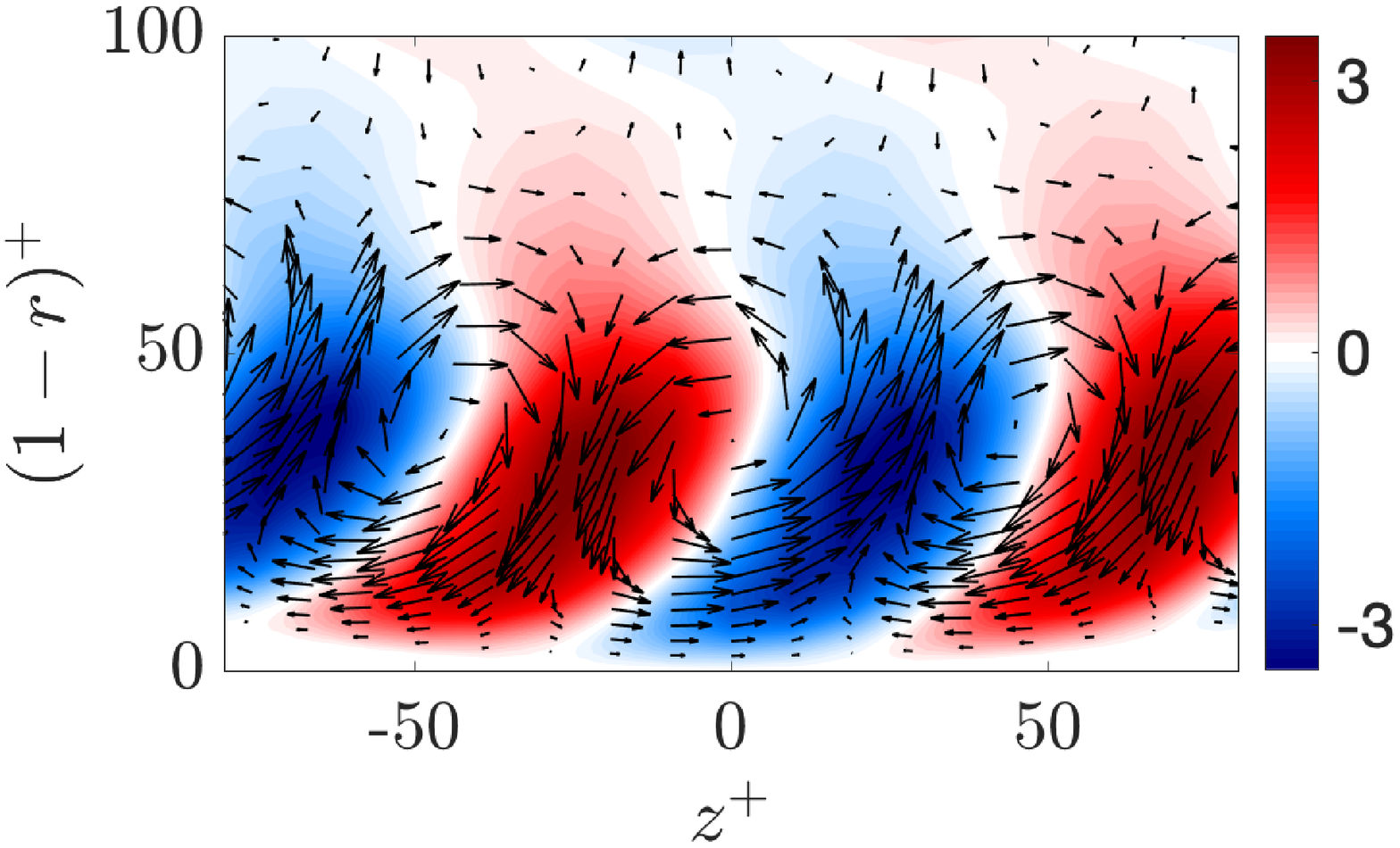}}
	\subfigure[Resolvent mode 1]{\includegraphics[width=0.47\textwidth,trim = 35 58 0 60,clip]{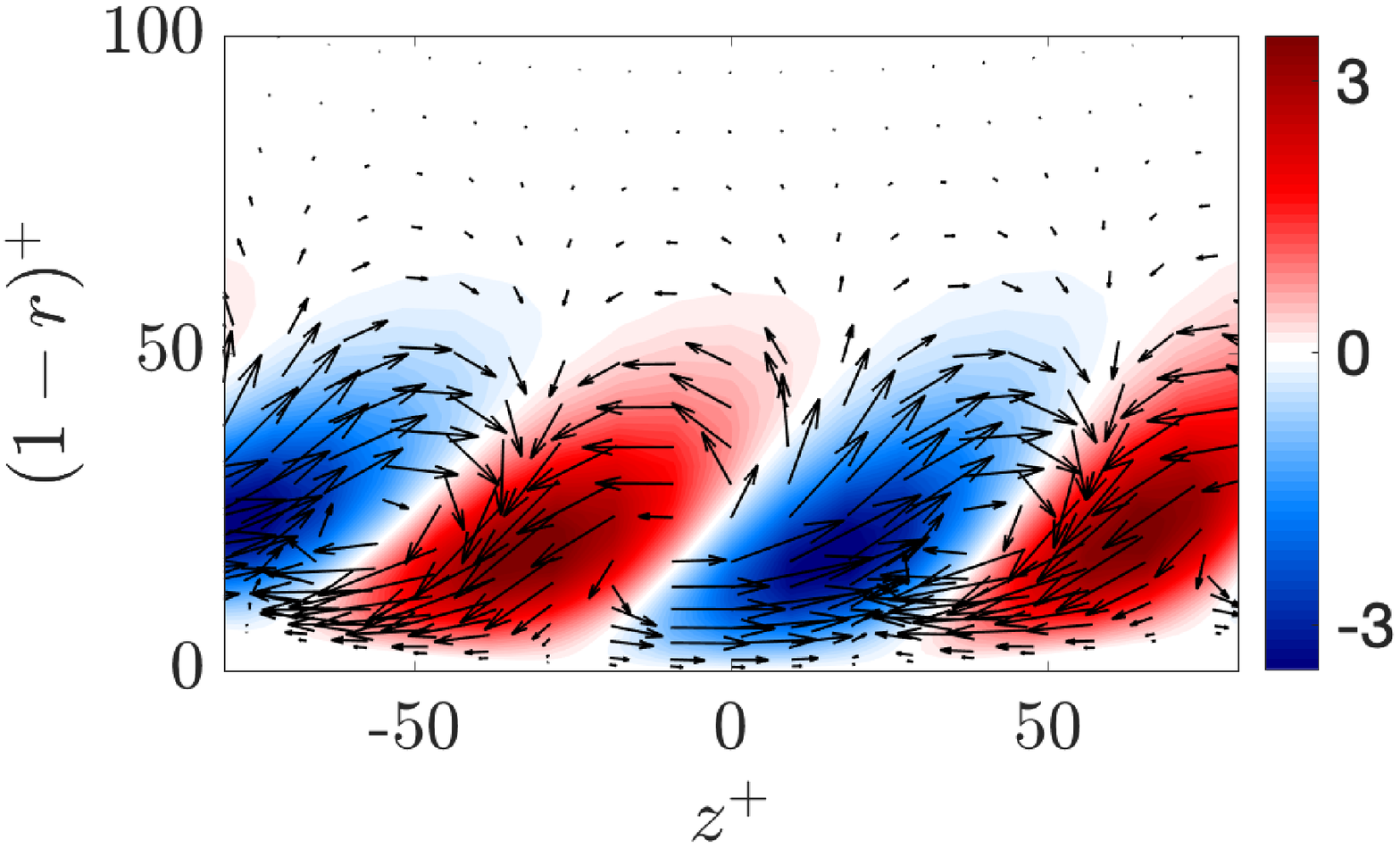}}
	
	\subfigure[SPOD mode 2]{\includegraphics[width=0.47\textwidth,trim = 35 58 0 60,clip]{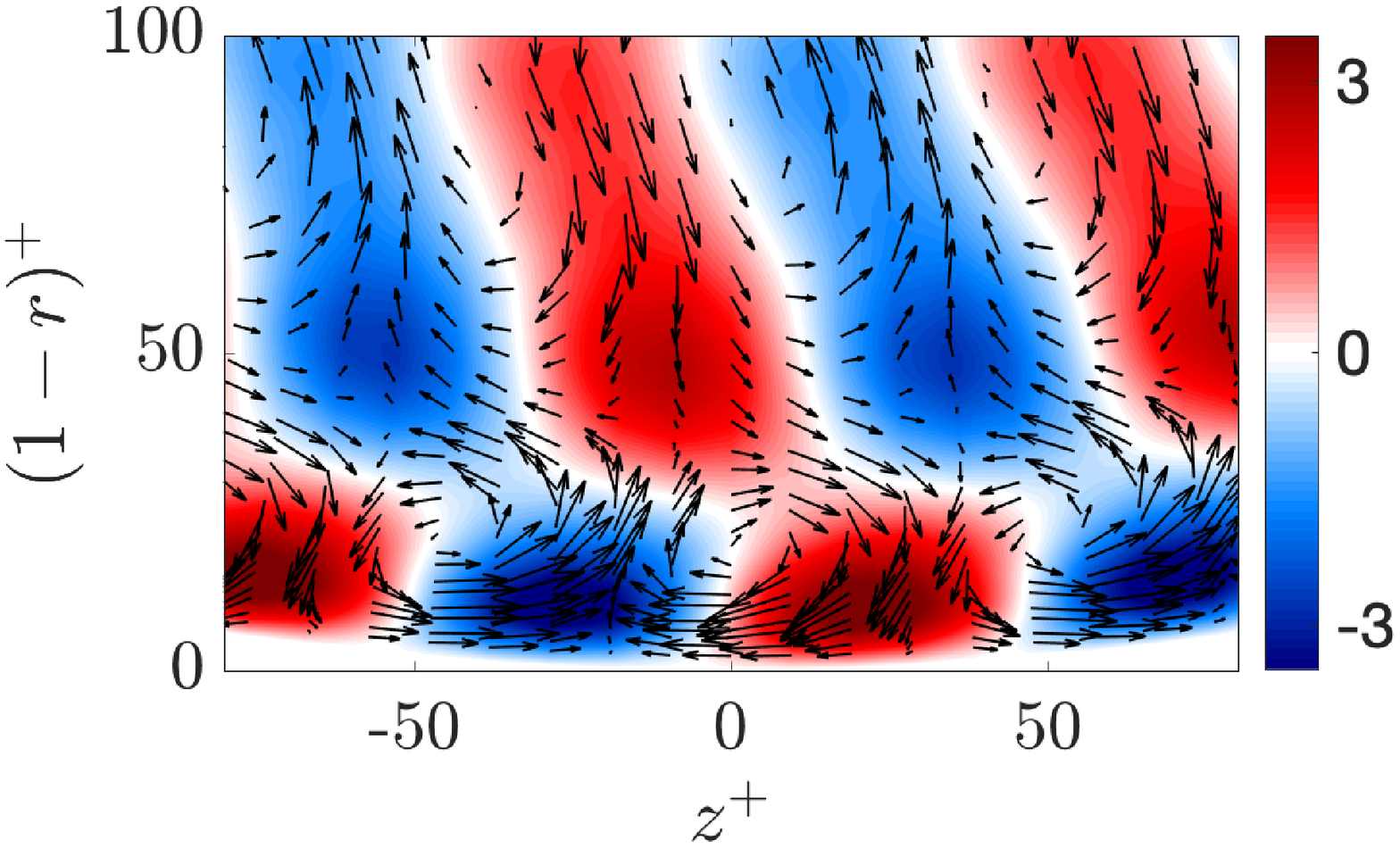}}
	\subfigure[Resolvent mode 2]{\includegraphics[width=0.47\textwidth,trim = 35 58 0 60,clip]{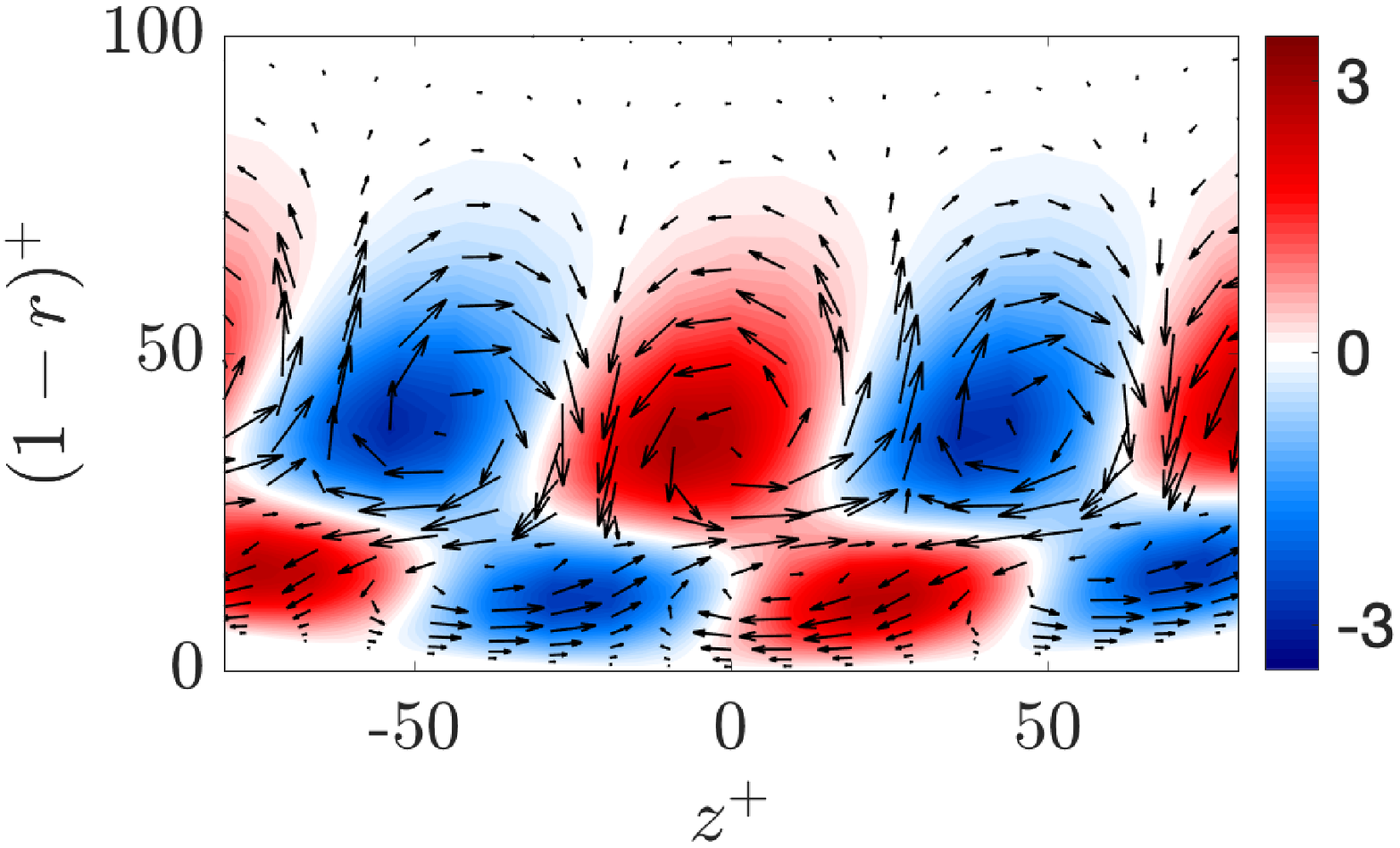}}
	
	\caption{Comparison between the first two SPOD modes and optimal and suboptimal responses from resolvent analysis using cross-stream view of the $v-w$ components of the vortices (arrows) and the $u$ component of the streak (red and blue contours) for $(\lambda_{x}^{+},\lambda_{z}^{+},\lambda_{t}^{+})\approx (1000,100,100)$ at $Re_{\tau}=550$. The axis ticks labels are scaled in inner units.}
	\label{fig:uvw_2D_550}
\end{figure}

Figures \ref{fig:uvw_2D_180} and \ref{fig:uvw_2D_550} show the first two SPOD modes compared with the optimal and suboptimal responses from resolvent analysis for $Re_{\tau}=180$ and $550$, respectively, considering $(\lambda_{x}^{+},\lambda_{z}^{+},\lambda_{t}^{+})\approx (1000,100,100)$, or the corresponding frequency $\omega^{+} = 2 \pi / \lambda_{t}^{+} \approx 0.06$. This  is representative of the near-wall cycle, corresponding to the peak wavenumber in the premultiplied spectra shown in Figure \ref{fig:energyspec} and to a phase speed of $c^+ \approx 10$, a value typical of buffer-layer disturbances. Notice that the vertical direction does not correspond to constant spacing in viscous length scale, due to polar system. The results for the first mode for both Reynolds numbers (Figures \ref{fig:uvw_2D_180} (a,b) and \ref{fig:uvw_2D_550} (a,b)) show that the velocity field is associated with streamwise vortices (shown with arrows) and accompanying low- and high-speed streaks (colors). Negative wall-normal fluctuations carry high-momentum fluid and create high-velocity streaks with $u' > 0$ (red contour lines), and the opposite occurs for the positive wall-normal disturbances, creating slow-velocity streaks with $u' < 0$ (blue contour lines), characterizing the lift-up mechanism \citep{ellingsen1975stability,landahl1980note}. The second mode (Figures \ref{fig:uvw_2D_180} (c,d) and \ref{fig:uvw_2D_550} (c,d)) shows a pattern with two streamwise vortices and two streaks as a function of radius, similar to the observations of \cite{hellstrom2016self} for higher-order POD modes of larger-scale structures at higher Reynolds number. Streamwise vortices and streaks are again arranged consistently with the lift-up mechanism. The results show that the optimal and suboptimal flow responses recover the leading SPOD modes for both Reynolds numbers, highlighting that the responses obtained using the linearized operator serve as a pertinent model for the dominant structures observed in the DNS for the analysed frequency-wavenumber combination, $(\lambda_{x}^{+},\lambda_{z}^{+},\lambda_{t}^{+})\approx (1000,100,100)$.

%

Our aim is to perform additional, detailed quantitative comparisons between the first SPOD mode from the DNS and the optimal response from resolvent analysis. In order to evaluate the agreement for several values of wavelengths $\lambda_{x}$ and $\lambda_{z}$ at a fixed frequency $\omega$, we define the metric:

\begin{equation}
\beta = \frac{ {\rm Real} (\langle \mathbf{\hat{q}}_{1_{SPOD}},\mathbf{\hat{q}}_{1_{res}} \rangle )}{||\mathbf{\hat{q}}_{1_{SPOD}}||||\mathbf{\hat{q}}_{1_{res}}||},
\end{equation}
where $\mathbf{\hat{q}}_{1_{SPOD}} = [\hat{u}_{1_{SPOD}}; \hat{v}_{1_{SPOD}}; \hat{w}_{1_{SPOD}}]$ is the first SPOD mode; $\mathbf{\hat{q}}_{1_{res}}= [\hat{u}_{1_{RES}}; \hat{v}_{1_{RES}}; \hat{w}_{1_{RES}}]$ is the optimal response from resolvent analysis; and $\beta$ is the projection of $\mathbf{\hat{q}}_{1_{SPOD}}$ into $\mathbf{\hat{q}}_{1_{res}}$; note that $\langle , \rangle$ denotes the $L_2$ inner product. Thus, $\beta=1$ indicates a perfect alignment between both vectors, and $\beta=0$ indicates that the modes are orthogonal. The SPOD and resolvent modes include all velocity components in order to include information about the phase.

\begin{figure}
	\subfigure[$\beta$, $\lambda_{t}^{+} \approx 100$]{\includegraphics[width=0.325\textwidth,trim = 0 0 0 0,clip]{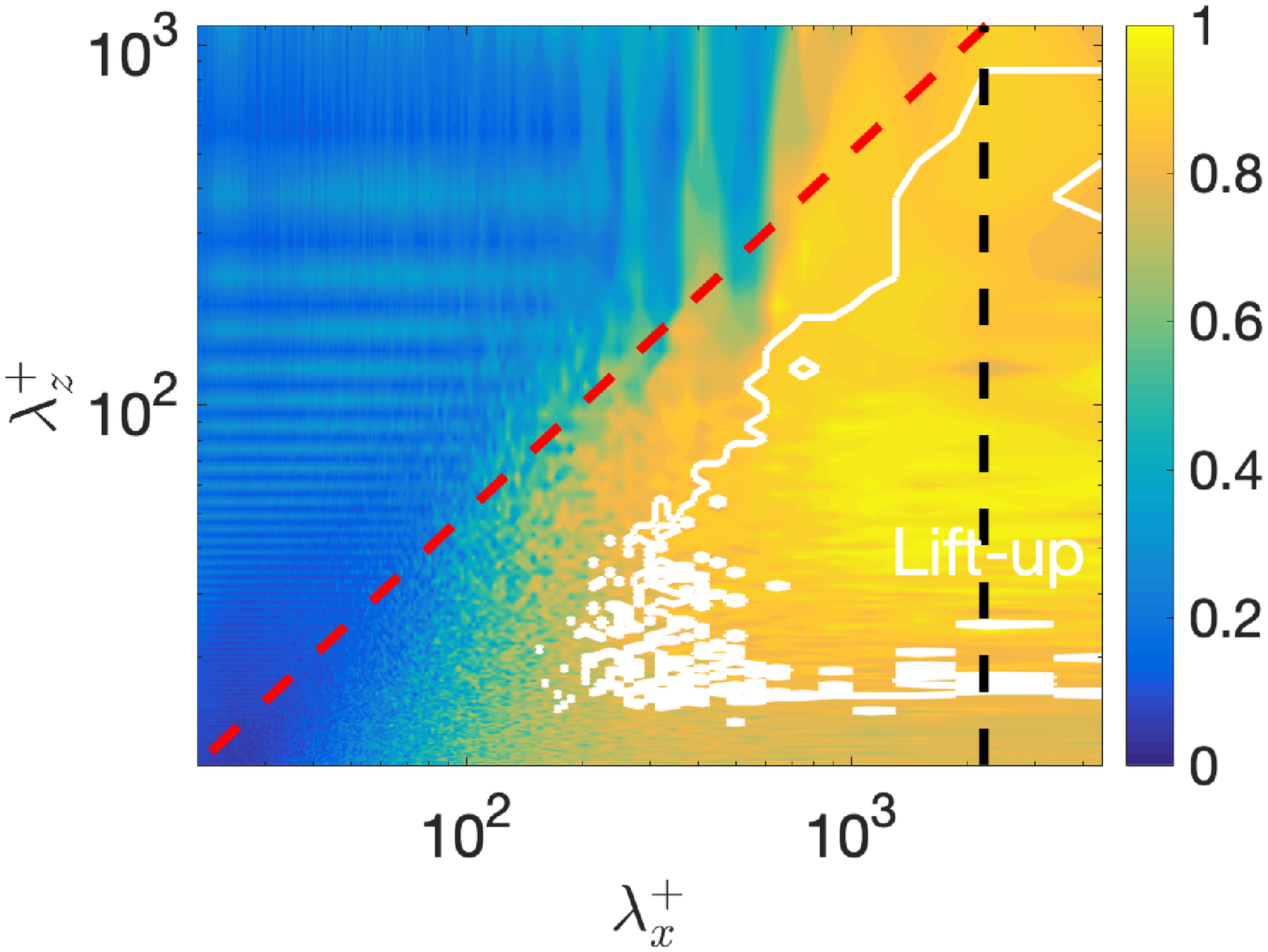}}
	\subfigure[$\beta$, $\lambda_{t}^{+} \approx 250$]{\includegraphics[width=0.325\textwidth,trim = 0 0 0 0,clip]{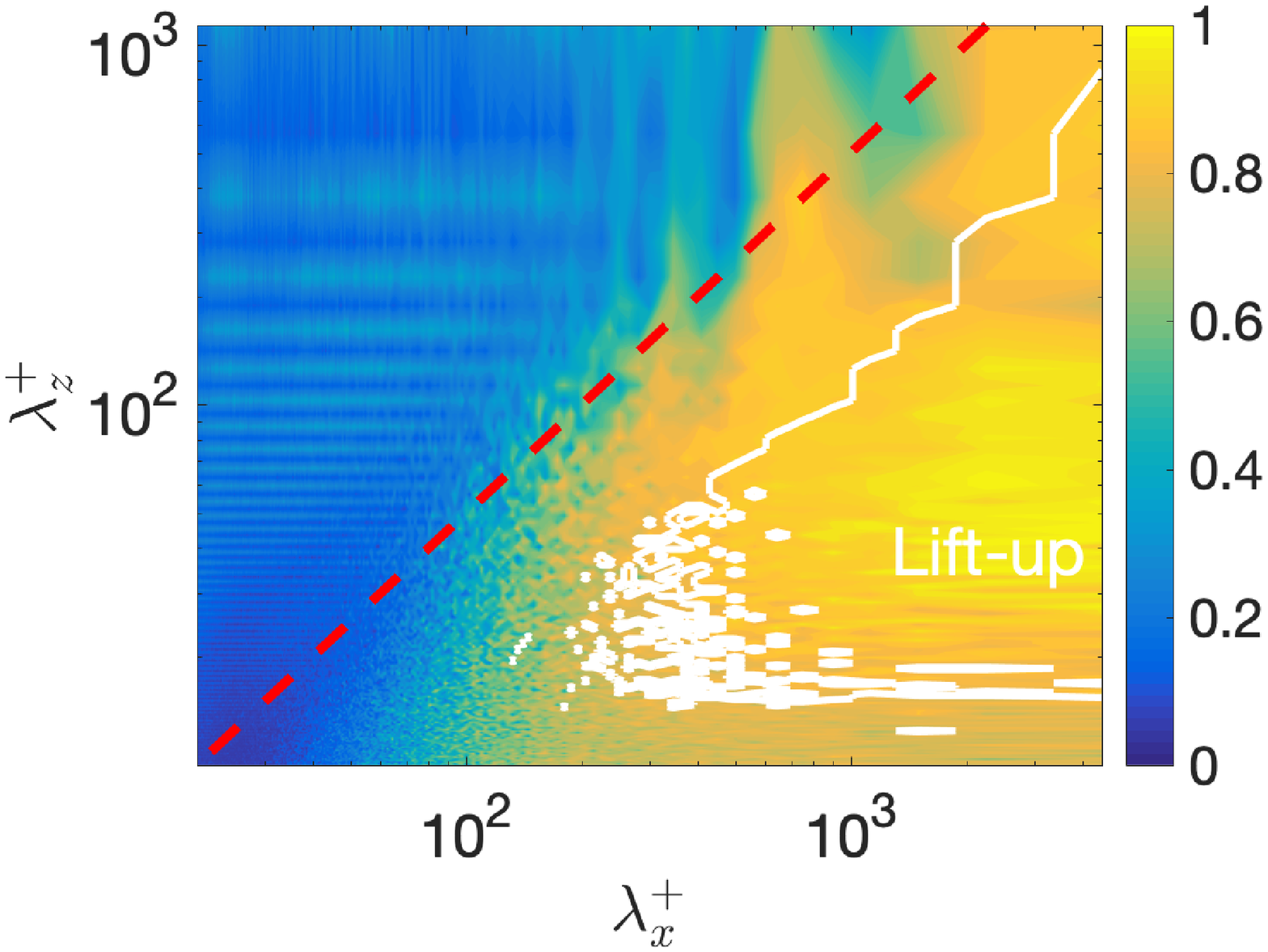}}
	\subfigure[$\beta$, $\lambda_{t}^{+} \approx 1500$]{\includegraphics[width=0.325\textwidth,trim = 0 0 0 0,clip]{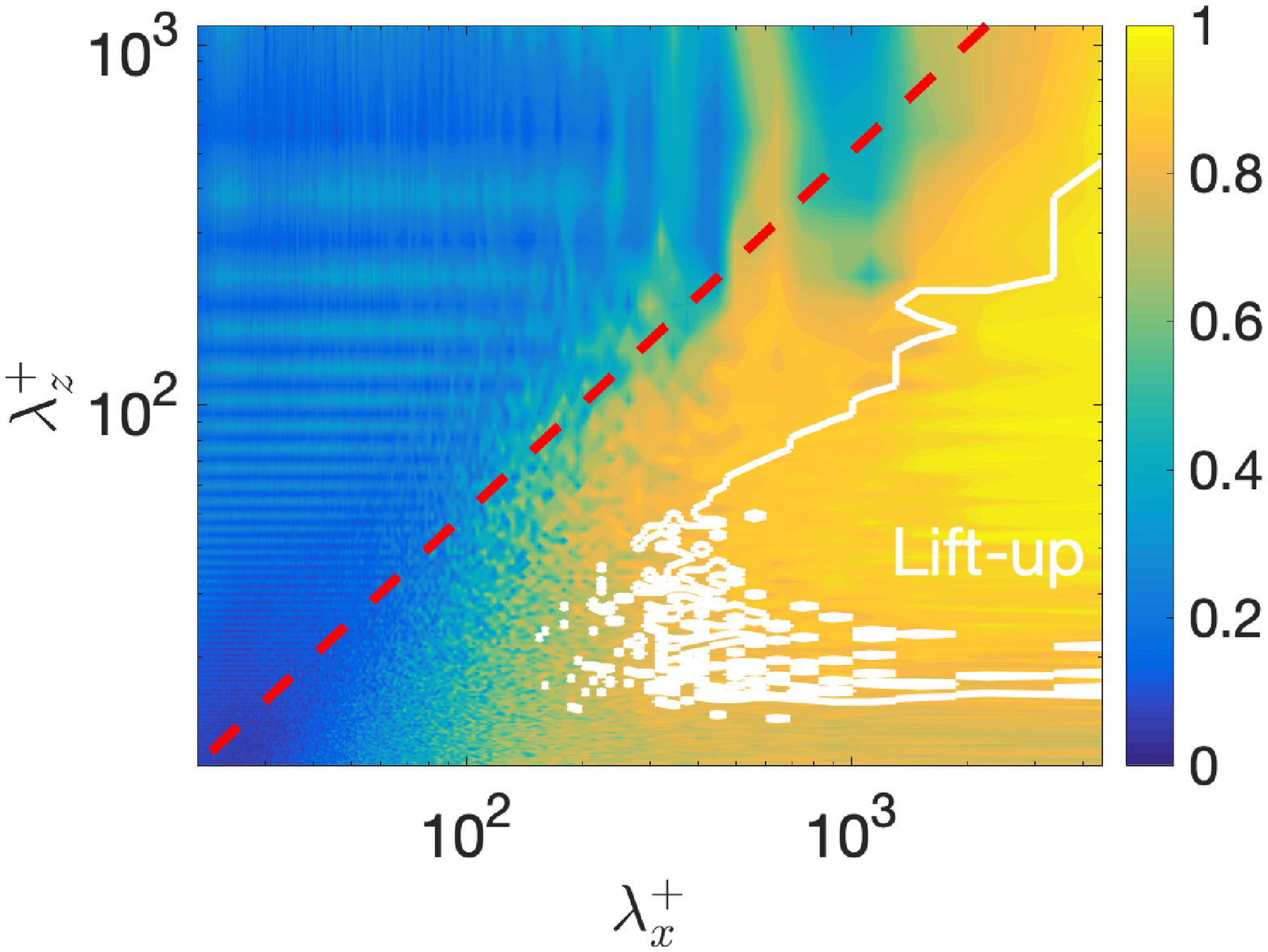}}

	\subfigure[$\beta_{(f_x=0)}$, $\lambda_{t}^{+} \approx 100$]{\includegraphics[width=0.325\textwidth,trim = 0 0 0 0,clip]{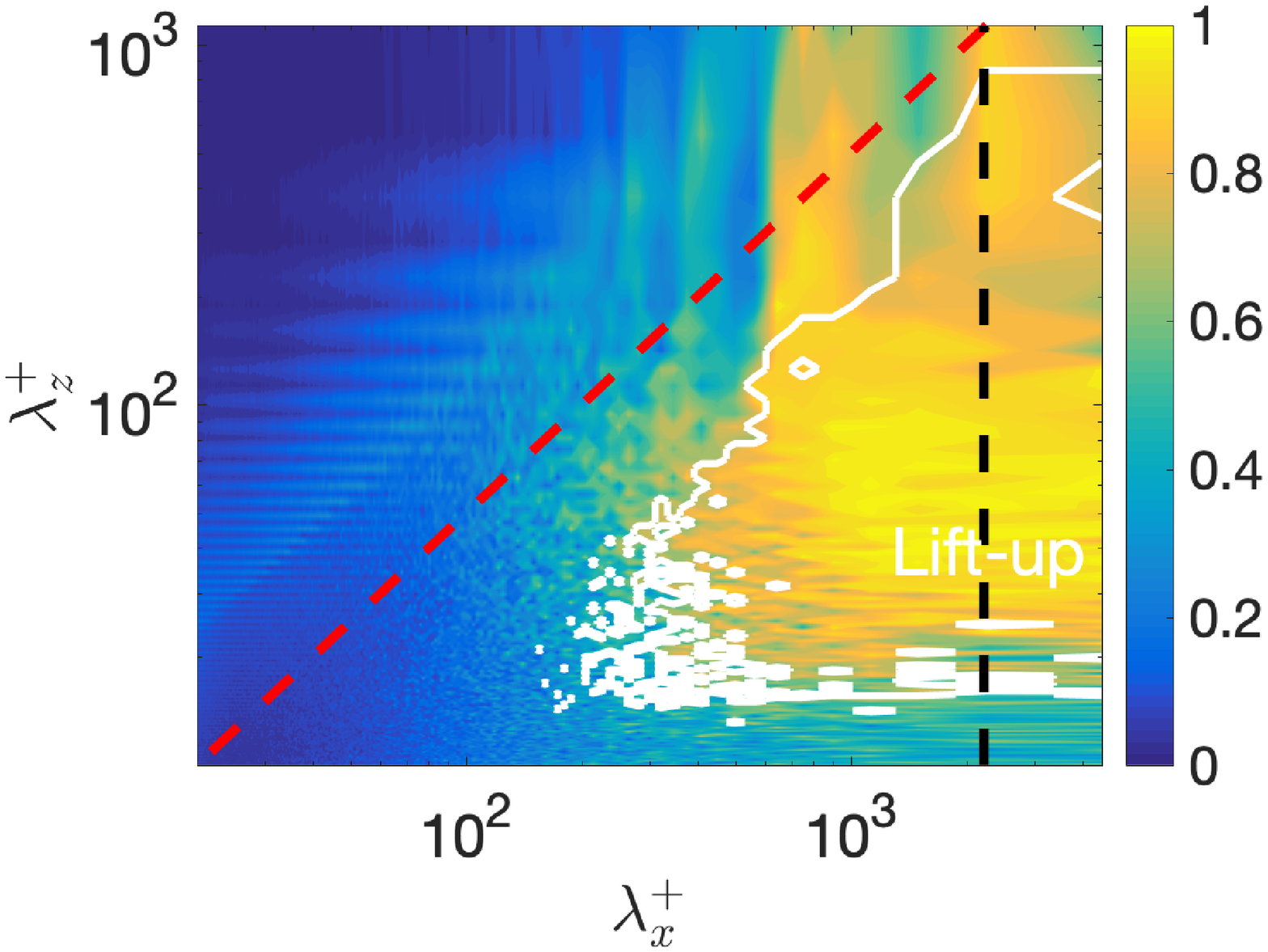}}
	\subfigure[$\beta_{(f_x=0)}$, $\lambda_{t}^{+} \approx 250$]{\includegraphics[width=0.325\textwidth,trim = 0 0 0 0,clip]{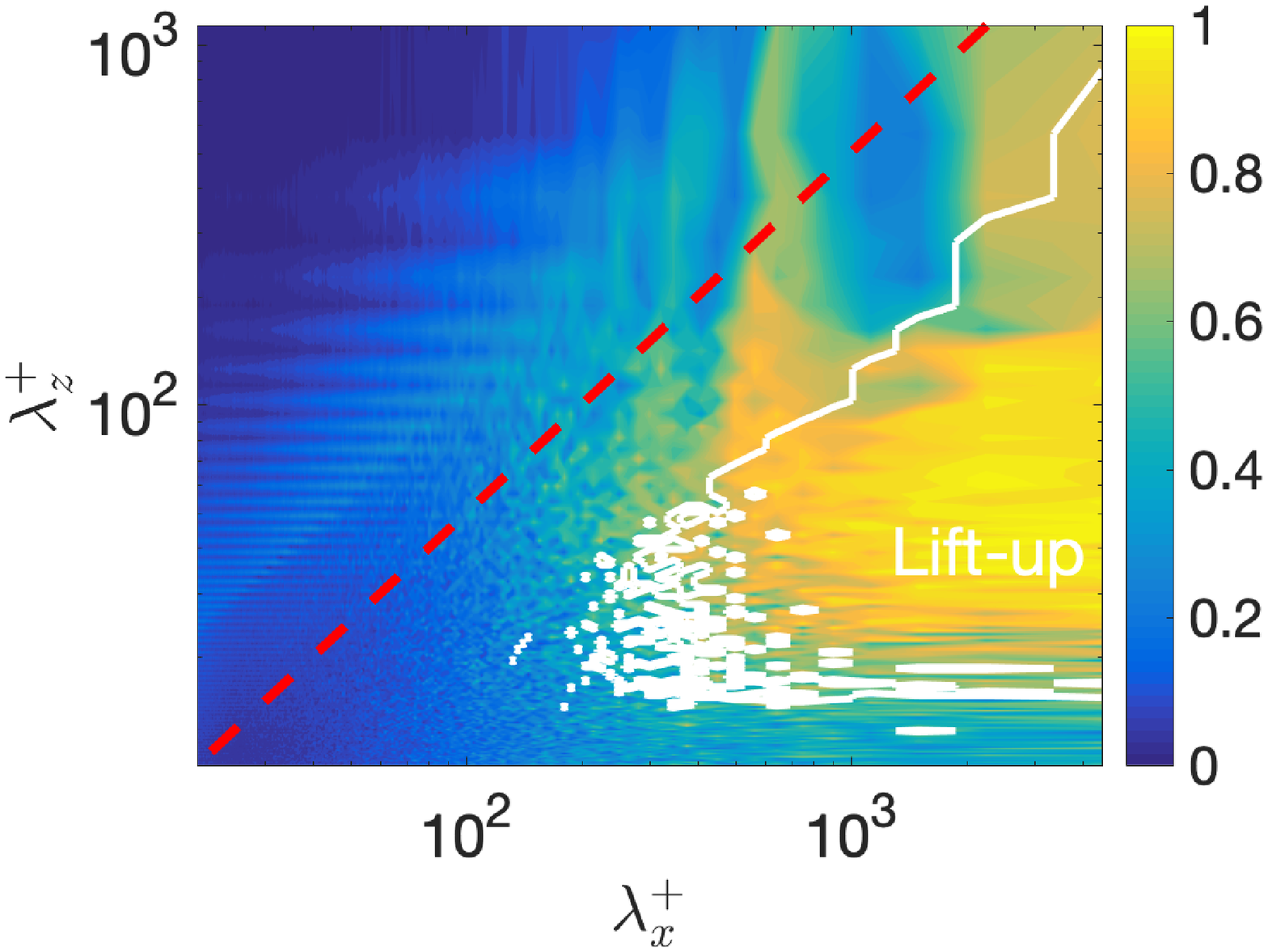}}
	\subfigure[$\beta_{(f_x=0)}$, $\lambda_{t}^{+} \approx 1500$]{\includegraphics[width=0.325\textwidth,trim = 0 0 0 0,clip]{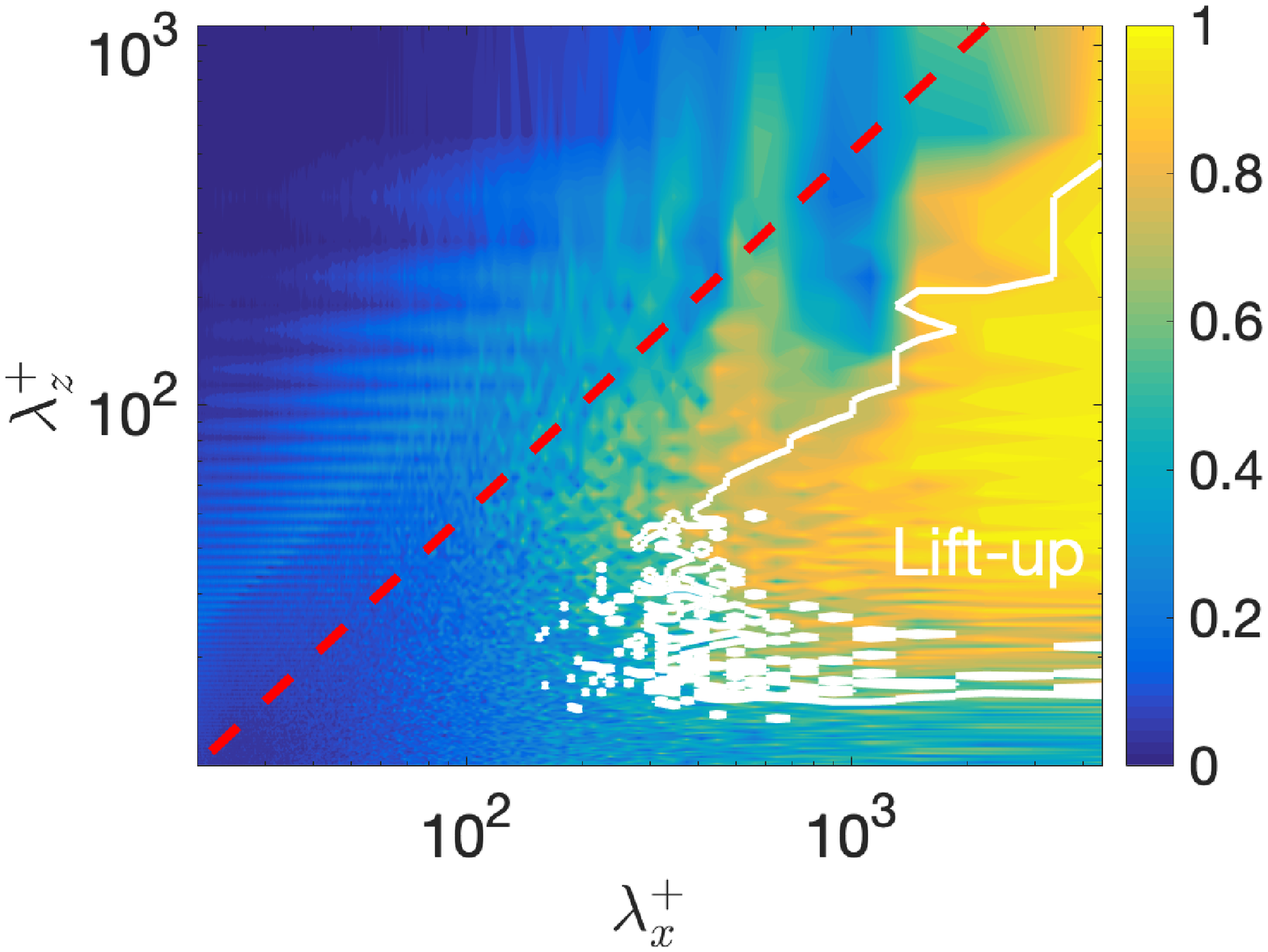}}

	\subfigure[$\log_{10}(\sigma_{1}/\sigma_{2})$, $\lambda_{t}^{+} \approx 100$]{\includegraphics[width=0.325\textwidth,trim = 0 0 0 0,clip]{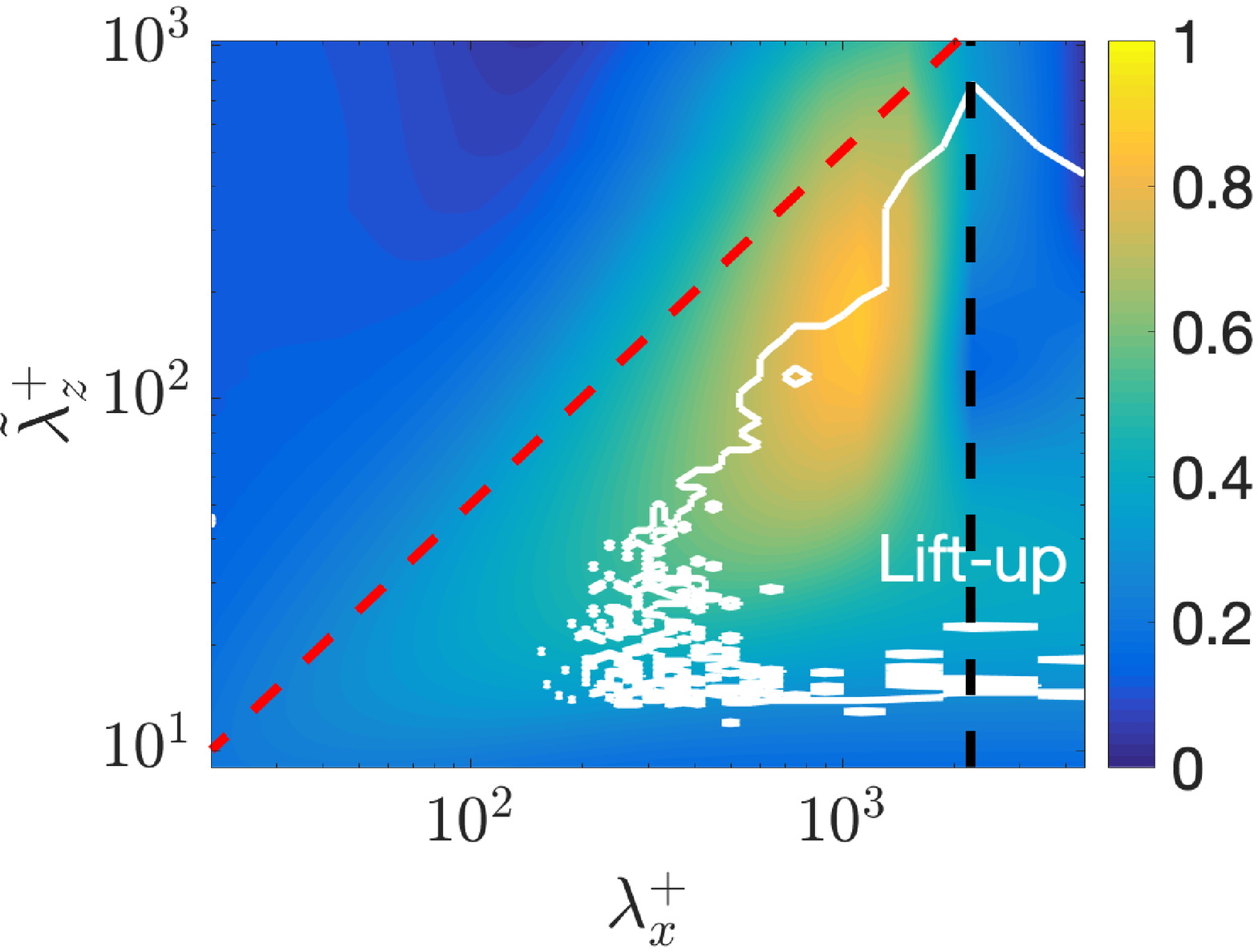}}
	\subfigure[$\log_{10}(\sigma_{1}/\sigma_{2})$, $\lambda_{t}^{+} \approx 250$]{\includegraphics[width=0.325\textwidth,trim = 0 0 0 0,clip]{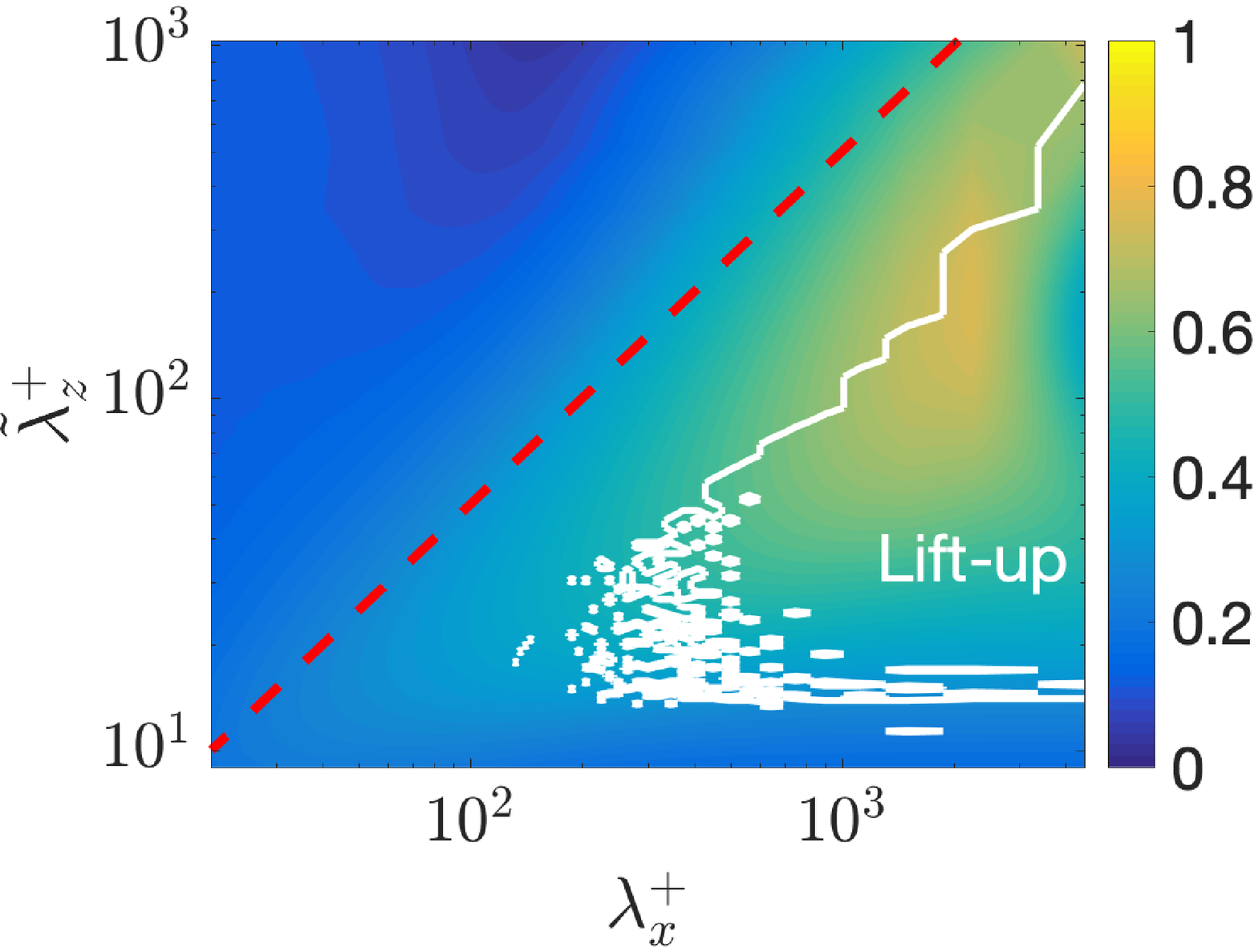}}
	\subfigure[$\log_{10}(\sigma_{1}/\sigma_{2})$, $\lambda_{t}^{+} \approx 1500$]{\includegraphics[width=0.325\textwidth,trim = 0 0 0 0,clip]{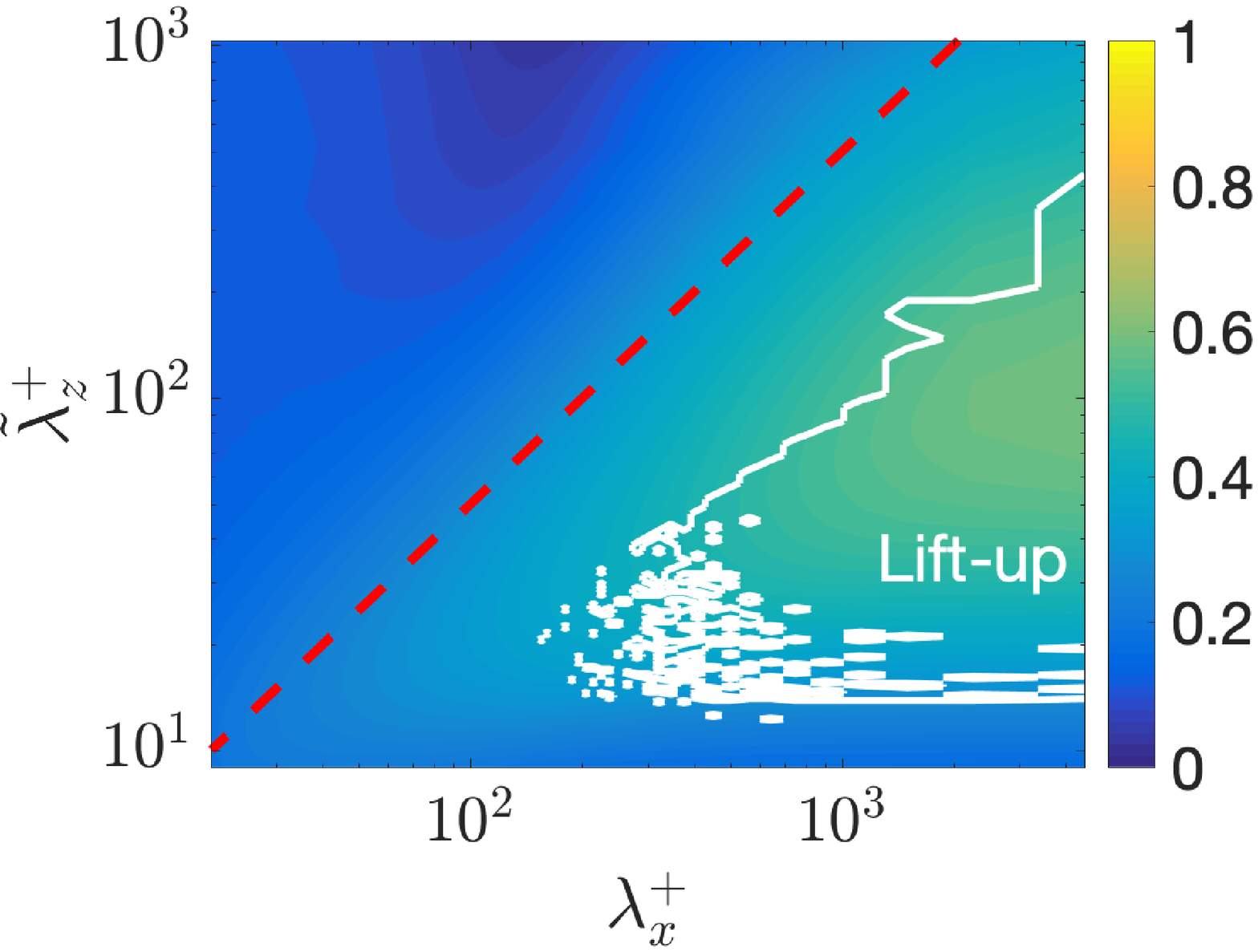}}

	\caption{(a,b,c) Agreement between first SPOD mode and optimal response from resolvent analysis characterized in terms of $\beta$. (d,e,f) Agreement between first SPOD and resolvent modes considering $f_x=0$, characterized in terms of $\beta_{(f_x=0)}$. (g,h,i) Ratio between optimal and suboptimal resolvent gains in logarithmic scale. Results for $Re_{\tau}=180$ and fixed frequencies: $\lambda_{t}^{+} \approx 100$, $250$ and $1500$ (from left to right). The region surrounded by the white line in all plots represents an indicator of lift-up mechanism from resolvent analysis. The red dashed line in all plots represents $\lambda_{x}^{+}=2 \lambda_{z}^{+}$.}
	\label{fig:beta180}
\end{figure}

\begin{figure}
	\subfigure[$\beta$, $\lambda_{t}^{+} \approx 100$]{\includegraphics[width=0.325\textwidth,trim = 0 0 0 0,clip]{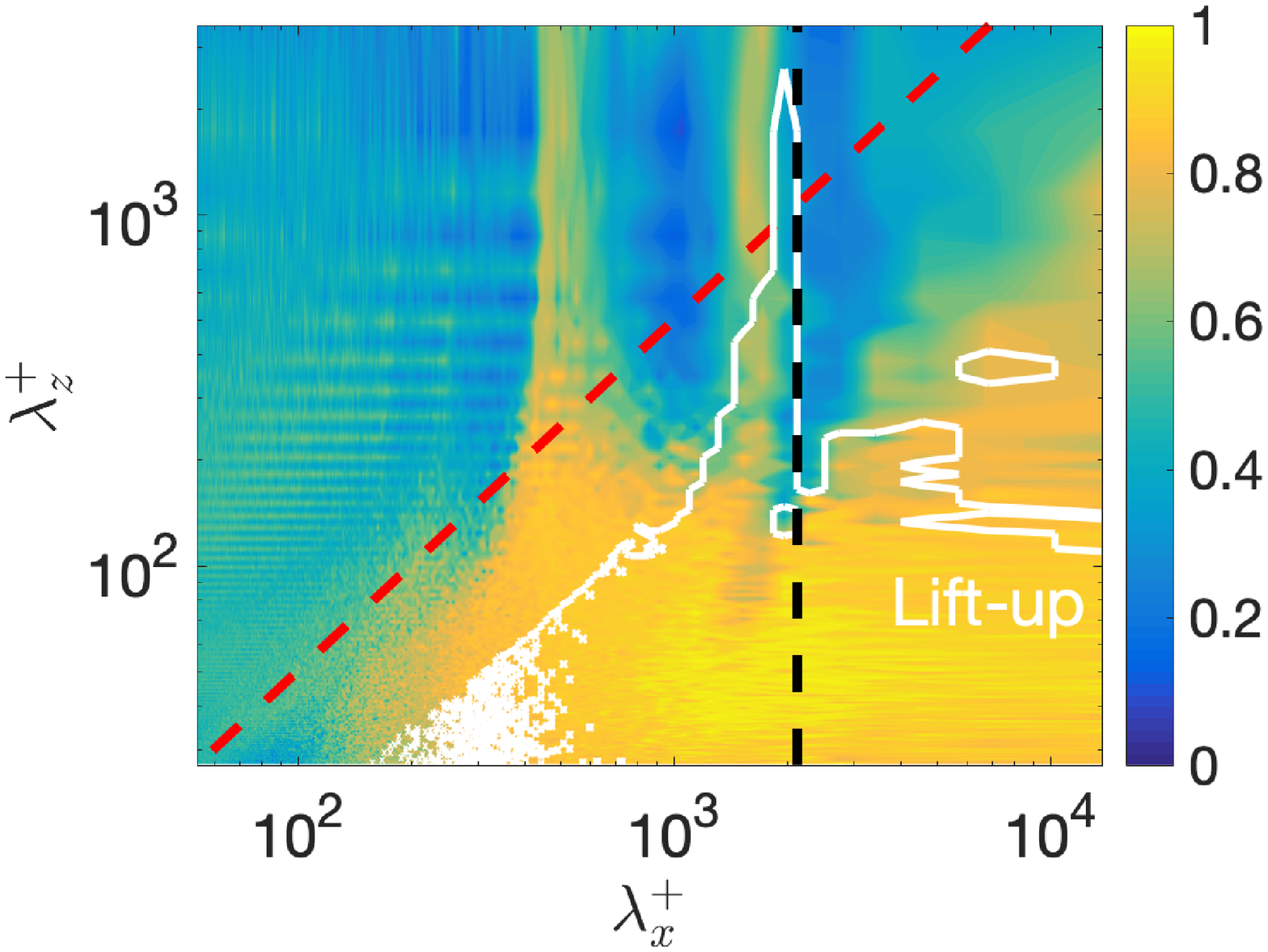}}
	\subfigure[$\beta$, $\lambda_{t}^{+} \approx 250$]{\includegraphics[width=0.325\textwidth,trim = 0 0 0 0,clip]{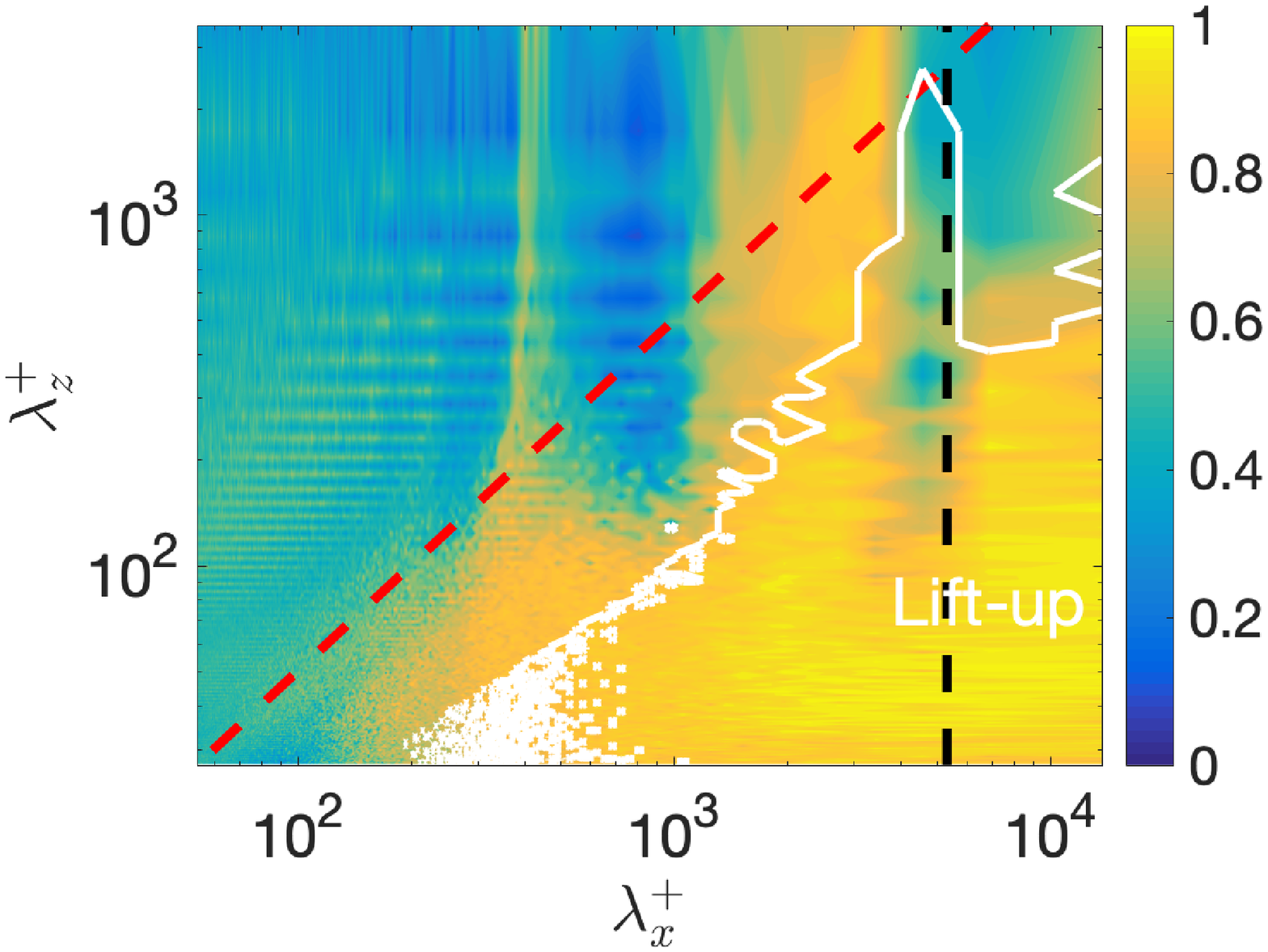}}
	\subfigure[$\beta$, $\lambda_{t}^{+} \approx 1000$]{\includegraphics[width=0.325\textwidth,trim = 0 0 0 0,clip]{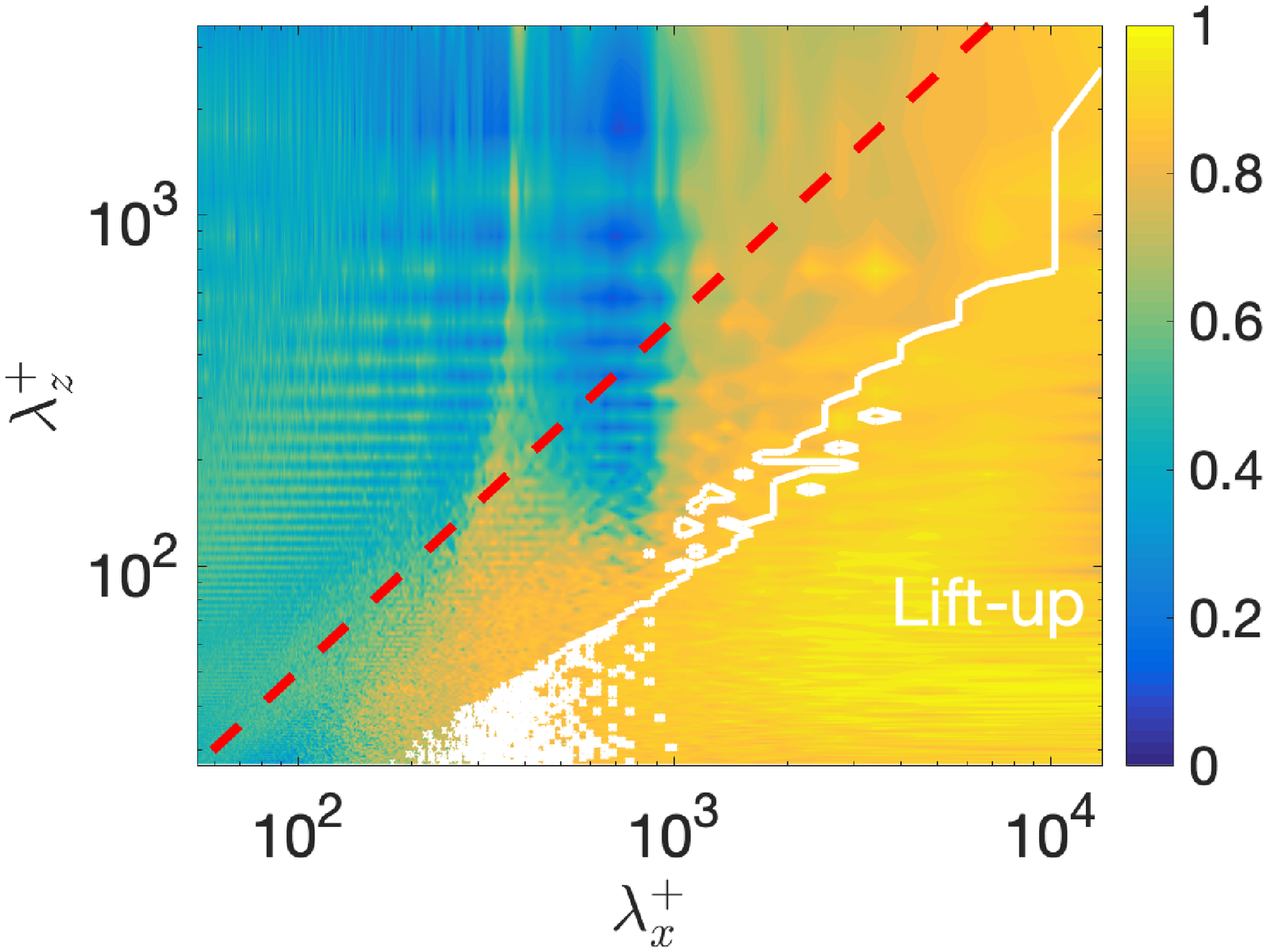}}

	\subfigure[$\beta_{(f_x=0)}$, $\lambda_{t}^{+} \approx 100$]{\includegraphics[width=0.325\textwidth,trim = 0 0 0 0,clip]{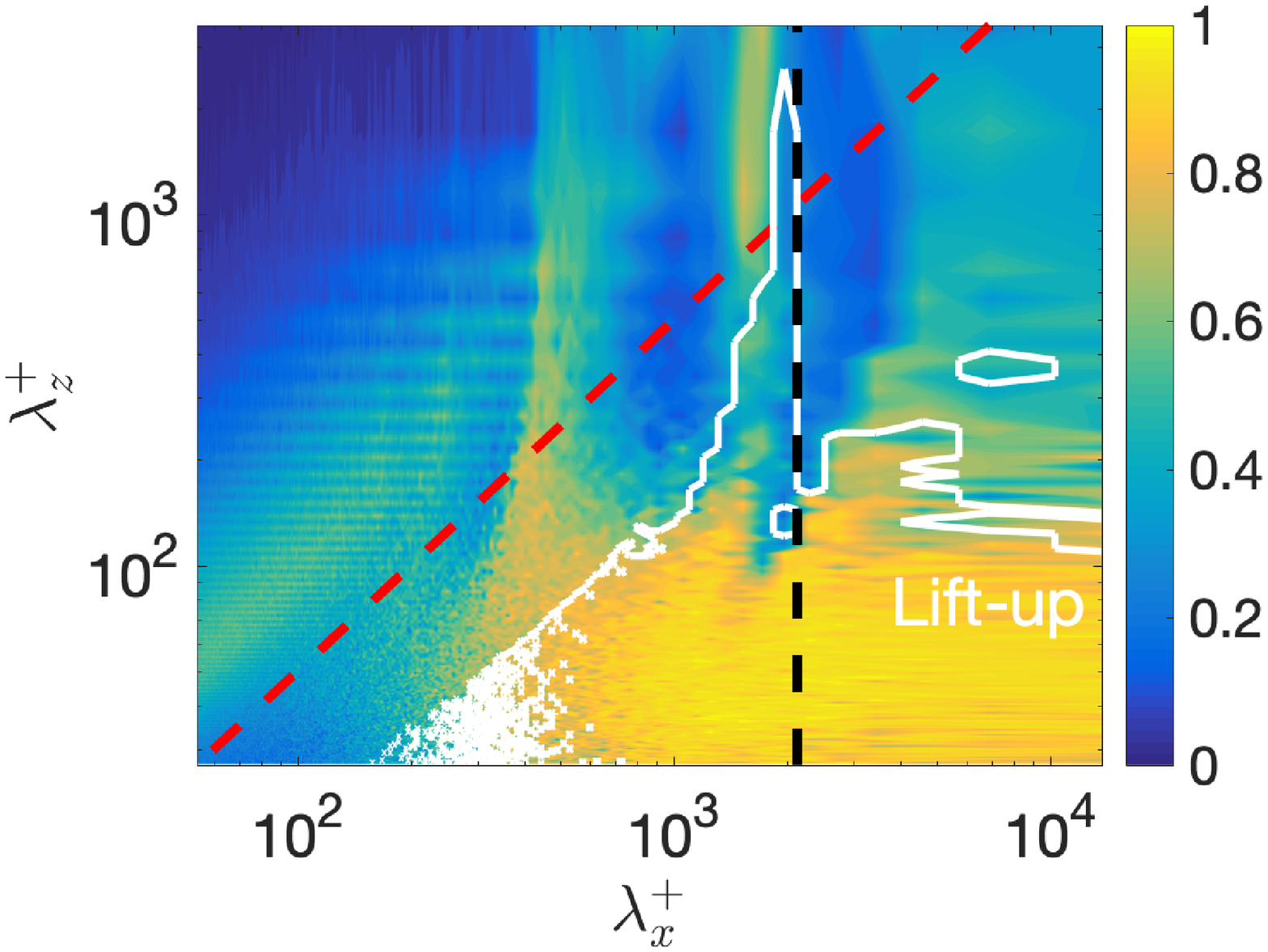}}
	\subfigure[$\beta_{(f_x=0)}$, $\lambda_{t}^{+} \approx 250$]{\includegraphics[width=0.325\textwidth,trim = 0 0 0 0,clip]{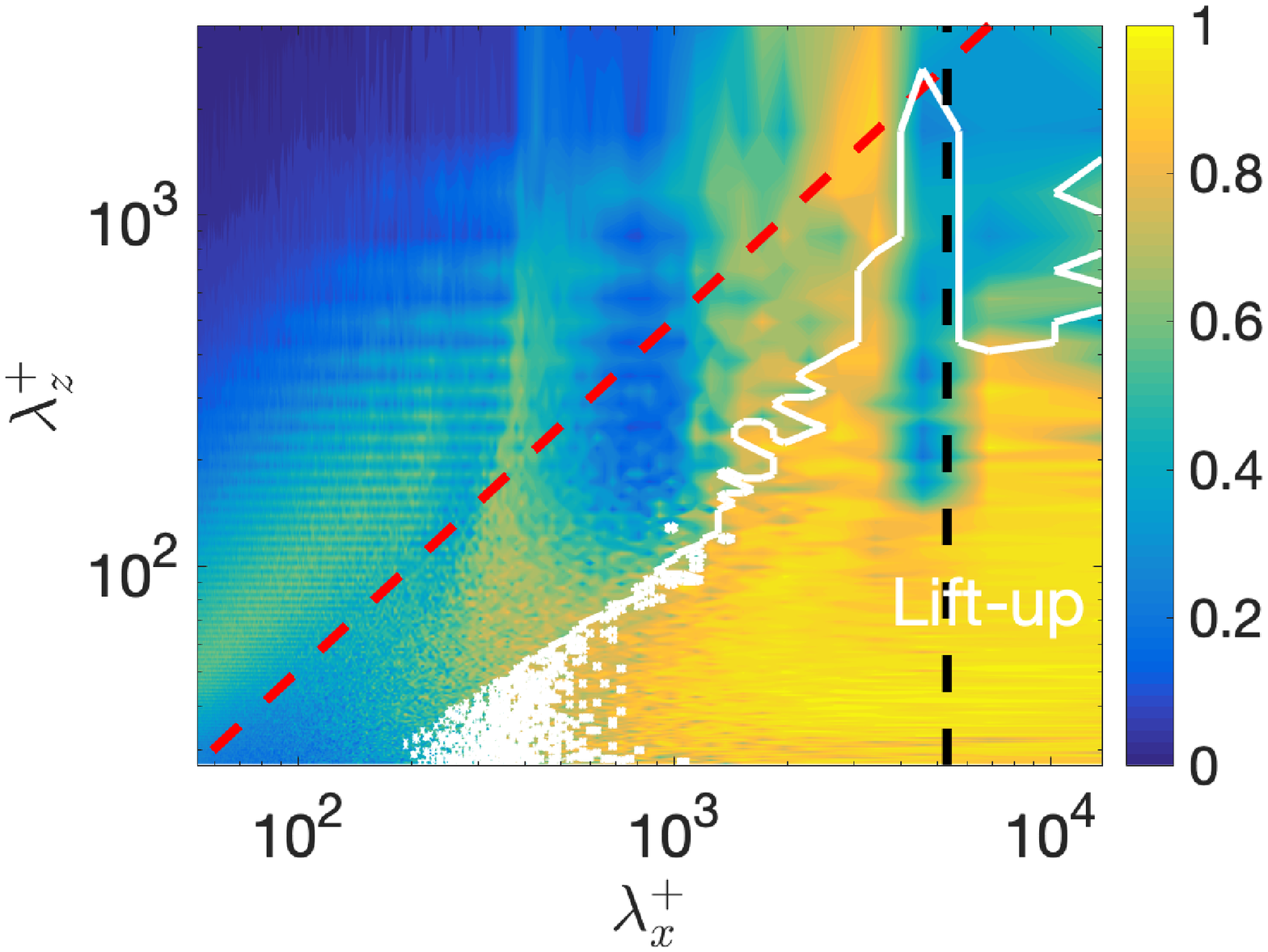}}
	\subfigure[$\beta_{(f_x=0)}$, $\lambda_{t}^{+} \approx 1500$]{\includegraphics[width=0.325\textwidth,trim = 0 0 0 0,clip]{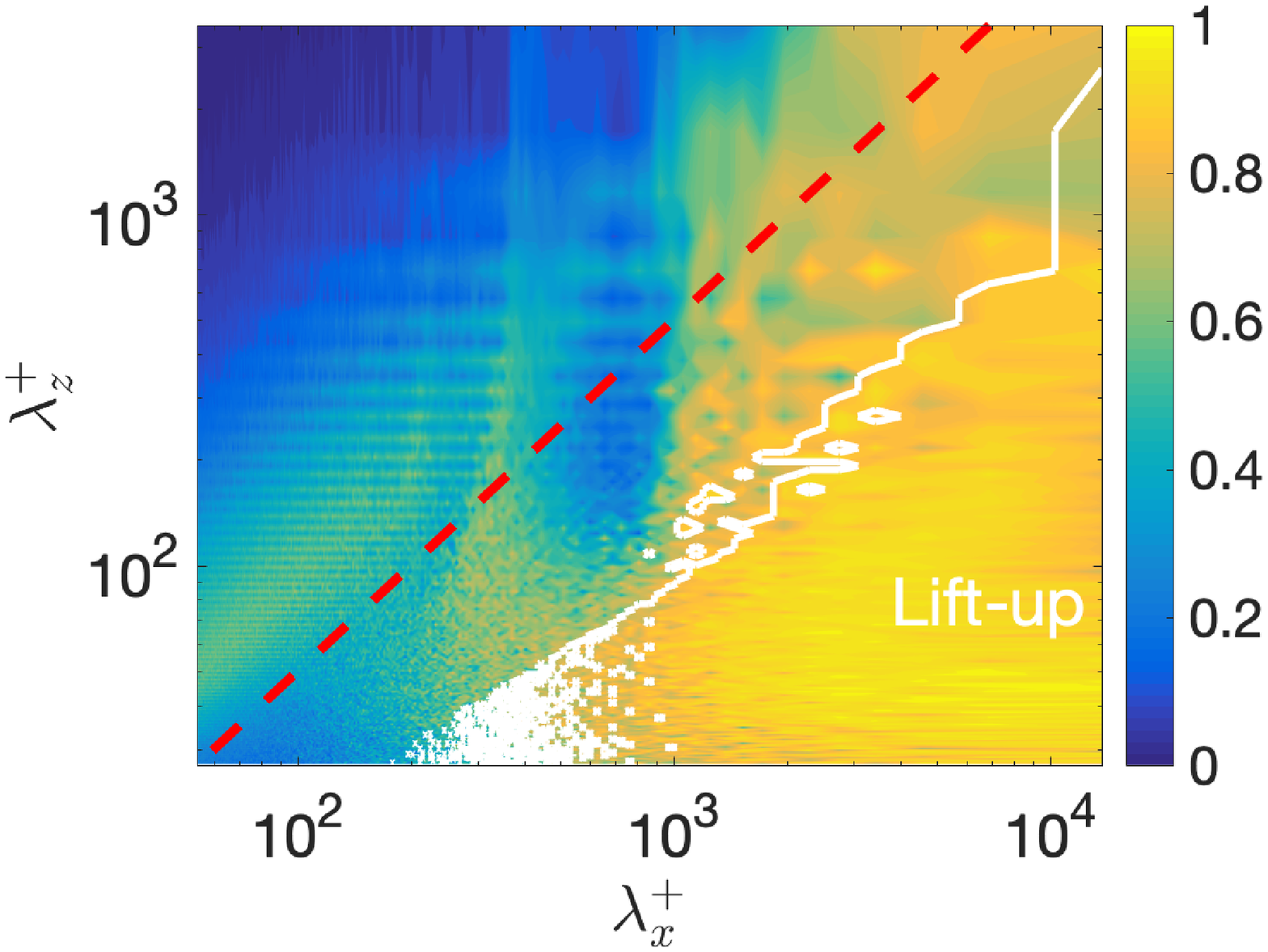}}

	\subfigure[$\log_{10}(\sigma_{1}/\sigma_{2})$, $\lambda_{t}^{+} \approx 100$]{\includegraphics[width=0.325\textwidth,trim = 0 0 0 0,clip]{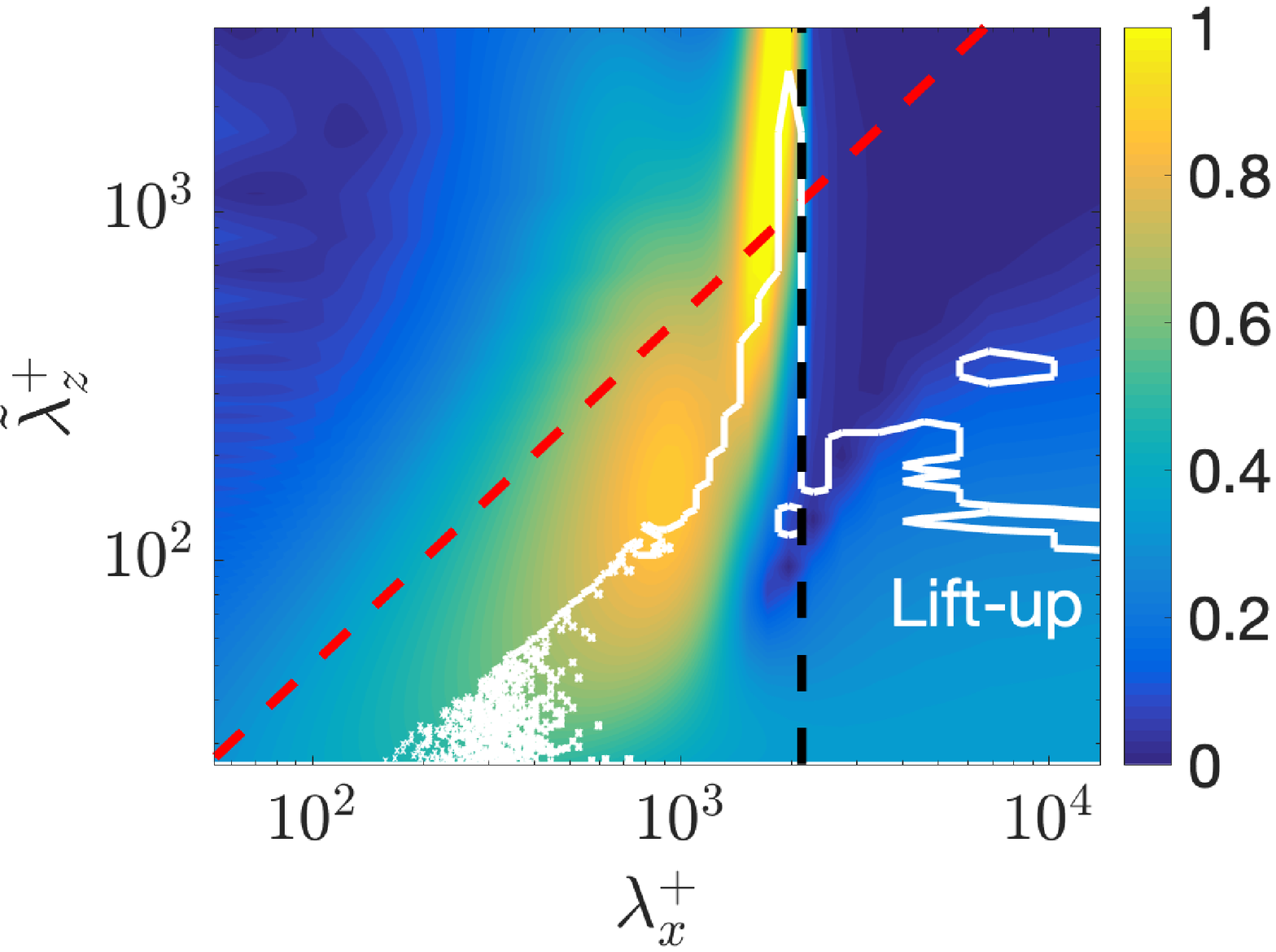}}
	\subfigure[$\log_{10}(\sigma_{1}/\sigma_{2})$, $\lambda_{t}^{+} \approx 250$]{\includegraphics[width=0.325\textwidth,trim = 0 0 0 0,clip]{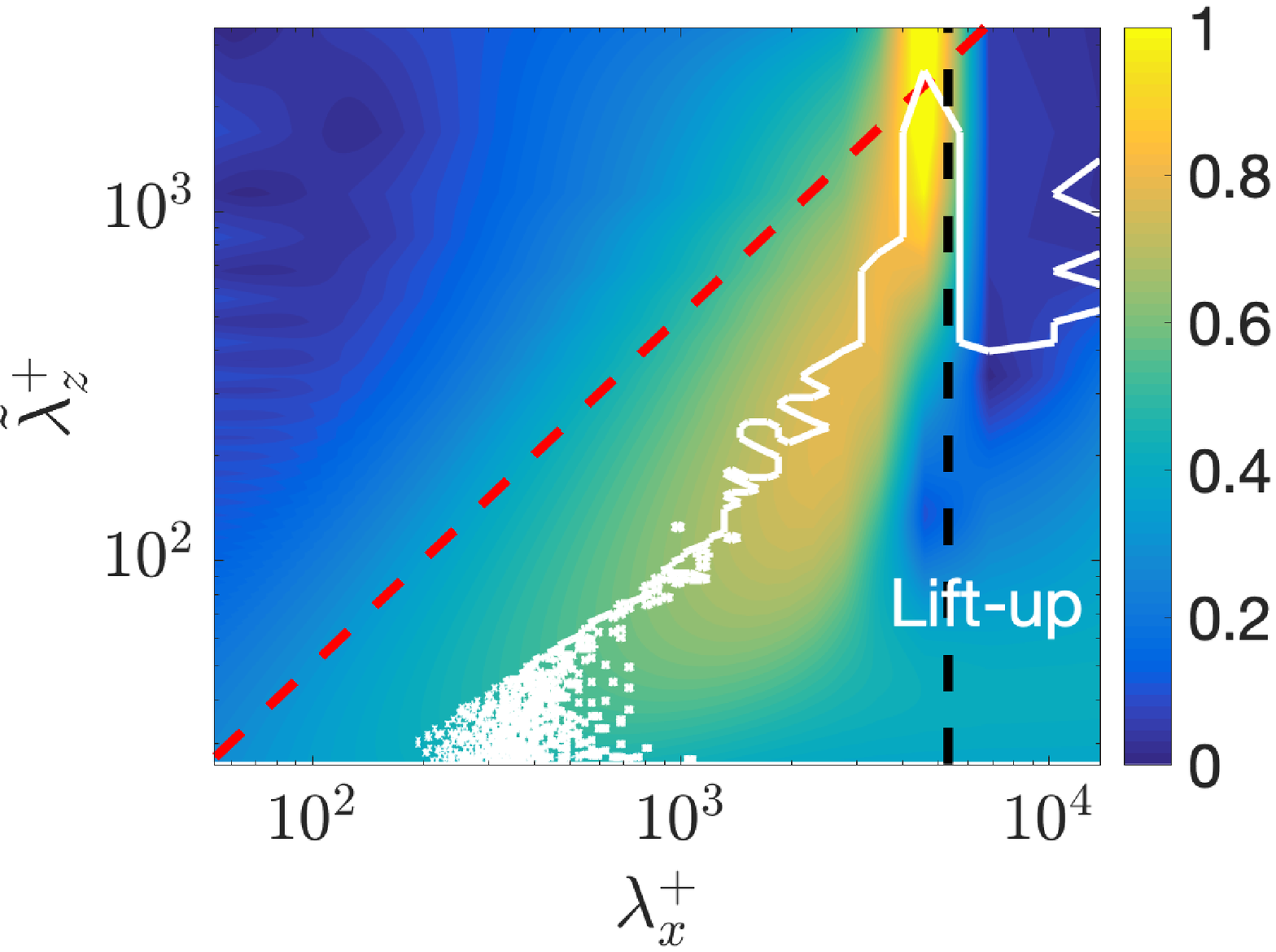}}
	\subfigure[$\log_{10}(\sigma_{1}/\sigma_{2})$, $\lambda_{t}^{+} \approx 1000$]{\includegraphics[width=0.325\textwidth,trim = 0 0 0 0,clip]{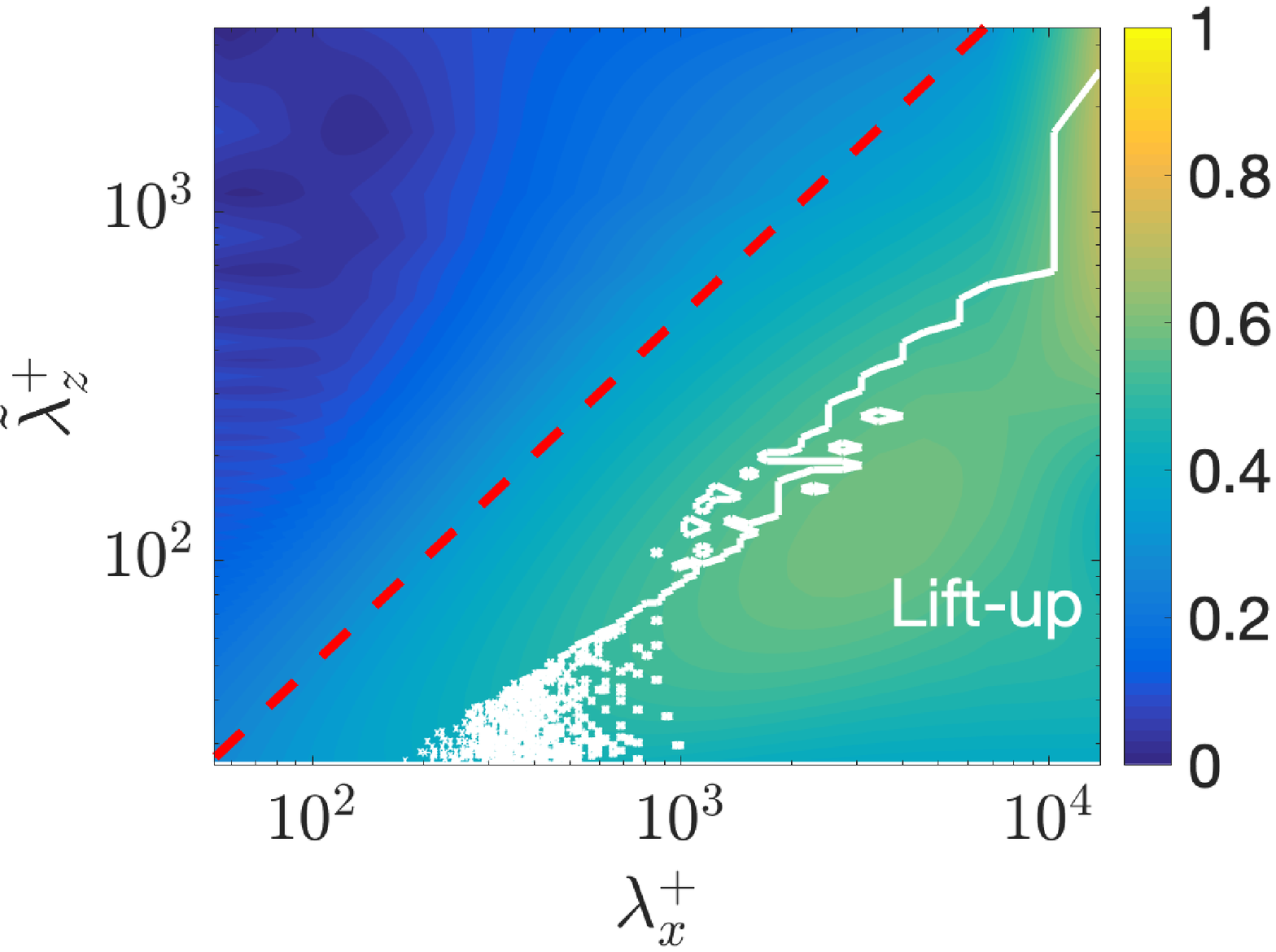}}

	\caption{(a,b,c) Agreement between first SPOD mode and optimal response from resolvent analysis characterized in terms of $\beta$. (d,e,f) Agreement between first SPOD and resolvent modes considering $f_x=0$, characterized in terms of $\beta_{(f_x=0)}$. (g,h,i) Ratio between optimal and suboptimal resolvent gains in logarithmic scale. Results for $Re_{\tau}=550$ and fixed frequencies: $\lambda_{t}^{+} \approx 100$, $250$ and $1000$ (from left to right). The region surrounded by the white line in all plots represents an indicator of lift-up mechanism from resolvent analysis. The red dashed line in all plots represents $\lambda_{x}^{+}=2 \lambda_{z}^{+}$.}
	\label{fig:beta550}
\end{figure}

Results of agreement between the first SPOD mode and the optimal response from resolvent analysis in terms of $\beta$ are shown in Figure \ref{fig:beta180} (a,b,c) for $Re_{\tau}=180$ at fixed frequencies correspondent to $\lambda_{t}^{+} \approx 100$, $250$ and $1500$; and in Figure \ref{fig:beta550} (a,b,c) for $Re_{\tau}=550$ at $\lambda_{t}^{+} \approx 100$, $250$ and $1000$. The frequencies are discretized by the application of the Welch method, and were chosen to be representative of the near-wall cycle $(\lambda_t^+ \approx 100)$ and of lower frequencies of larger structures. The red dashed line in all plots in Figures \ref{fig:beta180} and \ref{fig:beta550} represents the line $\lambda_{x}^{+}=2 \lambda_{z}^{+}$, such that below that line the structures are elongated in the streamwise direction with $\lambda_{x}^{+} > 2 \lambda_{z}^{+}$. The black dashed line represents $\lambda_{x}^{+}=U_{max}^{+} \lambda_{t}^{+}$, representing the limit where the phase velocity $c^{+}$ is equal to the velocity in the pipe center $U_{max}^{+}$. Results show for both Reynolds numbers that the coefficient $\beta$ is close to one for a large part of the parameter space, highlighting a significant region with very good agreement between the first SPOD and resolvent modes, most of it below the line $\lambda_{x}^{+} = 2 \lambda_{z}^{+} $, indicating that resolvent analysis leads to an accurate modeling of such turbulent structures in turbulent pipe flow for all the analysed frequencies. 

In order to explore features leading to better or worse agreement, $\beta \approx 1$ or $\beta \approx 0$, respectively, we evaluated for both friction Reynolds numbers the ratio between optimal and suboptimal resolvent gains in logarithmic scale $\log_{10}(\sigma_{1}/\sigma_{2})$, indicating regions where the optimal gain is much larger than the suboptimal; these results are shown in Figure \ref{fig:beta180} (g,h,i) for $Re_{\tau}=180$ and Figure \ref{fig:beta550} (g,h,i) for $Re_{\tau}=550$, for the considered frequencies in the preceding plots. We observe in general that regions where the first resolvent gain is much larger than the second correspond to the region of good SPOD-resolvent agreement, {\it i.e.} the triangular region below the red line $\lambda_{x}^{+}=2 \lambda_{z}^{+}$ and to the left of the vertical line marking $c^+ = U_{max}^{+}$. This behavior is observed even for some regions crossing the red line, for $\lambda_{x}^{+} < 2 \lambda_{z}^{+}$. This thus indicates that regions where the optimal response is much more dominant than the suboptimal may be accurately modelled considering the first SPOD mode. In this region the leading flow response predicted by resolvent analysis is much more amplified than suboptimal modes, which explains the agreement with leading modes from the DNS, as discussed in Section~\S\ref{sec:SPODres}. The region at the right of the vertical lines in the plots has low dominance of the leading resolvent mode, as shown in Figures \ref{fig:beta180} (g,h,i) and \ref{fig:beta550} (g,h,i), but nonetheless display good agreement with SPOD modes. The said regions are seen to lie to the right of the vertical line marking a phase speed equal to the mean velocity at the jet centerline. Such frequency and wavenumber combinations thus relate to disturbances with phase speed higher than the mean velocity anywhere in the pipe, and are \emph{a priori} not of much interest.

The analysis above highlights that a better agreement between leading SPOD and resolvent modes is observed when a certain mode dominance is verified by analysis of the linearized operator. We now investigate whether this dominance can be attributed to the lift-up mechanism. The region delimited by the white line in Figures \ref{fig:beta180} and \ref{fig:beta550} denotes the presence of the lift-up effect, purely from the resolvent analysis, at $Re_{\tau}=180$ and $550$, respectively. This region shows an indicator of lift-up mechanism, which is here considered to be relevant when the ratio of peak transverse and streamwise forcings, $\max(|f_y|/|f_x|)$ and $\max(|f_z|/|f_x|)$ (indicating streamwise vortices as optimal forcing) and the ratio of peak transverse and streamwise velocity components, $\max(|u|/|v|)$ and $\max(|u|/|w|)$ (indicating streaks of streamwise velocity as associated most amplified response) are simultaneously larger than 1. The regions satisfying these criteria are inside the ``lift-up" contour, or the white line, in Figures \ref{fig:beta180} and \ref{fig:beta550}. The result shows that the regions where the lift-up mechanism is present are close to those where $\beta$ is around 1, and also to parameters with dominance of the optimal forcing in resolvent analysis. The present results highlight that the lift-up mechanism is active for a wide range of frequencies and wavenumbers in turbulent pipe flow for both Reynolds numbers analysed here, with a strong amplification mechanism leading to structures that dominate the near-wall velocity field.

To further explore the lift-up effect, we performed the resolvent analysis neglecting the forcing in the streamwise direction (restricting $f_x=0$ using the $\mathcal{B}$ operator), which should retain the lift-up mechanism, as the streamwise forcing $f_x$ is not expected to be relevant in this case. We evaluated the agreement between first SPOD and resolvent mode for this case, denoted as $\beta_{(f_x=0)}$. Results are shown in Figures \ref{fig:beta180} (d,e,f) and \ref{fig:beta550} (d,e,f) for $Re_{\tau}=180$ and $550$, respectively. Similar agreement between SPOD and resolvent modes is obtained, despite the fact that the streamwise forcing was neglected. Parameters where good agreement is found also correspond to the region surrounded by the white line, defined by the lift-up indicator. Such results confirm that in the roughly triangular region, marking streamwise elongated structures with phase speeds lower than the centreline velocity, the lift-up mechanism is dominant, with transverse forcing components leading to streamwise vortices that in turn create amplified streaks. The relevance of such mechanism in resolvent analysis is seen by a large gain ratio, $\sigma_1 \ll \sigma_2$, and the dominance of the optimal response leads to good agreement with the leading SPOD mode from the DNS.

\begin{figure}
	\centering
	\subfigure[]{\includegraphics[width=0.49\textwidth,trim = 0 0 0 0,clip]{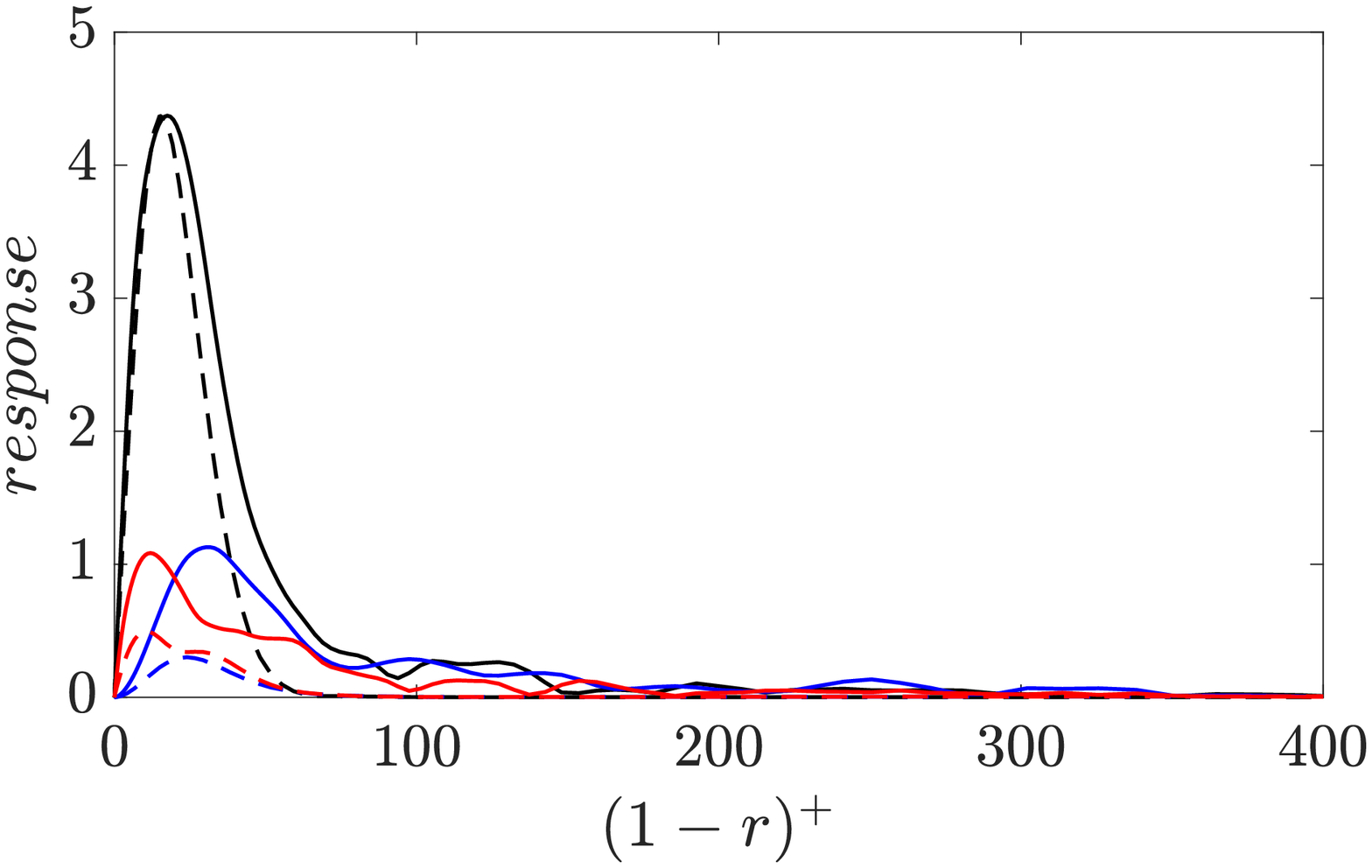}}
	\subfigure[]{\includegraphics[width=0.49\textwidth,trim = 0 0 0 0,clip]{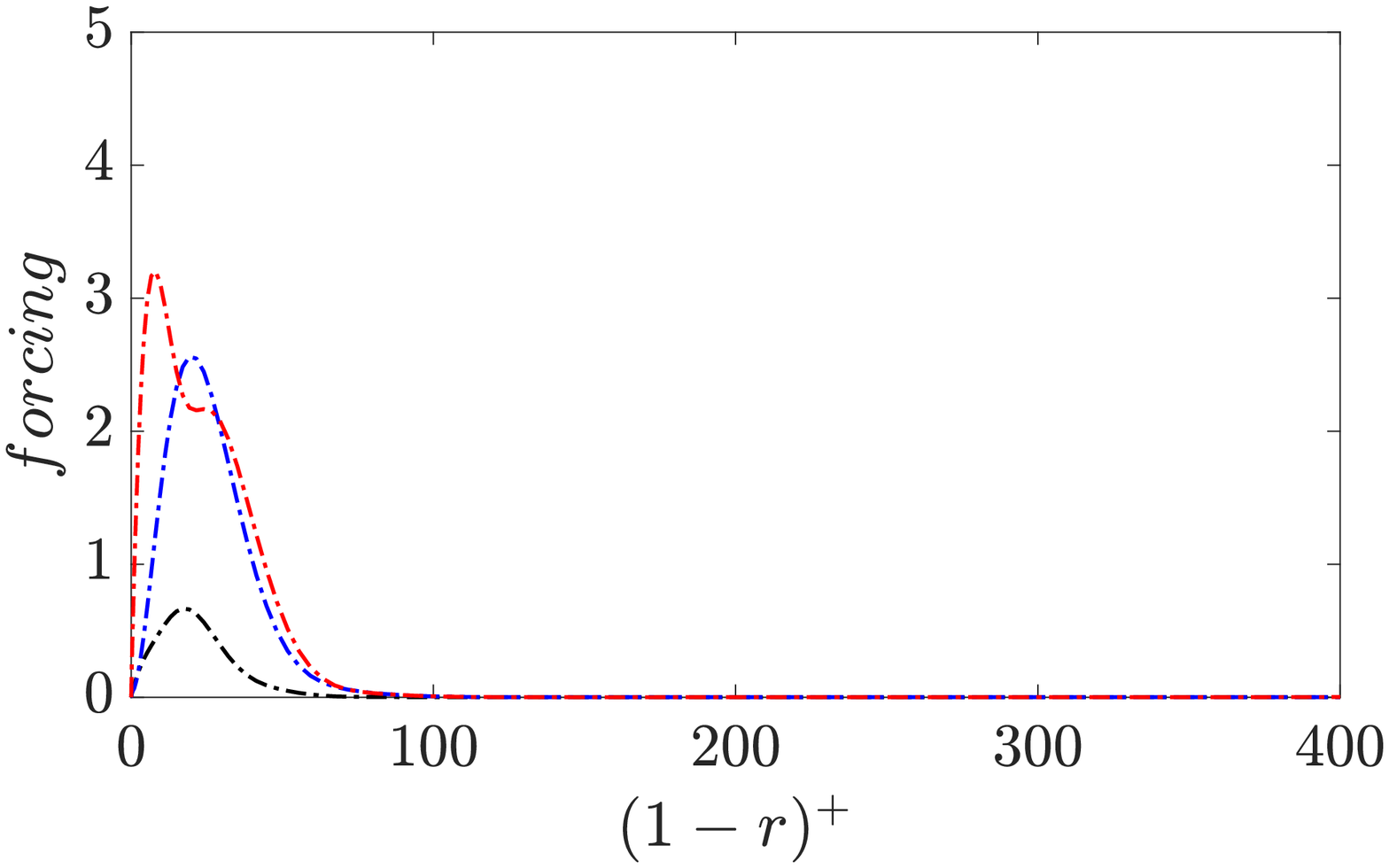}}

	\caption{Case of $\beta=0.95$ for $Re_{\tau}=550$ and combination $(\lambda_{x}^{+},\lambda_{z}^{+},\lambda_{t}^{+})\approx (1000,100,100)$, showing in (a) the comparison between first SPOD mode and the optimal response from resolvent analysis for the three velocity components and in (b) the associated forcing from resolvent analysis for the three components $(f_x,f_y,f_z)$.}
	\label{fig:Re550_case1}
\end{figure}

\begin{figure}
	\centering
	\subfigure[]{\includegraphics[width=0.49\textwidth,trim = 0 0 0 0,clip]{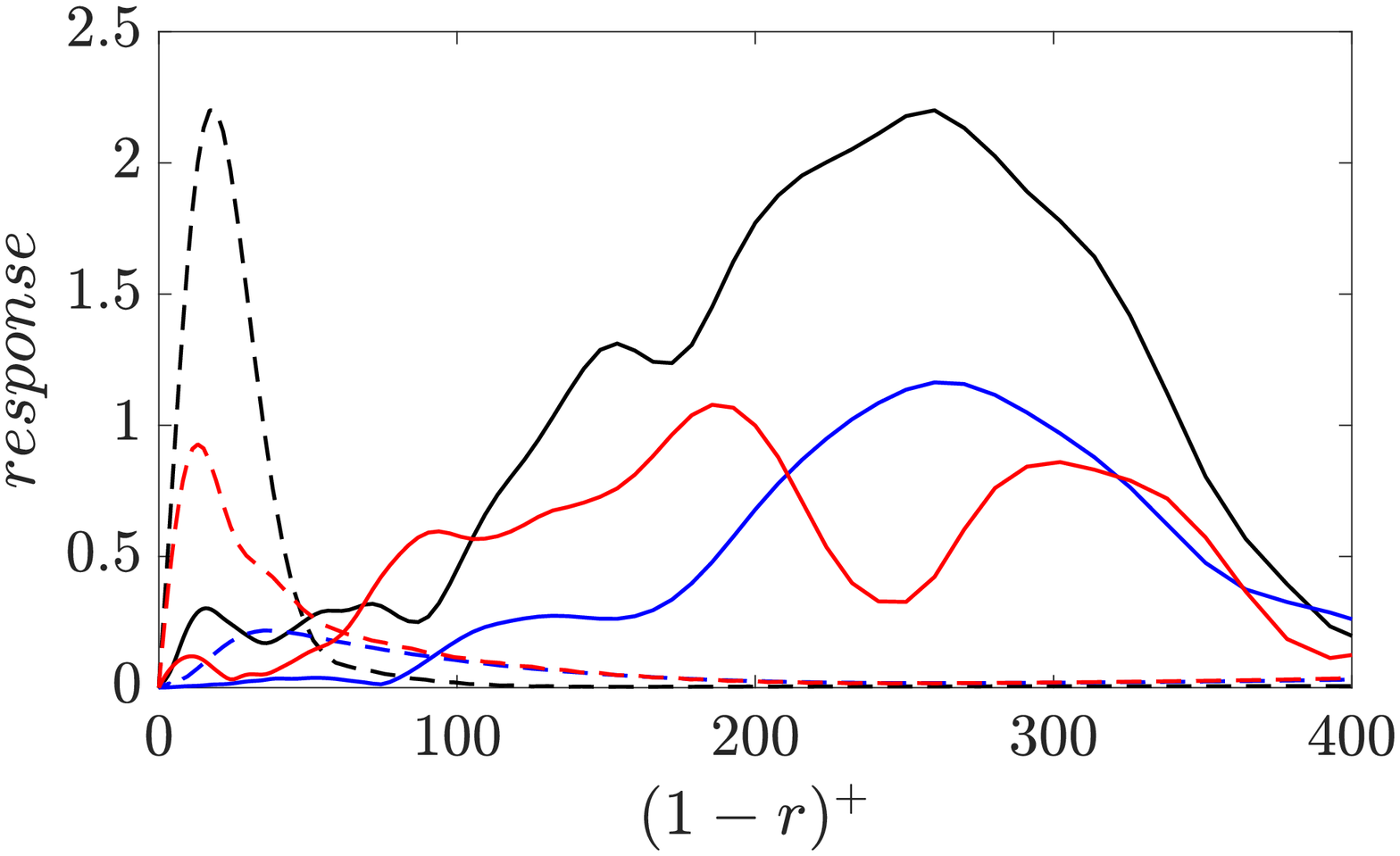}}
	\subfigure[]{\includegraphics[width=0.49\textwidth,trim = 0 0 0 0,clip]{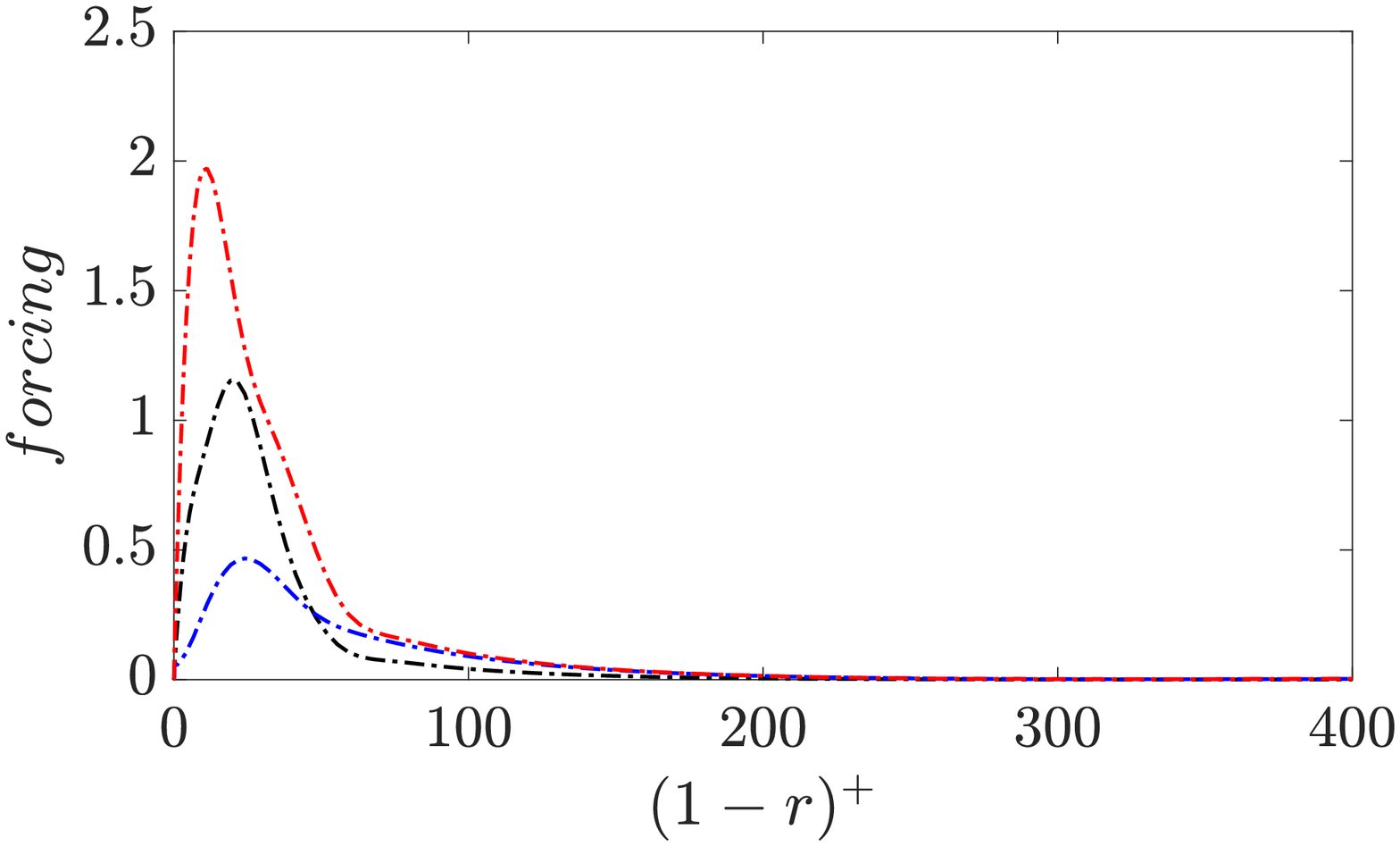}}
	
	\caption{Case of $\beta=0.11$ for $Re_{\tau}=550$ and combination $(\lambda_{x}^{+},\lambda_{z}^{+},\lambda_{t}^{+})\approx (1000,500,100)$, showing in (a) the comparison between first SPOD mode and the optimal response from resolvent analysis for the three velocity components and in (b) the associated forcing from resolvent analysis for the three components $(f_x,f_y,f_z)$.}
	\label{fig:Re550_case2}
\end{figure}

To see some sample results of the above analysis in more detail, Figure \ref{fig:Re550_case1} shows for $Re_{\tau}=550$ results of comparisons between SPOD and resolvent modes for the combination $(\lambda_{x}^{+},\lambda_{z}^{+},\lambda_{t}^{+})\approx (1000,100,100)$, which is representative of the signature of the near-wall cycle of streaks and streamwise vortices, with $\beta = 0.95$. In contrast, Figure \ref{fig:Re550_case2} shows the same kind of comparison, but for a larger azimuthal wavelength $(\lambda_{x}^{+},\lambda_{z}^{+},\lambda_{t}^{+})\approx (1000,500,100)$, which leads to a $\beta = 0.11$ and thus substantial discrepancies between SPOD and resolvent modes. Figures \ref{fig:Re550_case1} (a) and \ref{fig:Re550_case2} (a) show the first SPOD mode compared with the optimal response from resolvent analysis for the three velocity components $(u,v,w)$, for the cases with $\beta = 0.95$ and $\beta = 0.11$, respectively. We can see that the resolvent analysis reproduces very well the coherent structures obtained using SPOD for the case $(\lambda_{x}^{+},\lambda_{z}^{+},\lambda_{t}^{+})\approx (1000,100,100)$, which has a large ratio $\lambda_{x}^{+}/\lambda_{z}^{+}=10$, indicative of very elongated streaky structures. The agreement is better for the streamwise component $u$, while the $v$ and $w$ components have similar shapes in the resolvent and SPOD modes, but have larger amplitudes in the case of the SPOD. This indicates that the streaks are reproduced well, whereas the in-plane velocities responsible for the streamwise vortices are underpredicted by a factor of around three. The phases between velocity components are nonetheless matched, as shown in Figure \ref{fig:uvw_2D_550}; otherwise, a lower agreement metric would be obtained. On the other hand, the optimal response from the resolvent is not able to model the structure for the case $(\lambda_{x}^{+},\lambda_{z}^{+},\lambda_{t}^{+})\approx (1000,500,100)$: here, the ratio $\lambda_{x}^{+}/\lambda_{z}^{+} \approx 2$ is much lower, departing from the streaky disturbances typical of the lift-up mechanism. In particular, the spanwise wavelength of $\lambda_{z}^{+}=500$ implies that these structures are centered farther from the wall. This is evident from the fact that the SPOD modes exhibit significant energy in the range from $(1-r)^{+}=200$ to around $400$. On the other hand, but as in the case with $\beta=0.95$, the resolvent modes exhibit high energy mainly in the near-wall region, and are therefore unable to reproduce the flow dynamics in this case.

Figures \ref{fig:Re550_case1} (b) and \ref{fig:Re550_case2} (b) show the forcing from resolvent analysis for the three components $(f_x,f_y,f_z)$, in the cases with $\beta = 0.95 $ and $\beta = 0.11$, respectively. We can notice that when $\beta \approx 1$ the forcing corresponds to streamwise vortices, since $f_y$ and $f_z$ are simultaneously larger than $f_x$. On the other hand, when $\beta \approx 0$ the forcing can no longer be associated to such streamwise vortices, since here $f_x$ becomes much larger than $f_y$. These results indicate that the lift-up mechanism is not present in that case, and thus reinforce the conclusion that resolvent analysis is an adequate reduced-order model to reproduce streaky structures associated with the lift-up mechanism. The observed concordances were generally for structures with peak in the buffer layer for both Reynolds numbers.

\section{Conclusions}
\label{sec:conclusions}

In the present study we used signal processing of a DNS, based on SPOD, to identify near-wall coherent structures in a turbulent pipe flow for friction Reynolds numbers $Re_\tau=180$ and $550$. In order to model such structures, a theoretical approach, {\it i.e.} resolvent analysis, was used. The homogeneous directions of this flow allow the evaluation of SPOD and resolvent analysis over a range of streamwise and azimuthal wavenumbers and frequencies. The mean flow was used as a basis for the computation of resolvent modes; optimal responses were considered as the most likely structures to be excited by non-linear terms in the Navier--Stokes system, particularly when the gain of the optimal forcing is much larger than for suboptimal ones~\citep{beneddine2016conditions,cavalieri2019wave}. Coherent structures in the flow were extracted using SPOD, and we carried out thorough quantitative comparisons between leading response modes from the resolvent analysis and the SPOD eigenfunctions.

For both Reynolds numbers, the results show good agreement between SPOD and resolvent, mostly for $2 \lambda_{z}^{+} \le \lambda_{x}^{+}$. These are parameters related to streaky structures, with aspect ratio (streamwise over azimuthal extent) larger than 2. We evaluated the ratio between first and second SPOD eigenvalues, as well the ratio between optimal and suboptimal gain from resolvent analysis, and observed that the regions where those ratios have larger values correspond to cases where the agreement between SPOD and resolvent modes are good. 

We also explored the physical reasons behind this agreement by introducing an indicator of the lift-up mechanism using the optimal forcing and associated response from resolvent analysis. Such a mechanism is considered as active when the forcing is related to streamwise vortices and the associated responses to streaks. The results show a clear lift-up effect for wavenumbers and frequencies with good agreement between SPOD and resolvent modes. Also, the observed concordances were generally for structures with peak in the buffer layer.

In conclusion, based on our results, it can be stated that the resolvent analysis provides a simplified model leading to an accurate representation of coherent structures mostly for the cases where the lift-up mechanism is present, with the optimal forcing corresponding to transverse components shaped as streamwise vortices, and the associated response corresponding to streaks. Such structures are observed for a broad range of frequencies and wavenumbers, which indicates that the lift-up effect occurs over a wide range of scales in turbulent pipe flow. Earlier studies in turbulent channel flow have applied transient growth \citep{del2006linear} and resolvent analysis  \citep{mckeon2010critical}, extracting structures consistent with observations from simulation and experiment. Here, the analysis is made for a broad range of scales, showing the relevance of lift-up for the studied structures and highlighting the pertinence of linearized models.


It is not surprising to find streamwise vortices leading to streaks in wall-bounded turbulence, since this has been considered as an important part of the dynamics of such flows for some time \citep{kline1967structure,landahl1980note,hamilton1995regeneration}. The present results highlight the relevance of this mechanism for most of the parameters considered in turbulent pipe flow, which can be understood by the clear dominance of the optimal forcing, with the shape of streamwise vortices, in leading to amplified flow responses of streaky shape. Lift-up thus is naturally selected as the preferred mechanism giving rise to streaky structures in turbulent pipe flow.

The use of resolvent analysis for such a wide range of turbulent scales is relevant. For parallel flows, in particular, the resolvent operator only needs to be discretised in the radial direction, and forcing and response modes can be obtained in fast computations. This allows simple predictions of the dominant structures in turbulent flows. Moreover, reconstructions of flow fluctuations from a limited number of sensors are also possible using the resolvent operator \cite{towne2020resolvent}, which opens possibilities for closed-loop control of turbulent flows. As the lift-up effect studied here is also the basis of bypass transition in boundary layers~\citep{andersson1999optimal, brandt2014lift}, extension of control methods used in bypass transition~\citep{sasaki2019role} is a promising direction for the control of wall-bounded turbulence.

\section*{Acknowledgments}
The authors acknowledge the financial support received from Conselho Nacional de Desenvolvimento Cient\'{i}fico e Tecnol\'{o}gico, CNPq, under grant No. 310523/2017-6, and by CAPES through the PROEX program. We also acknowledge funding from CISB.

Declaration of Interests. The authors report no conflict of interest.

\bibliographystyle{jfm}
\bibliography{paper_pipe_flow_2020}

\newcommand{\noop}[1]{}
\begin{thebibliography}{44}
\expandafter\ifx\csname natexlab\endcsname\relax\def\natexlab#1{#1}\fi
\def\au#1{#1} \def\ed#1{#1} \def\yr#1{#1}\def\at#1{#1}\def\jt#1{\textit{#1}}
  \def\bt#1{#1}\def\bvol#1{\textbf{#1}} \def\vol#1{#1} \def\pg#1{#1}
  \def\publ#1{#1}\def\arxiv#1{#1}\def\org#1{#1}\def\st#1{\textit{#1}}

\bibitem[Abreu {\em et~al.\/}(2017)Abreu, Cavalieri \& Wolf]{abreu2017coherent}
{\sc \au{Abreu, L.~I.}, \au{Cavalieri, A. V.~G.} \& \au{Wolf, W.~R.}} \yr{2017}
  Coherent hydrodynamic waves and trailing-edge noise.  \bt{In {\em 23rd
  AIAA/CEAS Aeroacoustics Conference\/}},  \pg{p. 3173}.

\bibitem[Andersson {\em et~al.\/}(1999)Andersson, Berggren \&
  Henningson]{andersson1999optimal}
{\sc \au{Andersson, P.}, \au{Berggren, M.} \& \au{Henningson, D.}} \yr{1999}
  \at{Optimal disturbances and bypass transition in boundary layers}.
  \jt{Physics of Fluids}  \bvol{11}~(1),  \pg{134--150}.

\bibitem[Beneddine {\em et~al.\/}(2016)Beneddine, Sipp, Arnault, Dandois \&
  Lesshafft]{beneddine2016conditions}
{\sc \au{Beneddine, S.}, \au{Sipp, D.}, \au{Arnault, A.}, \au{Dandois, J.} \&
  \au{Lesshafft, L.}} \yr{2016}  \at{Conditions for validity of mean flow
  stability analysis}.  \jt{Journal of Fluid Mechanics}  \bvol{798},
  \pg{485--504}.

\bibitem[Brandt(2014)]{brandt2014lift}
{\sc \au{Brandt, L.}} \yr{2014}  \at{The lift-up effect: the linear mechanism
  behind transition and turbulence in shear flows}.  \jt{European Journal of
  Mechanics-B/Fluids}  \bvol{47},  \pg{80--96}.

\bibitem[Cavalieri {\em et~al.\/}(2019)Cavalieri, Jordan \&
  Lesshafft]{cavalieri2019wave}
{\sc \au{Cavalieri, A.V.G.}, \au{Jordan, P.} \& \au{Lesshafft, L.}} \yr{2019}
  \at{Wave-packet models for jet dynamics and sound radiation}.  \jt{Applied
  Mechanics Reviews}  \bvol{71}~(2),  \pg{020802}.

\bibitem[Del~Alamo \& Jimenez(2006)]{del2006linear}
{\sc \au{Del~Alamo, J.C.} \& \au{Jimenez, J.}} \yr{2006}  \at{Linear energy
  amplification in turbulent channels}.  \jt{Journal of Fluid Mechanics}
  \bvol{559},  \pg{205--213}.

\bibitem[Eitel-Amor {\em et~al.\/}(2014)Eitel-Amor, {\"O}rl{\"u} \&
  Schlatter]{eitel2014simulation}
{\sc \au{Eitel-Amor, Georg}, \au{{\"O}rl{\"u}, Ramis} \& \au{Schlatter,
  Philipp}} \yr{2014}  \at{Simulation and validation of a spatially evolving
  turbulent boundary layer up to $re_{\theta}=8300$}.  \jt{International
  Journal of Heat and Fluid Flow}  \bvol{47},  \pg{57--69}.

\bibitem[El~Khoury {\em et~al.\/}(2013)El~Khoury, Schlatter, Noorani, Fischer,
  Brethouwer \& Johansson]{el2013direct}
{\sc \au{El~Khoury, G.K.}, \au{Schlatter, P.}, \au{Noorani, A.}, \au{Fischer,
  P.F.}, \au{Brethouwer, G.} \& \au{Johansson, A.V.}} \yr{2013}  \at{Direct
  numerical simulation of turbulent pipe flow at moderately high reynolds
  numbers}.  \jt{Flow, Turbulence and Combustion}  \bvol{91}~(3),
  \pg{475--495}.

\bibitem[Ellingsen \& Palm(1975)]{ellingsen1975stability}
{\sc \au{Ellingsen, T.} \& \au{Palm, E.}} \yr{1975}  \at{Stability of linear
  flow}.  \jt{The Physics of Fluids}  \bvol{18}~(4),  \pg{487--488}.

\bibitem[Farrell \& Ioannou(1993)]{farrell1993stochastic}
{\sc \au{Farrell, Brian~F} \& \au{Ioannou, Petros~J}} \yr{1993}  \at{Stochastic
  forcing of the linearized navier--stokes equations}.  \jt{Physics of Fluids
  A: Fluid Dynamics}  \bvol{5}~(11),  \pg{2600--2609}.

\bibitem[Farrell \& Ioannou(2012)]{farrell2012dynamics}
{\sc \au{Farrell, Brian~F} \& \au{Ioannou, Petros~J}} \yr{2012}  \at{Dynamics
  of streamwise rolls and streaks in turbulent wall-bounded shear flow}.
  \jt{Journal of Fluid Mechanics}  \bvol{708},  \pg{149--196}.

\bibitem[Fischer {\em et~al.\/}(2008)Fischer, Lottes \& Kerkemeier]{fischer}
{\sc \au{Fischer, P.F.}, \au{Lottes, J.W.} \& \au{Kerkemeier, S.G.}} \yr{2008}
  Nek5000: Open source spectral element cfd solver. available at:
  http://nek5000.mcs.anl.gov.

\bibitem[Gupta {\em et~al.\/}(1971)Gupta, Laufer \& Kaplan]{gupta1971spatial}
{\sc \au{Gupta, A.K.}, \au{Laufer, J.} \& \au{Kaplan, R.E.}} \yr{1971}
  \at{Spatial structure in the viscous sublayer}.  \jt{Journal of Fluid
  Mechanics}  \bvol{50}~(3),  \pg{493--512}.

\bibitem[Hall \& Sherwin(2010)]{hall2010streamwise}
{\sc \au{Hall, Philip} \& \au{Sherwin, Spencer}} \yr{2010}  \at{Streamwise
  vortices in shear flows: harbingers of transition and the skeleton of
  coherent structures}.  \jt{Journal of Fluid Mechanics}  \bvol{661},
  \pg{178--205}.

\bibitem[Hamilton {\em et~al.\/}(1995)Hamilton, Kim \&
  Waleffe]{hamilton1995regeneration}
{\sc \au{Hamilton, J.M.}, \au{Kim, J.} \& \au{Waleffe, F.}} \yr{1995}
  \at{Regeneration mechanisms of near-wall turbulence structures}.  \jt{Journal
  of Fluid Mechanics}  \bvol{287},  \pg{317--348}.

\bibitem[Hellstr{\"o}m {\em et~al.\/}(2016)Hellstr{\"o}m, Marusic \&
  Smits]{hellstrom2016self}
{\sc \au{Hellstr{\"o}m, L.H.O.}, \au{Marusic, I.} \& \au{Smits, A.J.}}
  \yr{2016}  \at{Self-similarity of the large-scale motions in turbulent pipe
  flow}.  \jt{Journal of Fluid Mechanics}  \bvol{792}.

\bibitem[Hoyas \& Jim{\'e}nez(2006)]{hoyas2006scaling}
{\sc \au{Hoyas, S.} \& \au{Jim{\'e}nez, J.}} \yr{2006}  \at{Scaling of the
  velocity fluctuations in turbulent channels up to $re_{\tau}=2003$}.
  \jt{Physics of Fluids}  \bvol{18}~(1),  \pg{011702}.

\bibitem[Hutchins \& Marusic(2007)]{hutchins2007evidence}
{\sc \au{Hutchins, N.} \& \au{Marusic, I.}} \yr{2007}  \at{Evidence of very
  long meandering features in the logarithmic region of turbulent boundary
  layers}.  \jt{Journal of Fluid Mechanics}  \bvol{579},  \pg{1--28}.

\bibitem[Hwang \& Cossu(2010)]{hwang2010linear}
{\sc \au{Hwang, Y.} \& \au{Cossu, C.}} \yr{2010}  \at{Linear non-normal energy
  amplification of harmonic and stochastic forcing in the turbulent channel
  flow}.  \jt{Journal of Fluid Mechanics}  \bvol{664},  \pg{51--73}.

\bibitem[Jim{\'e}nez(2018)]{jimenez2018coherent}
{\sc \au{Jim{\'e}nez, Javier}} \yr{2018}  \at{Coherent structures in
  wall-bounded turbulence}.  \jt{Journal of Fluid Mechanics}  \bvol{842}.

\bibitem[Jovanovi{\'c} \& Bamieh(2005)]{jovanovic2005componentwise}
{\sc \au{Jovanovi{\'c}, Mihailo~R} \& \au{Bamieh, Bassam}} \yr{2005}
  \at{Componentwise energy amplification in channel flows}.  \jt{Journal of
  Fluid Mechanics}  \bvol{534},  \pg{145--183}.

\bibitem[Kline {\em et~al.\/}(1967)Kline, Reynolds, Schraub \&
  Runstadler]{kline1967structure}
{\sc \au{Kline, S.~J.}, \au{Reynolds, W.~C.}, \au{Schraub, F.~A.} \&
  \au{Runstadler, P.~W.}} \yr{1967}  \at{The structure of turbulent boundary
  layers}.  \jt{Journal of Fluid Mechanics}  \bvol{30}~(4),  \pg{741--773}.

\bibitem[Landahl(1980)]{landahl1980note}
{\sc \au{Landahl, MT}} \yr{1980}  \at{A note on an algebraic instability of
  inviscid parallel shear flows}.  \jt{Journal of Fluid Mechanics}
  \bvol{98}~(2),  \pg{243--251}.

\bibitem[Lesshafft {\em et~al.\/}(2019)Lesshafft, Semeraro, Jaunet, Cavalieri
  \& Jordan]{lesshafft2019resolvent}
{\sc \au{Lesshafft, Lutz}, \au{Semeraro, Onofrio}, \au{Jaunet, Vincent},
  \au{Cavalieri, Andr{\'e}~VG} \& \au{Jordan, Peter}} \yr{2019}
  \at{Resolvent-based modeling of coherent wave packets in a turbulent jet}.
  \jt{Physical Review Fluids}  \bvol{4}~(6),  \pg{063901}.

\bibitem[Lumley(1967)]{lumley1967structure}
{\sc \au{Lumley, John~Leask}} \yr{1967}  \at{The structure of inhomogeneous
  turbulent flows}.  \jt{Atmospheric Turbulence and Radio Wave Propagation} .

\bibitem[Lumley(1970)]{lumley1970stochastic}
{\sc \au{Lumley, John~L}} \yr{1970}  \bt{Stochastic tools in turbulence. volume
  12. applied mathematics and mechanics}. {\em Tech. Rep.\/}.  \org{Penn State
  University Department of Aerospace Engineering}.

\bibitem[Marusic {\em et~al.\/}(2017)Marusic, Baars \&
  Hutchins]{marusic2017scaling}
{\sc \au{Marusic, I.}, \au{Baars, W.J.} \& \au{Hutchins, N.}} \yr{2017}
  \at{Scaling of the streamwise turbulence intensity in the context of
  inner-outer interactions in wall turbulence}.  \jt{Physical Review Fluids}
  \bvol{2}~(10),  \pg{100502}.

\bibitem[McKeon {\em et~al.\/}(2013)McKeon, Sharma \&
  Jacobi]{mckeon2013experimental}
{\sc \au{McKeon, BJ}, \au{Sharma, AS} \& \au{Jacobi, Ian}} \yr{2013}
  \at{Experimental manipulation of wall turbulence: a systems approach}.
  \jt{Physics of Fluids}  \bvol{25}~(3),  \pg{031301}.

\bibitem[McKeon \& Sharma(2010)]{mckeon2010critical}
{\sc \au{McKeon, B.~J.} \& \au{Sharma, A.~S.}} \yr{2010}  \at{A critical-layer
  framework for turbulent pipe flow}.  \jt{Journal of Fluid Mechanics}
  \bvol{658},  \pg{336--382}.

\bibitem[Monty {\em et~al.\/}(2009)Monty, Hutchins, Ng, Marusic \&
  Chong]{monty2009comparison}
{\sc \au{Monty, J.P.}, \au{Hutchins, N.}, \au{Ng, H.C.H.}, \au{Marusic, I.} \&
  \au{Chong, M.S.}} \yr{2009}  \at{A comparison of turbulent pipe, channel and
  boundary layer flows}.  \jt{Journal of Fluid Mechanics}  \bvol{632},
  \pg{431--442}.

\bibitem[Morra {\em et~al.\/}(2019)Morra, Sasaki, Hanifi, Cavalieri \&
  Henningson]{morra2019realizable}
{\sc \au{Morra, P.}, \au{Sasaki, K.}, \au{Hanifi, A.}, \au{Cavalieri, A.V.G.}
  \& \au{Henningson, D.}} \yr{2019}  \at{A realizable data-driven approach to
  delay bypass transition with control theory}.  \jt{arXiv preprint
  arXiv:1902.05049} .

\bibitem[Nogueira {\em et~al.\/}(2020)Nogueira, Morra, Martini, Cavalieri \&
  Henningson]{nogueira2020forcing}
{\sc \au{Nogueira, Petr{\^o}nio~AS}, \au{Morra, Pierluigi}, \au{Martini,
  Eduardo}, \au{Cavalieri, Andr{\'e}~VG} \& \au{Henningson, Dan~S}} \yr{2020}
  \at{Forcing statistics in resolvent analysis: application in minimal
  turbulent couette flow}.  \jt{arXiv preprint arXiv:2001.02576} .

\bibitem[Picard \& Delville(2000)]{picard2000pressure}
{\sc \au{Picard, C.} \& \au{Delville, J.}} \yr{2000}  \at{Pressure velocity
  coupling in a subsonic round jet}.  \jt{International Journal of Heat and
  Fluid Flow}  \bvol{21}~(3),  \pg{359--364}.

\bibitem[Pickering {\em et~al.\/}(2019)Pickering, Rigas, Schmidt, Sipp \&
  Colonius]{pickering2019eddy}
{\sc \au{Pickering, Ethan}, \au{Rigas, Georgios}, \au{Schmidt, Oliver},
  \au{Sipp, Denis} \& \au{Colonius, Tim}} \yr{2019}  \at{Eddy viscosity for
  resolvent analysis of turbulent jets}.  \jt{Bulletin of the American Physical
  Society}  \bvol{64}.

\bibitem[Sasaki {\em et~al.\/}(2019)Sasaki, Morra, Cavalieri, Hanifi \&
  Henningson]{sasaki2019role}
{\sc \au{Sasaki, K.}, \au{Morra, P.}, \au{Cavalieri, A.V.G.}, \au{Hanifi, A.}
  \& \au{Henningson, D.}} \yr{2019}  \at{On the role of actuation for the
  control of streaky structures in boundary layers}.  \jt{arXiv preprint
  arXiv:1902.04923} .

\bibitem[Schmidt \& Colonius(2020)]{schmidt2020guide}
{\sc \au{Schmidt, Oliver~T} \& \au{Colonius, Tim}} \yr{2020}  \at{Guide to
  spectral proper orthogonal decomposition}.  \jt{AIAA Journal}  \pg{pp.
  1--11}.

\bibitem[Schoppa \& Hussain(2002)]{schoppa2002coherent}
{\sc \au{Schoppa, W} \& \au{Hussain, Fazle}} \yr{2002}  \at{Coherent structure
  generation in near-wall turbulence}.  \jt{Journal of fluid Mechanics}
  \bvol{453},  \pg{57--108}.

\bibitem[Sirovich(1987)]{sirovich1987turbulence}
{\sc \au{Sirovich, Lawrence}} \yr{1987}  \at{Turbulence and the dynamics of
  coherent structures. i. coherent structures}.  \jt{Quarterly of Applied
  Mathematics}  \bvol{45}~(3),  \pg{561--571}.

\bibitem[Smith \& Metzler(1983)]{smith1983characteristics}
{\sc \au{Smith, CR} \& \au{Metzler, SP}} \yr{1983}  \at{The characteristics of
  low-speed streaks in the near-wall region of a turbulent boundary layer}.
  \jt{Journal of Fluid Mechanics}  \bvol{129},  \pg{27--54}.

\bibitem[Smits {\em et~al.\/}(2011)Smits, McKeon \& Marusic]{smits2011high}
{\sc \au{Smits, A.J.}, \au{McKeon, B.J.} \& \au{Marusic, I.}} \yr{2011}
  \at{High--reynolds number wall turbulence}.  \jt{Annual Review of Fluid
  Mechanics}  \bvol{43},  \pg{353--375}.

\bibitem[Towne {\em et~al.\/}(2020)Towne, Lozano-Dur{\'a}n \&
  Yang]{towne2020resolvent}
{\sc \au{Towne, Aaron}, \au{Lozano-Dur{\'a}n, Adri{\'a}n} \& \au{Yang, Xiang}}
  \yr{2020}  \at{Resolvent-based estimation of space--time flow statistics}.
  \jt{Journal of Fluid Mechanics}  \bvol{883}.

\bibitem[Towne {\em et~al.\/}(2018)Towne, Schmidt \&
  Colonius]{towne2018spectral}
{\sc \au{Towne, Aaron}, \au{Schmidt, Oliver~T} \& \au{Colonius, Tim}} \yr{2018}
   \at{Spectral proper orthogonal decomposition and its relationship to dynamic
  mode decomposition and resolvent analysis}.  \jt{Journal of Fluid Mechanics}
  \bvol{847},  \pg{821--867}.

\bibitem[Trefethen {\em et~al.\/}(1993)Trefethen, Trefethen, Reddy \&
  Driscoll]{trefethen1993hydrodynamic}
{\sc \au{Trefethen, Lloyd~N}, \au{Trefethen, Anne~E}, \au{Reddy, Satish~C} \&
  \au{Driscoll, Tobin~A}} \yr{1993}  \at{Hydrodynamic stability without
  eigenvalues}.  \jt{Science}  \bvol{261}~(5121),  \pg{578--584}.

\bibitem[Waleffe(1997)]{waleffe1997self}
{\sc \au{Waleffe, Fabian}} \yr{1997}  \at{On a self-sustaining process in shear
  flows}.  \jt{Physics of Fluids}  \bvol{9}~(4),  \pg{883--900}.

\end{thebibliography}

\end{document}